\documentclass[runningheads]{llncs}
\usepackage{graphicx}
\usepackage{subcaption}
\usepackage{xcolor}
\usepackage{booktabs}
\newcommand{\Fig}[1]{Fig.~\ref{fig:#1}}
\newcommand{\Figure}[1]{Figure~\ref{fig:#1}}

\newcommand{\Sec}[1]{Sec.~\ref{sec:#1}}
\newcommand{\Section}[1]{Section~\ref{sec:#1}}

\usepackage{amsmath}
\usepackage{bbold}
\renewcommand\vec{\mathbf}
\newcommand{\mat}{\mathbf}

\DeclareMathOperator*{\argmin}{argmin}

\newcommand{\real}{\mathbb{R}}
\newcommand{\complex}{\mathbb{C}}

\usepackage{enumitem}
\setlist[itemize]{noitemsep, topsep=0pt}
\setlist[enumerate]{nosep}

\usepackage{multirow}
\usepackage{graphicx}

\usepackage{sidecap}

\usepackage{float}

\usepackage[flushleft]{threeparttable}

\usepackage{amssymb}
\usepackage{bm}

\usepackage{pifont}
\newcommand{\xmark}{\ding{55}}%
\newcommand{\crossmark}{\xmark}
\usepackage{amssymb}

\usepackage[linesnumbered,ruled,vlined]{algorithm2e}
\usepackage {algpseudocode}
\usepackage{algorithmicx}
\usepackage{algcompatible}

\usepackage{hyperref}

\begin{document}
\title{Deep End-to-end Adaptive k-Space Sampling, Reconstruction, and Registration \\for Dynamic MRI}
\titlerunning{E2E Adaptive Sampling, Reconstruction, and Registration for Dynamic MRI}

%
\author{George Yiasemis \inst{1,2},
Jan-Jakob Sonke\inst{1,2},
Jonas Teuwen\inst{1,2,3,}\thanks{Corresponding author. Email: \url{j.teuwen@nki.nl}}\\
  }

\authorrunning{G. Yiasemis et al.}
%
\institute{Netherlands Cancer Institute, Amsterdam, Netherlands \\
\and
University of Amsterdam,  Amsterdam, Netherlands\\
\and
Radboud University Medical Center, Nijmegen, Netherlands\\
}

\newcommand*{\thischapter}{.}
\maketitle              
\begin{abstract}
    Dynamic MRI enables a range of clinical applications, including cardiac function assessment, organ motion tracking, and radiotherapy guidance. However, fully sampling the dynamic $k$-space data is often infeasible due to time constraints and physiological motion such as respiratory and cardiac motion. This necessitates undersampling, which degrades the quality of reconstructed images. Poor image quality not only hinders visualization  but also impairs the estimation of deformation fields, crucial for registering dynamic (moving) images to a static reference image. This registration enables tasks such as motion correction, treatment planning, and quantitative analysis, particularly in applications like cardiac imaging and MR-guided radiotherapy. To overcome the challenges posed by undersampling and motion, we introduce an end-to-end deep learning (DL) framework that integrates adaptive dynamic $k$-space sampling, reconstruction, and registration. Our approach begins with a  DL-based adaptive sampling strategy, optimizing dynamic $k$-space acquisition to capture the most relevant data for each specific case. This is followed by a DL-based reconstruction module that produces images optimized for accurate deformation field estimation from the undersampled moving data. Finally, a registration module estimates the deformation fields aligning the reconstructed dynamic images with a static reference. The proposed framework is independent of specific reconstruction and registration modules allowing for plug-and-play integration of these components. The entire framework is jointly trained using a combination of supervised and unsupervised loss functions, enabling end-to-end optimization for improved performance across all components. Through controlled experiments and ablation studies, we validate each component, demonstrating that each choice contributes to robust motion estimation from undersampled dynamic data.

\end{abstract}

\newpage

\section{Introduction}
\label{sec:chapter8:sec1}

\begin{figure}[!hbt]
    \centering
    \includegraphics[width=\columnwidth]{\thischapter/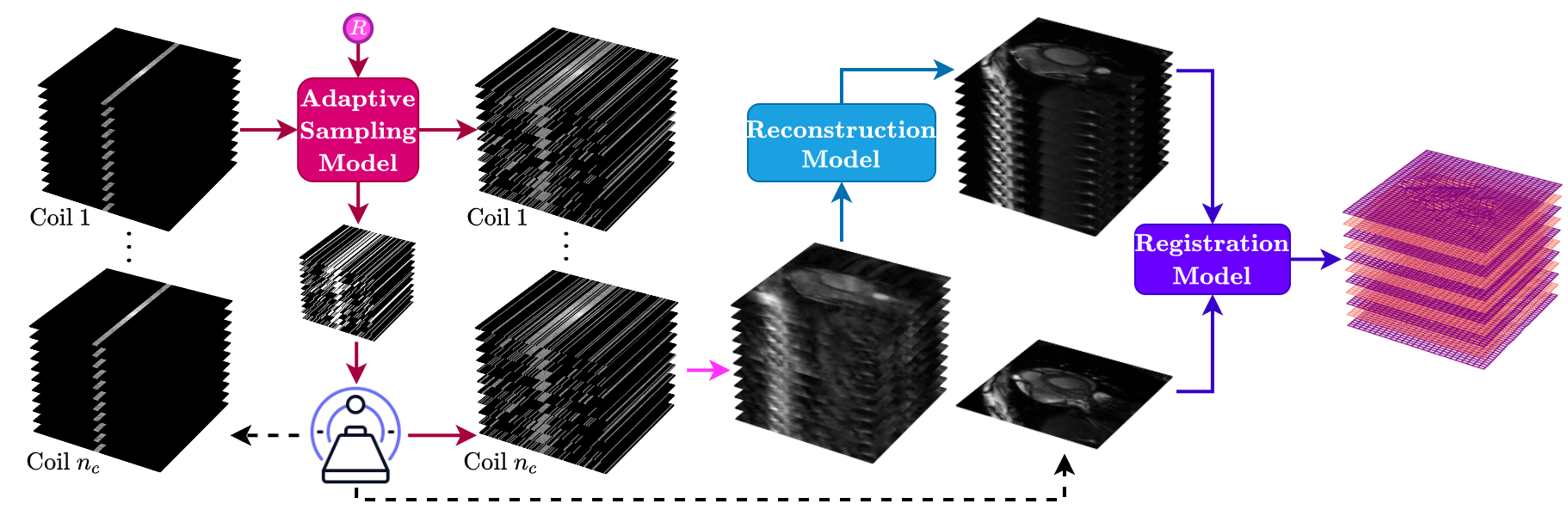}
    \caption{
    Overview of our proposed pipeline: Utilizing initial dynamic (moving) $k$-space data, an adaptive sampling model generates a pattern tailored to a specific acceleration factor. This pattern is used to obtain undersampled moving $k$-space measurements, which are then fed into a reconstruction model to recover a high-fidelity moving image. Finally, the reconstructed dynamic image and a reference image are input into a registration model to predict motion patterns as registration fields.
    }
    \label{fig:chapter8:overview}
\end{figure}

Magnetic Resonance Imaging (MRI) is an essential imaging modality due to its non-invasive nature, absence of ionizing radiation, and exceptional soft tissue contrast.  Dynamic MRI serves a wide range of applications, including real-time tasks such as cardiac function assessment, motion tracking of various organs, and adaptive treatments in radiotherapy, making it indispensable in both diagnostic and treatment guidance. However, a major limitation of dynamic MRI is its inherently slow acquisition process, which becomes particularly challenging in real-time scenarios and is further exacerbated by voluntary or involuntary physiological motions such as respiration \cite{bertelsen2022effect} and heartbeat \cite{ismail2022cardiac}.

Such motion can interfere with the ability to fully sample the $k$-space—the frequency domain data needed to reconstruct high-fidelity images in dynamic scenarios. To achieve artifact-free image reconstruction, the Nyquist sampling criterion (NC) \cite{1697831} must be met, which specifies the minimum amount of data required to avoid aliasing and loss of fine detail. Although fully sampling the $k$-space per NC ensures spatial accuracy, it is often impossible. Thus, accelerating MRI acquisition by undersampling the $k$-space (below NC) becomes the only feasible approach. Even if dynamic acquisition is attainable in some cases, such as ECG-triggered cardiac MRI acquisition \cite{cmrxrecon}, undersampling can decrease operational costs and allow for more 
MRI scans.

Advancements in deep learning (DL) have markedly enhanced the reconstruction of undersampled MRI \cite{Pal2022-ce,chen2022ai,bioengineering10030334}, demonstrating significant advantages over traditional methods like compressed sensing (CS) \cite{4472246,gamper2008compressed}.  These methods leverage data-driven, often convolution-based, models to reconstruct high-quality images from limited $k$-space data, achieving superior reconstructions at high accelerations, and more efficient computations than CS.

Beyond reconstruction, DL methods have emerged for adaptively (under)sampling the MRI acquisition showing improved performance over non-adaptive learned, fixed or random sampling schemes. Although most methods have been explored in the static case, only one DL approach has been devised specifically for dynamic MRI \cite{yiasemis2024endtoendadaptivedynamicsubsampling}.

Nonetheless, undersampling the $k$-space inherently compromises image quality, which complicates the preservation of fine spatial details and may limit the accuracy of subsequent motion estimation and registration processes— essential for dynamic MRI applications. In cardiac MRI, for example, registration aligns images across phases to support motion correction, segmentation, etc, for comprehensive functional and structural assessment. \cite{Khalil2018}. In adaptive radiotherapy, MRI-based registration allows precise tumor motion tracking, thereby enabling accurate dose delivery while minimizing exposure to adjacent healthy tissues \cite{HUNT2018711}.

Several traditional and DL-based methods exist for image registration in medical imaging, with the goal of aligning a moving image to a reference frame and estimating motion \cite{kostelec2003image,balakrishnan2019voxelmorph,chen2021deep,chen2022transmorph,sun2024medical}. Within the scope of accelerated MRI, some approaches have combined registration with a reconstruction step by co-optimizing both processes \cite{corona2021variational,terpstra2021real,TERPSTRA2022102509}, while others employ learned motion models that bypass direct reconstruction 
\cite{yang2024robust}.

Motivated by the need for precise motion estimation in undersampled dynamic MRI, we introduce the first end-to-end DL-based framework that integrates adaptive undersampling, reconstruction, and registration for dynamic MRI. An overview of our method is depicted in \Fig{chapter8:overview}. Our contributions are as follows:

\begin{itemize}
    \item We introduce a novel end-to-end pipeline that integrates adaptive undersampling, reconstruction, and registration for dynamic MRI—a comprehensive approach that, to our knowledge, has not been explored in any similar DL-based or traditional optimization framework.
    \item We present a detailed methodology covering all components of our approach, ensuring reproducibility for each module. We will also make our code available. 
    \item We rigorously evaluate our framework under various configurations. We validate our approach on a cardiac cine dataset and demonstrate its generalizability on out-of-distribution data (aorta dataset).
\end{itemize}

The rest of the paper is organized as follows: \Section{chapter8:sec2} covers background and related work, \Section{chapter8:sec3} details our methodology, \Section{chapter8:sec4} presents  our experimental setup and results, and \Section{chapter8:sec5} offers a discussion and conclusions.
\section{Background and Related work}
\label{sec:chapter8:sec2}

\subsection{Dynamic MRI Reconstruction}
\label{sec:chapter8:sec2.1}

Dynamic MRI reconstruction aims to recover a moving image $\vec{x} \in \mathbb{C}^{n \times n_t}$ from acquired dynamic $k$-space data, where $n = n_x \times n_y$ denotes the spatial dimensions, $n_c$ represents the number of scanner coils, and $n_t$ the temporal dimension of the sequence.
\noindent
Given fully sampled data $\vec{y} \in \mathbb{C}^{n \times n_c \times n_t}$, the underlying image sequence can be obtained by applying the inverse Fast Fourier Transform $\mathcal{F}^{-1}$, followed by the root-sum-of-squares (RSS) as follows:

\begin{equation}
\begin{gathered}
    \vec{x}_{\cdot, \tau} := \text{RSS} \circ \mathcal{F}^{-1}(\vec{y}_{\cdot, \cdot, \tau}), \quad \tau = 1, \cdots, n_t,\\
    \text{RSS}(\vec{w}) = \big( \sum_{k=1}^{n_c} |\vec{w}_{\cdot, k}|^2 \big)^{1/2} \in \real^{n}, \quad \vec{w} \in \complex^{n \times n_c}.
\end{gathered}
\end{equation}
Here, $|\vec{a}| := \sqrt{\text{real}(\vec{a})^2 + \text{imag}(\vec{a})^2}$, denotes the modulus operator.
To reduce the MRI acquisition time, undersampling is applied to the $k$-space: $\tilde{\vec{y}}^{\mat{M}} := \mat{M} (\vec{y})$. Undersampling is characterized by a binary mask operator $\mat{M} = \left(\mat{M}_1,\cdots\mat{M}_{n_t}\right) \in \{0,1\}^{n \times n_t}$, which zeros-out non-acquired $k$-space samples as follows:

\begin{equation}
\begin{gathered}
    \tilde{\vec{y}}^{\mat{M}}_{i,k,\tau} =  \big(\mat{M} (\vec{y})\big)_{i,k,\tau} := \begin{cases} 
    \vec{y}_{i, k,\tau} & \text{if } (\mat{M}_{\tau})_i = 1 \\
    0, & \text{if } (\mat{M}_{\tau})_i = 0,
\end{cases}\\
    i=1,\cdots,n, \quad k=1,\cdots,n_c, \quad \tau = 1, \cdots, n_t.
\end{gathered}
\end{equation}
The resulting forward model for each frame $\tau$ of the sequence  is described by the forward operator $ \mathcal{T}_{\mat{M}_{\tau}, \mat{S}_{\tau}}$:

\begin{equation}
    \tilde{\vec{y}}^{\mat{M}}_{\cdot,\cdot,\tau} \, = \,  \mathcal{T}_{\mat{M}_{\tau}, \mat{S}_{\tau}}(\vec{x}_{\tau}) \, := \,   \mat{M}_{\tau} \circ \mathcal{F} \circ \mathcal{C}_{\mat{S}_{\tau}} (\vec{x}_{\cdot,\tau}),
    \label{eq:chapter8:forward}
\end{equation}
where $\mathcal{C}_{\mat{S}_{\tau}}$ denotes the coil sensitivity-encoding operator, which decomposes an image into individual coil images using coil sensitivity profiles $\mathbf{S}_{\tau} \in \mathbb{C}^{n^2 \times n_c}$ that represent the spatial sensitivity of each coil. This is expressed as:

\begin{equation}
    \mathcal{C}_{\mat{S}_{\tau}}(\vec{z}) \, = \, \left(\mat{S}_{\tau}^1\vec{z},\cdots,\mat{S}_{\tau}^{n_c}\vec{z}\right), \quad \text{for} \quad \vec{z} \in \complex^{n}.
\end{equation}
A reconstruction for \eqref{eq:chapter8:forward} can be formulated as a solution to a regularized least squares optimization problem:

\begin{equation}
    \argmin_{\vec{x}' \in \mathbb{C}^{n \times n_t}} \sum_{\tau=1}^{n_t} \frac{1}{2}\left| \left| \mathcal{T}_{\mat{M}_{\tau}, \mat{S}_{\tau}} (\vec{x}_{\cdot, \tau}') - \tilde{\vec{y}}_{\cdot, \cdot, \tau}^{\mat{M}} \right| \right|_2^2 + \mathcal{H}(\vec{x}'),
\label{eq:chapter8:inverse_problem}
\end{equation}
where $\mathcal{H}$ denotes an arbitrary regularization functional, which introduces prior domain knowledge. 

\subsection{Deep Learning-based MRI Reconstruction}

Deep learning methods have been widely applied to address accelerated MRI, bypassing the need for handcrafted priors and computationally intensive optimization techniques in CS. DL methods primarily fall into two categories: direct mapping and unrolled optimization approaches \cite{chen2022ai}. Direct mapping DL methods take undersampled images as input and learn to output reconstructed images, a straightforward approach that typically requires a substantial amount of training data. In contrast, unrolled optimization methods are inspired by CS and leverage algorithms like gradient descent \cite{Hammernik2017,Sriram2020,Yiasemis_2022_CVPR}, primal-dual descent \cite{adler2018learned}, conjugate gradient \cite{aggarwal2018modl}, and ADMM \cite{yiasemis2025vsharp} to iteratively solve \eqref{eq:chapter8:inverse_problem}. 

The availability of open-access $k$-space datasets from challenges such as fastMRI \cite{knoll2020advancing}, Multi-Coil MRI Reconstruction \cite{10.3389/fnins.2022.919186}, and the recent CMRxRecon 2023 and 2024 \cite{cmrxrecon2023,wang2024cmrxrecon2024multimodalitymultiviewkspace}, has empowered the community to pursue data-driven approaches. Consequently, numerous DL-based reconstruction techniques have been developed, primarily for static MRI acquisitions \cite{eo2018kiki,Sriram2020,Yiasemis_2022_CVPR}, with a recent growing expansion into dynamic settings \cite{yoo2021time,waddington2022real,yiasemis2024deep,lonning2024dynamic}.

\subsection{Adaptive MRI Acquisition}

Several methods have been proposed to replace standard sampling patterns (e.g., equidistant, random uniform, Gaussian \cite{yiasemis2023retrospective}) in DL-based MRI reconstruction with learned approaches. These methods include training dataset-based sampling optimization \cite{Bahadir2019,zhang2020extending,shor2023multi} and adaptive, case-specific patterns \cite{bakker2022on,gautam2024patientadaptive} jointly learned with reconstruction networks. While most focus on static MRI, \cite{yiasemis2024endtoendadaptivedynamicsubsampling} extends adaptive sampling to dynamic (2D + time) imaging through an end-to-end approach for adaptive dynamic undersampling and reconstruction (E2E-ADS-Recon).

Unlike previous methods optimized for single acceleration factors, this approach is proposed to be flexible across various acceleration factors. It includes two possible configurations for adapted undersampling: a phase-specific approach, where distinct patterns are generated for each temporal phase, and a unified approach, applying a single pattern across all phases.

\subsection{Deformable Image Registration}
\label{sec:chapter8:sec2.3}

Deformable (non-rigid) registration aims to align a moving image  $\vec{z}_{\text{mov}} \in \mathbb{R}^{n \times n_t}$, to a fixed, static reference image $\vec{z}_{\text{ref}} \in \mathbb{R}^n$, by estimating spatial deformation fields $\boldsymbol{\phi} = \left(\boldsymbol{\phi}_{1}, \cdots, \boldsymbol{\phi}_{n_t} \right) \in \mathbb{R}^{2 \times n \times n_t}$. 

The displacement field $\boldsymbol{\phi}_{\tau}$ maps the coordinates of $\vec{z}_{\text{ref}}$  to those of $(\vec{z}_{\text{mov}})_{\tau}$, achieved through a warping operation, $\mathcal{W}: \mathbb{R}^{n } \times \mathbb{R}^{2 \times n } \rightarrow \mathbb{R}^{n }$\cite{balakrishnan2019voxelmorph}, enabling the registration of $\vec{z}_{\text{mov}}$:

\begin{equation}
    \vec{z}_{\text{reg}} = \Big\{ \mathcal{W} \big( \left(\vec{z}_{\text{mov}}\right)_{\cdot, \tau}, \boldsymbol{\phi}_{\tau}\big)
    \Big\}_{\tau=1}^{n_t} \in \real^{n \times n_t}.
    \label{eq:chapter8:reg_loss_definition}
\end{equation}

The objective is to determine an optimal deformation field, $\boldsymbol{\phi}^*$, that aligns the moving image with the reference accurately. This is often posed as an unsupervised minimization problem using a similarity measure $\mathcal{L}_{\text{sim}}$:

\begin{equation}
    \boldsymbol{\phi}^* := \arg\min_{\boldsymbol{\phi}} \, \mathcal{L}_{\text{sim}} \left(\vec{z}_{\text{reg}}, \vec{z}_{\text{ref}}\right).
\end{equation}

Several approaches also leverage complementary (supervised) tasks to enhance the registration process, incorporating segmentation accuracy losses when ground truth delineations of target organs are available \cite{hering2018enhancinglabeldrivendeepdeformable,balakrishnan2019voxelmorph}. 

\section{Methods}
\label{sec:chapter8:sec3}

\subsection{Deep Learning Framework}
\label{sec:chapter8:subsec3.1}

\subsubsection{Sensitivity Profiles Estimation}
\label{sec:chapter8:subsubsec3.1.1}

Following \cite{Sriram2020}, we begin by estimating sensitivity profiles for the multi-coil data, denoted as $\tilde{\vec{S}} \in \mathbb{C}^{n^2 \times n_c \times n_t}$, using a fully-sampled central region of the $k$-space, $\tilde{\vec{y}}_{\text{mov}}^{\mat{M}^\text{acs}}$ corresponding to low frequencies—the autocalibration signal (ACS). Consistent with \cite{Sriram2020}, these estimates are then input into a DL-based model, specifically a two-dimensional U-Net \cite{Ronneberger2015}, denoted as $\mathcal{S}_{\boldsymbol{\sigma}}$, which is trained to refine them:

\begin{equation}
    \mat{S}_{\tau}^{k} :=  \mathcal{S}_{\boldsymbol{\sigma}}(\tilde{\mat{S}}_{\tau}^{k}), \quad \tau = 1, \cdots, n_t, \quad k = 1, \cdots, n_c.
\end{equation}
The refined sensitivity profiles are subsequently normalized to satisfy 

\begin{equation}
    \sum_{k=1}^{n_c} (\mat{S}_{\tau}^{k})^* \mat{S}_{\tau}^{k} = \mathbf{I}_{n} \in \real^{n\times n}.
\end{equation}

\subsubsection{Adaptive Sampling Model}
\label{sec:chapter8:subsubsec3.1.2}
\begin{figure}[!ht]
    \centering
    \includegraphics[width=\columnwidth]{\thischapter/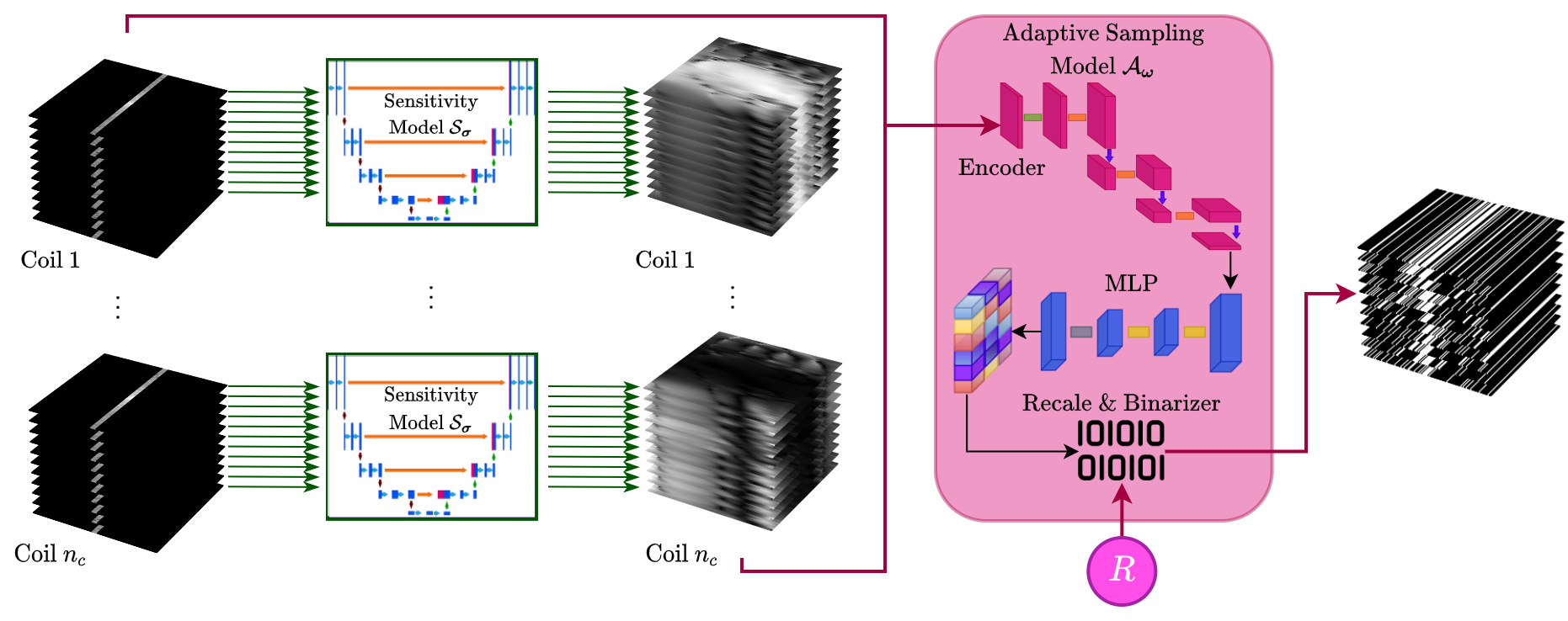}
    \caption{Adaptive Sampling Model ($\mathcal{A}_{\boldsymbol{\omega}}$) framework. Sensitivity profiles from the sensitivity model ($\mathcal{S}_{\boldsymbol{\sigma}}$) and initial undersampled moving data ($\tilde{\vec{y}}_{\text{mov}}^{\mat{M}^0}$) are processed by a U-Net-like encoder followed by a MLP, generating sampling probabilities. These are rescaled for the specified acceleration $R$ and binarized via a straight-through estimator creating an adapted binary dynamic sampling mask. Here we assume $\tilde{\vec{y}}_{\text{mov}}^{\mat{M}^0} = \tilde{\vec{y}}_{\text{mov}}^{\mat{M}^\text{acs}}$.}
    \label{fig:chapter8:ads_model}
\end{figure}

An adaptive dynamic sampling (ADS) module, inspired by the E2E-ADS-Recon framework \cite{yiasemis2024endtoendadaptivedynamicsubsampling}, denoted as $\mathcal{A}_{\boldsymbol{\omega}}$, takes as input initially undersampled moving data $\tilde{\vec{y}}_{\text{mov}}^{\mat{M}^0}$, acquired with an initial mask $\mat{M}^0$, along with sensitivity profiles $\vec{S}$ and a target acceleration factor $R$. ADS consists of cascades of 3D U-Net-like encoders and multi-layer perceptrons (MLPs) to generate sampling selection probabilities, which are binarized using a straight-through estimator \cite{bengio2013estimating,yin2021end} to match the sampling budget and enable adaptive sampling according to the target acceleration $R$. 

This module outputs a binary mask operator  $\mat{M} = \left(\mat{M}_1, \cdots, \mat{M}_{n_t}\right)$, which guides the subsequent dynamic image reconstruction:
\begin{equation}
    \mat{M} := \mathcal{A}_{\omega} (\tilde{\vec{y}}_{\text{mov}}^{\mat{M}^0}; \mat{S}, R).
\end{equation}
By default, the ACS data $\tilde{\vec{y}}_{\text{mov}}^{\mat{M}^0} = \tilde{\vec{y}}_{\text{mov}}^{\mat{M}^{\text{acs}}}$ initializes $\mathcal{A}_{\omega}$, unless otherwise specified.

\subsubsection{Reconstruction Model}
\label{sec:chapter8:subsubsec3.1.3}

Our proposed pipeline (see \Sec{chapter8:subsec3.2}) is agnostic to the choice of reconstruction network, supporting any DL-based dynamic reconstruction model. For the main experiments, we use the variable Splitting Half-quadratic ADMM algorithm for Reconstruction of inverse-Problems (vSHARP) \cite{yiasemis2025vsharp}, a model that demonstrated competitive performance in both reconstruction quality and speed at the CMRxRecon challenge 2023 \cite{cmrxrecon2023}. The vSHARP model unrolls a DL-based optimization over $T$ iterations. Denoted as $ \mathcal{V}_{\boldsymbol{\theta}}$, it accepts an undersampled dynamic image $\tilde{\vec{x}}_{\text{mov}}^{\mat{M}}$ and sensitivity maps $\mat{S}$ to yield a refined dynamic reconstruction:

\begin{equation}
\begin{gathered}
        \hat{\vec{x}}_{\text{mov}} = \mathcal{V}_{\boldsymbol{\theta}}( \tilde{\vec{x}}_{\text{mov}}^{\mat{M}}; \mat{S}), \, (\tilde{\vec{x}}_{\text{mov}}^{\mat{M}})_{\cdot, \tau} = \mathcal{R}_{\mat{S}_{\tau}} \circ \mathcal{F}^{-1}(\tilde{\vec{y}}_{\text{mov}}^{\mat{M}})_{\cdot,\cdot,\tau}.
\end{gathered}
\label{eq:chapter8:reduce}
\end{equation}
The operator $\mathcal{R}_{\mat{S}_{\tau}}: \complex^{n \times n_c} \rightarrow \complex^{n}$ represents the coil-combining operator using $\mat{S}_{\tau}$:
\begin{equation}
   \mathcal{R}_{\mat{S}_{\tau}}(\vec{v}) := \sum_{k=1}^{n_c} (\mat{S}_{\tau}^{k})^* \vec{v}_{\cdot, k} \in \complex^{n} , \quad \text{for} \quad \vec{v} \in \complex^{n \times n_c}.
\end{equation}
Additional details on the reconstruction network are provided in Appendix \ref{sec:chapter8:appendix:appendix1}.

\subsubsection{Registration - Motion Estimation Model}
\label{sec:chapter8:subsubsec3.1.4}
\begin{figure}[!h]
    \centering
    \includegraphics[width=\columnwidth]{\thischapter/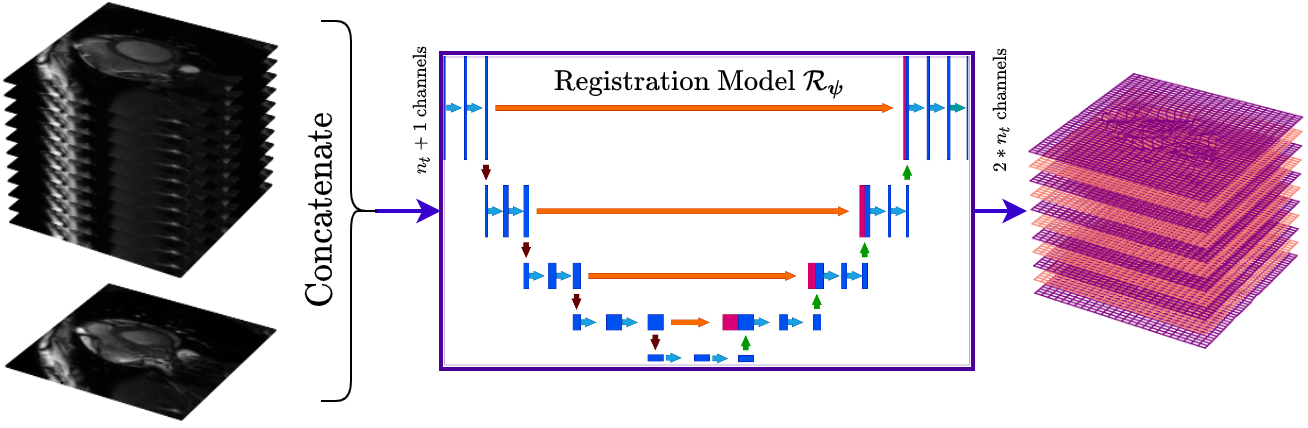}
    \caption{A 2D U-Net plays the role of the registration model. It takes as input the reconstructed moving and reference images, concatenated along the channel dimension. The registration model outputs a sequence of registration fields corresponding to the displacements/motion between the moving and reference images.}
    \label{fig:chapter8:reg_model}
\end{figure}

As with \Sec{chapter8:subsubsec3.1.3}, any deformable image registration method—DL-based or otherwise—can be employed to estimate deformation fields for each temporal phase of the moving image relative to a reference image. For the main experiments, we use a lightweight approach by using a two-dimensional U-Net \cite{Ronneberger2015}, denoted as $\mathcal{R}_{\boldsymbol{\psi}}$, which processes a concatenated along the channel dimension input of the (reconstructed) moving image and the reference image $\vec{x}_{\text{ref}} \in \mathbb{R}^{n}$, rather than using, for instance, a 3D U-Net. This setup configures the initial layer to accept $n_t + 1$ input channels, with the model outputting  $2 \times n_t$ channels, representing the deformation fields for each temporal phase of the moving image relative to the reference image:
\begin{equation}
    \boldsymbol{\phi} = \mathcal{R}_{\boldsymbol{\psi}}(\vec{w}), \quad \vec{w} = \big[|\hat{\vec{x}}_{\text{mov}}|, \vec{x}_{\text{ref}}\big] \in \real^{n\times (n_t + 1)}.
\end{equation}

\subsection{End-to-End Framework}
\label{sec:chapter8:subsec3.2}

\begin{figure*}[!hbt]
    \centering
    \includegraphics[width=0.9\textwidth]{\thischapter/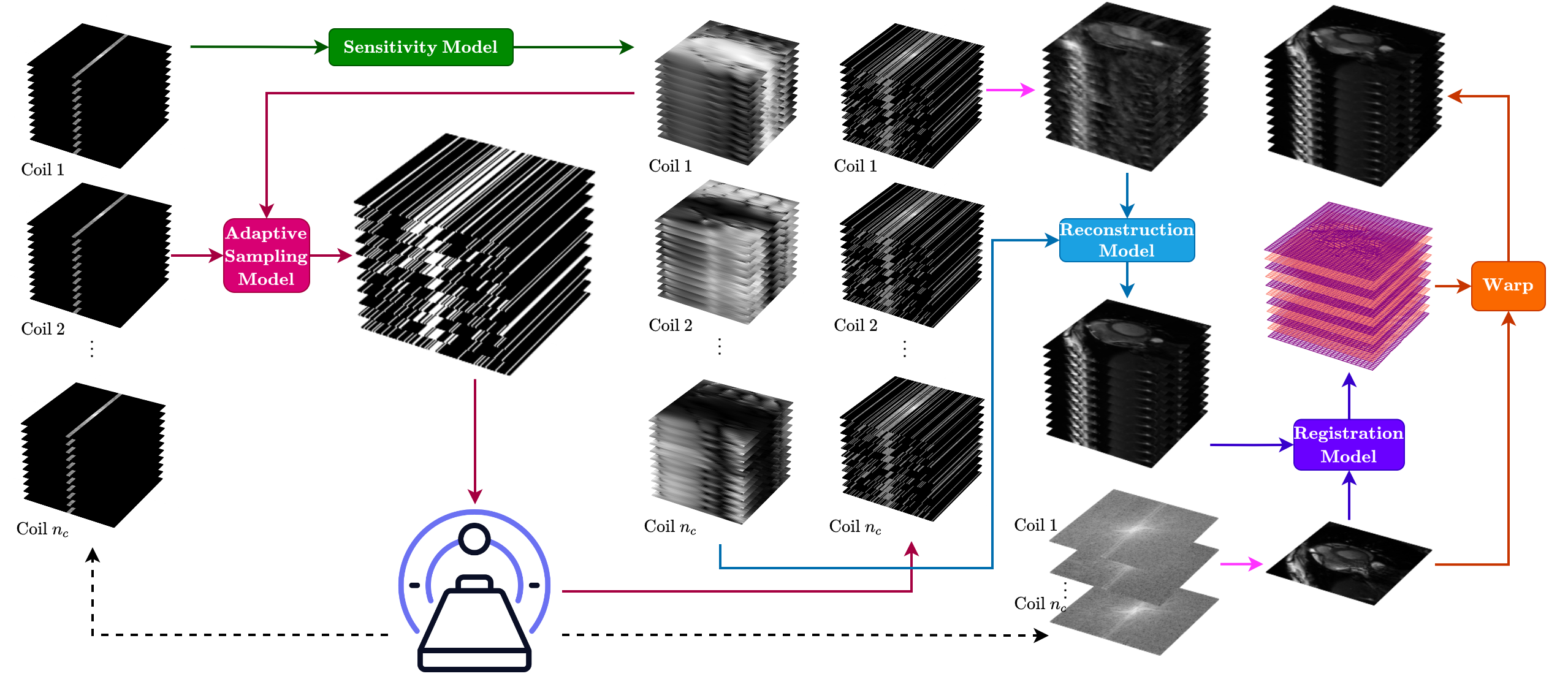}
     \caption{End-to-end pipeline of the proposed method. The process starts with coil sensitivity profile estimation from ACS data. Next, the adaptive sampling model selects optimized sampling patterns based on initial moving data (here equivalent to ACS data) and sensitivities. The adaptively acquired data are then input into the reconstruction model. Finally, the registration model outputs deformation fields that align the reconstructed sequence to a reference accounting for motion. A registered output (through warping) is used for loss computation.}
    \label{fig:chapter8:diagram}
\end{figure*}

Our end-to-end framework incorporates sensitivity estimation, adaptive sampling, reconstruction, and registration in a sequential pipeline, as outlined in the previous sections. First, sensitivity profiles $\mat{S}$ are estimated and refined using  $\mathcal{S}_{\boldsymbol{\sigma}}$. These refined sensitivity maps $\mat{S}$, together with the initial undersampled $k$-space data $\tilde{\vec{y}}_{\text{mov}}^{\mat{M}^0}$, inform the adaptive sampling model $\mathcal{A}_{\boldsymbol{\omega}}$, which generates dynamic sampling patterns $\mat{M}$ based on the specified acceleration factor $R$. The adaptively sampled moving data $\tilde{\vec{y}}_{\text{mov}}^{\mat{M}}$ is then processed by the reconstruction model $\mathcal{V}_{\boldsymbol{\theta}}$, to obtain a high-quality dynamic image. Finally, the registration model $\mathcal{R}_{\boldsymbol{\psi}}$ calculates deformation fields to align the reconstructions with a reference frame, correcting for motion and ensuring temporal consistency across frames. The full pipeline is presented in Algorithm \ref{alg:e2ealgo} and illustrated in \Figure{chapter8:diagram}.
\setlength{\textfloatsep}{5pt} 
\begin{algorithm}[!htb]
\footnotesize	
\caption{End-to-end Adaptive Sampling, Reconstruction and Registration for Dynamic MRI}
\label{alg:e2ealgo}
\DontPrintSemicolon
\SetAlgoLined
\KwIn{$\tilde{\vec{y}}_{\text{mov}}^{\mat{M}^0}, \, \tilde{\vec{y}}_{\text{mov}}^{\mat{M}^\text{acs}} \in \complex^{n \times n_c \times n_t}$, $R$, $\vec{y}_{\text{ref}} \in \complex^{n \times n_c}$}
\KwOut{$\boldsymbol{\phi}\in \real^{2 \times n \times n_t}$, [Optional] $\vec{x}_{\text{reg}} \in  \real^{n \times n_t} $ 
}
\smallskip
\For{$\tau \gets 1$ \KwTo $n_t$}{
    \For{$k \gets 1$ \KwTo $n_c$}{
        Estimate $\tilde{\mat{S}}_{\tau}^{k}$ from $(\tilde{\vec{y}}_{\text{mov}}^{\mat{M}^\text{acs}})_{\cdot,k.\tau}$ as in \cite{Sriram2020}\;     
        $\mat{S}_{\tau}^{k} \gets \mathcal{S}_{\boldsymbol{\sigma}}(\tilde{\mat{S}}_{\tau}^{k})$  \tcp*{\scriptsize Refine Sensitivities}
         Normalize s.t. $\sum_{k\gets1}^{n_c} (\mat{S}_{\tau}^{k})^* \mat{S}_{\tau}^{k} \gets \mathbf{I}_{n} \in \real^{n \times n}$ \;
    }
}
\smallskip
$\mat{M} \gets \mathcal{A}_{\omega} (\tilde{\vec{y}}_{\text{mov}}^{\mat{M}^0}; \mat{S}, R)$ \hfill \tcp*{\scriptsize Adapt sampling pattern}
\smallskip
$\tilde{\vec{y}}_{\text{mov}}^{\mat{M}} \gets \mat{M} (\vec{y}_{\text{mov}})$ \tcp*{\scriptsize Sample moving data with $\mat{M}$}
\smallskip
\For{$\tau \gets 1$ \KwTo $n_t$}{
    $(\tilde{\vec{x}}_{\text{mov}}^{\mat{M}})_{\cdot, \tau} \gets \mathcal{R}_{\mat{S}_{\tau}} \circ \mathcal{F}^{-1}(\tilde{\vec{y}}_{\text{mov}}^{\mat{M}})_{\cdot,\cdot,\tau}$
}
\smallskip
$\hat{\vec{x}}_{\text{mov}} \gets \mathcal{V}_{\boldsymbol{\theta}}( \tilde{\vec{x}}_{\text{mov}}^{\mat{M}} ; \mat{S})$ \tcp*{\scriptsize Reconstruct moving image}
\smallskip
$\vec{x}_{\text{ref}} \gets \text{RSS} \circ \mathcal{F}^{-1}(\vec{y}_{\text{ref}})$\;
\smallskip
$\vec{w} \gets \Big[|\hat{\vec{x}}_{\text{mov}}|, \vec{x}_{\text{ref}} \Big]$ \tcp*{\scriptsize Concatenate along channel dim}
\smallskip
$\boldsymbol{\phi} \gets \mathcal{R}_{\boldsymbol{\psi}}(\vec{w})$ \tcp*{\scriptsize Compute registration field}
\smallskip 
[Optional]  \For{$\tau \gets 1$ \KwTo $n_t$}{
    $(\vec{x}_{\text{reg}})_{\cdot, \tau} \gets  \mathcal{W} \big( \left|\hat{\vec{x}}_{\text{mov}}\right|_{\cdot, \tau}, \boldsymbol{\phi}_{\tau}\big) $ \tcp*{\scriptsize Warp mov. image}
}
\end{algorithm}
\vspace{-10pt}

\subsection{Training Loss Function}
\label{sec:chapter8:subsec3.3}
With the end-to-end framework established, the primary objective is to optimize the registration task, enhancing the accuracy of the deformation fields. Although the reconstruction task is secondary, we expect that improving the image reconstruction indirectly supports better motion estimation by providing high-fidelity inputs for alignment.

Let $\Tilde{\vec{y}}_{\text{mov}}^{\mat{M}} \in \complex^{n \times n_c \times n_t } $ represent the undersampled moving data, acquired using the predicted sampling set $\mat{M}$ from $\mathcal{A}_{\omega}$. Also let $\vec{y}_{\text{mov}} \in \complex^{n \times n_c  \times n_t } $ and $\vec{y}_{\text{ref}} \in \complex^{ n \times n_c}$ denote the fully sampled (moving) and reference $k$-space measurements. The total loss we aim to optimize is given by:
\begin{equation}
        \mathcal{L}: = \alpha \mathcal{L}_{\text{rec}} + \beta \mathcal{L}_{\text{reg}},
        \label{eq:chapter8:total_loss}
\end{equation}
where $\alpha, \beta > 0$ weight the reconstruction and registration losses. The reconstruction loss is defined as:
\begin{equation}
    \begin{gathered}
      \mathcal{L}_{\text{rec}} = \mathcal{L}_{\text{ssim}}\left(|\hat{\vec{x}}_{\text{mov}}|, {\vec{x}}_{\text{mov}} \right) + \mathcal{L}_{1}(|\hat{\vec{x}}_{\text{mov}}|, {\vec{x}}_{\text{mov}}),
    \end{gathered}
    \label{eq:chapter8:recon_loss}
\end{equation}
with $\hat{\vec{x}}_{\text{mov}} = \mathcal{V}_{\boldsymbol{\theta}}(\Tilde{\vec{y}}_{\text{mov}}^{\mat{M}}; \mat{S})$ representing the predicted moving image and $\vec{x}_{\text{mov}} = \text{RSS} \circ \mathcal{F}(\vec{y}_{\text{mov}})$ denoting the ground truth. Here, $\mathcal{L}_{\text{ssim}} := \mathcal{L}_{\text{ssim2D}} + \mathcal{L}_{\text{ssim3D}}$ combines 2D and 3D structural similarity index measure (SSIM) losses \cite{yiasemis2025vsharp}, calculated over individual slices and across the sequence, respectively.

For the registration loss $\mathcal{L}_{\text{reg}}$,  we focus on aligning the registered reconstructed moving and reference images while ensuring a smooth deformation field. This includes a similarity term for image alignment and a smoothness term to constrain the registration field:

\begin{equation}
      \mathcal{L}_{\text{reg}} = \mathcal{L}_{\text{sim}}\left(
      \vec{x}_{\text{reg}}, \left\{{\vec{x}}_{\text{ref}}\right\}_{t=1}^{n_t} \right) + \mathcal{L}_{\text{smooth}}(\boldsymbol{\phi}),
    \label{eq:chapter8:reg_loss}
\end{equation}
where $\vec{x}_{\text{reg}} = \big\{ \mathcal{W} \big( \left| \hat{\vec{x}}_{\text{mov}} \right|_{\cdot, \tau}, \boldsymbol{\phi}_{\tau}\big) \big\}_{\tau=1}^{n_t} \in \real^{n \times n_t}$ and $\vec{x}_{\text{ref}} = \text{RSS} \circ \mathcal{F}(\vec{y}_{\text{ref}}) \in \real^{n}$. Note that $\left\{{\vec{x}}_{\text{ref}}\right\}_{t=1}^{n_t}  \in \real^{n \times n_t} $ denotes ${\vec{x}}_{\text{ref}}$ repeated $n_t$ times. For $\mathcal{L}_{\text{sim}}$, we apply the same loss functions as in \eqref{eq:chapter8:recon_loss}, and $\boldsymbol{\phi} = \mathcal{R}_{\boldsymbol{\psi}}(\big[|\hat{\vec{x}}_{\text{mov}}|, \vec{x}_{\text{ref}}\big])$ denotes the registration field prediction. The smoothness term, inspired by diffusion regularization \cite{commowick2008efficient,balakrishnan2019voxelmorph}, penalizes abrupt changes in $\boldsymbol{\phi}$:
\begin{equation}
    \mathcal{L}_{\text{smooth}}(\boldsymbol{\phi}) = \frac{1}{2 n n_t} \sum_{\tau}^{n_t} \sum_{\vec{p}} \left| \frac{\partial {\boldsymbol{\phi}_{\tau}}(\vec{p})}{\partial u_1} \right| + \left| \frac{\partial {\boldsymbol{\phi}_{\tau}}(\vec{p})}{\partial u_2} \right|.
\end{equation}

\section{Experiments}
\label{sec:chapter8:sec4}

\subsection{Datasets}
\label{sec:chapter8:sec4.1}

We used the CMRxRecon 2023 cardiac cine dataset \cite{cmrxrecon2023,Wang2021}, which comprises 472 scans with fully sampled, ECG-triggered multi-coil ($n_c=10$) $k$-space data, totaling 3,185 2D sequences, with the cardiac cycle segmented into 12 temporal frames. For external validation, we used the aorta dataset from CMRxRecon 2024 \cite{wang2024cmrxrecon2024multimodalitymultiviewkspace,wang2025universallearningbasedmodelcardiac}, also ECG-triggered with $n_c=10$ coils and 12-frame cycles. The inference subset comprises 111 scans and 1,332 sequences.

\subsection{Undersampling} 
Undersampling was applied retrospectively in all experiments, with fully sampled data used for loss calculation and evaluation. A 4\% center fraction within the ACS was retained. Training used arbitrary acceleration factors of $4\times$, $6\times$, or $8\times$, and all three were evaluated during inference.

\subsection{Selection of Registration Reference}
Each sequence contained 12 segments of the cardiac cycle. We selected the $6^{\text{th}}$ segment as registration reference $\vec{y}_{\text{ref}} \in \complex^{n \times n_c}$, corresponding to the end-systolic (contracted) phase, since this phase can be consistently triggered by ECG \cite{MADA2015148}, providing a stable anatomical structure. The remaining segments ($n_t = 11$) served as the moving image target $\vec{y}_{\text{mov}} \in \complex^{n \times n_c \times n_t}$, which were retrospectively undersampled to create the undersampled moving input, $\tilde{\vec{y}}_{\text{mov}}^{\mat{M}}$.

\subsection{Evaluation}
We assess estimated motion quality by evaluating the similarity of the registered moving images (warped reconstructed images using predicted deformation fields) to the reference image. We employ three image quality metrics: SSIM, PSNR, and NMSE. For brevity, definitions are provided in in Appendix \ref{sec:chapter8:appendix:appendix2}. For each metric $m$, we averaged results across all phases of the sequence:
\begin{equation}
    \overline{m} = \frac{1}{n_t} \sum_{\tau=1}^{n_t} m\Big(\mathcal{W}\big(|\hat{\vec{x}}_{\text{mov}}|_{\cdot,\tau}, \boldsymbol{\phi}_{\tau} \big), {\vec{x}}_{\text{ref}}\Big).
\end{equation}
\subsection{Training and Optimization Details}
Models were implemented in PyTorch \cite{paszke2019pytorch} and trained on single NVIDIA A100 or H100 GPUs with the Adam optimizer (no weight decay) \cite{zhang2018improved}, a batch size of 1, and a learning rate schedule with a 10k step size and 0.8 decay. After a 2k-iteration linear warm-up to reach a 3e-3 learning rate, training continued for 52k iterations. The best checkpoint, based on validation SSIM, was used for inference.

\subsection{Ablation and Component Analysis} 
Our proposed pipeline is the first to integrate adaptive sampling, reconstruction, and registration into an end-to-end framework for dynamic MRI, making direct comparisons with existing baselines impossible. Therefore, we evaluate each key component within our approach through ``controlled'' comparisons and ablation studies.

\paragraph{Baseline Setup} 
Our baseline setup includes:
\begin{itemize}
    \item Sensitivity Model: four scales (16, 32, 64, 128 filters).
    \item Adaptive Sampling Module: Identical to \cite{yiasemis2024endtoendadaptivedynamicsubsampling} in hyperparameters, but with one cascade instead of two. Unless specified otherwise, adaptive sampling is phase-specific. We employ ACS data for initialization ($\tilde{\vec{y}}_{\text{mov}}^{\mat{M}^0} = \tilde{\vec{y}}_{\text{mov}}^{\mat{M}^\text{acs}}$).
    \item Reconstruction Model: vSHARP with $T = 10$ optimization steps utilizing 3D U-Nets with four scales (16, 32, 64, and 128 filters) and $T_x = 6$ data consistency steps. Other parameters match \cite{yiasemis2025vsharp}.
    \item Registration Model: A 2D U-Net (see \Section{chapter8:subsubsec3.1.4}) with four scales (16, 32, 64, and 128 filters).
    \item End-to-end training loss calculation with $\alpha=\beta=1$, reflecting the expectation that improving reconstruction quality will contribute to reliable registration by reducing artifacts and inconsistencies that could otherwise impair motion estimation.
\end{itemize}
In each experiment, only the evaluated component is modified, keeping all other settings constant to isolate its effect.

\paragraph{Registration Component}
To assess our lightweight registration approach (concatenating moving and reference images along the channel dimension in a 2D U-Net), we compare it against DL-based alternatives and traditional motion estimation methods: 
\begin{enumerate}[label=\arabic{enumi})]
    \item A standard registration method - VoxelMorph \cite{balakrishnan2019voxelmorph}, using a 2D U-Net with four scales (16, 32, 64, 128 filters).
    \item A vision transformer-based registration method following TransMorph \cite{chen2022transmorph}
    \item Optical flow with iterative Lucas-Kanade (ILK) \cite{1529706} and TV-$L_1$ \cite{tvl1} solvers from \verb|scikit-image| \cite{van2014scikit}
    \item DEMONS registration algorithm \cite{thirion1998image} provided by \verb|SimpleITK| \cite{simpleitk}
\end{enumerate}
See Appendix \ref{sec:chapter8:appendix:appendix2} for the choice of hyperparameters.

\paragraph{Reconstruction Component} To evaluate the choice of reconstruction network (vSHARP) and showcase the pipeline’s modularity, we also test a state-of-the-art model, the End-to-End Variational Network (VarNet) \cite{Sriram2020} (extended to 3D), configured with 10 cascades and 3D U-Nets as regularizers (4 scales: 32, 64, 128, 256 filters).

\paragraph{Adaptive Sampling Component}
\begin{itemize}
    \item \textbf{Initialization} The ADS module relies on initial data  $\tilde{\vec{y}}_{\text{mov}}^{\mat{M}^0}$ to guide the acquisition. In \cite{yiasemis2024endtoendadaptivedynamicsubsampling}, equispaced initialization improved reconstruction quality; we evaluate this for registration quality as well (initial data, include ACS and accelerated at $(R-4)\times$ where $R$ is the target acceleration).
    \item \textbf{Phase-specific vs Unified Sampling} Phase-specific experiments generate a unique adaptive pattern for each temporal phase, whereas unified settings produce a single pattern for all phases.
    \item \textbf{Learned vs Fixed Sampling}
    We compare our pipeline by replacing the adaptive sampling module with non-adaptive dataset-optimized \cite{zhang2020extending}, fixed non-adaptive equispaced \cite{yiasemis2023retrospective} (distinct pattern per frame in phase-specific experiments; same for all frames in unified settings) and $k$t-equispaced \cite{tsao2003k} (temporally interleaved trajectory, applicable only to phase-specific) sampling schemes.  Offset of patterns was randomly selected during training and fixed (per case) during inference.
\end{itemize}

\paragraph{Weighting Parameters for Joint Loss}
To investigate how task weighting impacts registration quality, we test varying weighting parameters in \eqref{eq:chapter8:total_loss}. This allows us to examine the extent to which additional emphasis on the registration task influences the final motion estimation.

\paragraph{Joint vs. Decoupled Training}
To evaluate task interactions, we adopt a decoupled loss approach. For the registration loss $\mathcal{L}_{\text{reg}}$, only the registration model parameters ($\boldsymbol{\psi}$) are updated, while holding all others constant (frozen). Conversely, for the reconstruction loss $\mathcal{L}_{\text{rec}}$, only the parameters $\boldsymbol{\sigma}$, $\boldsymbol{\omega}$, and $\boldsymbol{\theta}$ of the sensitivity, sampling, and reconstruction models are optimized, while $\boldsymbol{\psi}$ remains fixed. This decoupling allows each task to be evaluated independently, compared to our original joint, end-to-end training.

\paragraph{Choice of Loss Function} We test two variants:

\begin{enumerate}
    \item using only \(\mathcal{L}_{\text{reg}} = \mathcal{L}_{\text{sim}} = \mathcal{L}_{\text{ssim2D}}\),
    \item using only \(\mathcal{L}_{\text{reg}} = \mathcal{L}_{\text{sim}} = \mathcal{L}_{1}\).
\end{enumerate}

\subsection{Experimental Results}
We present average quantitative results for each experiment across both datasets (cine - seen during training, aorta - unseen), displayed as line graphs (function of acceleration factor) with $\pm 0.1$ standard deviations, along with average inference times (s). Detailed results are also in tabular form in Appendix \ref{sec:chapter8:appendix:appendix3}. Examples of qualitative results are provided in \Figure{chapter8:example}, with further examples in Appendix \ref{sec:chapter8:appendix:appendix3}.

\begin{figure}[!ht]
    \centering
    \begin{subfigure}{\columnwidth}
        \centering
        \includegraphics[width=\columnwidth]{\thischapter/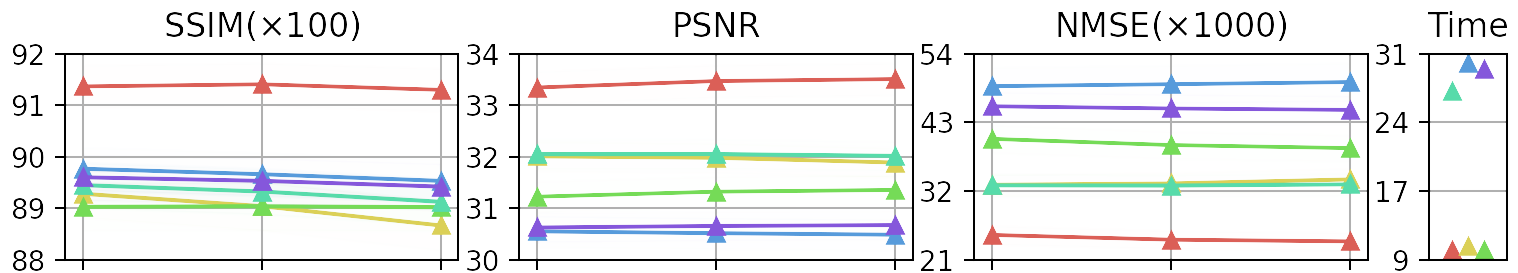}
    \end{subfigure}
    \vspace{0.0pt} 
    \hfill
    \begin{subfigure}{\columnwidth}
        \centering
        \includegraphics[width=\columnwidth]{\thischapter/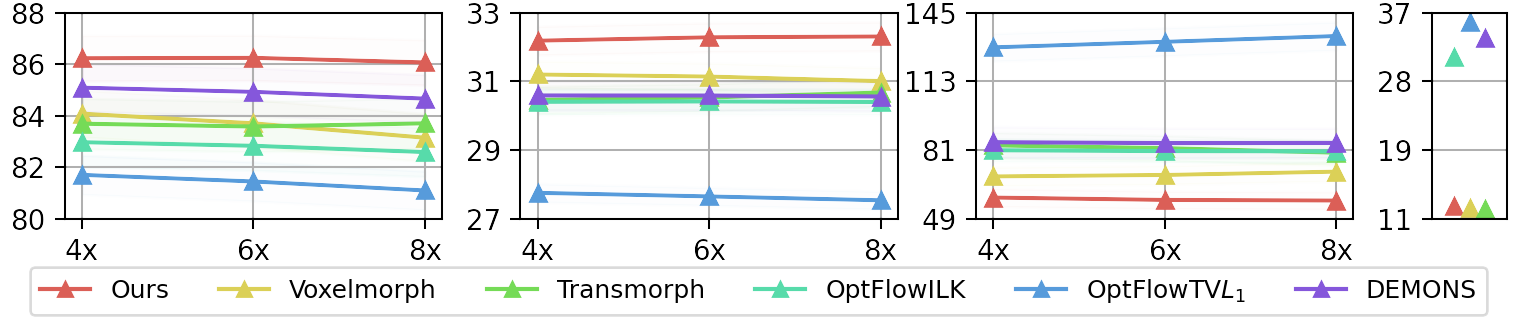}
    \end{subfigure}
    \caption{Comparison of registration performance across different methods. Top: Cine. Bottom: Aorta (not seen during training).}
    \label{fig:chapter8:metrics_registration}
\end{figure}

\begin{figure}[!ht]
    \centering
    \begin{subfigure}{\columnwidth}
        \centering
        \includegraphics[width=\columnwidth]{\thischapter/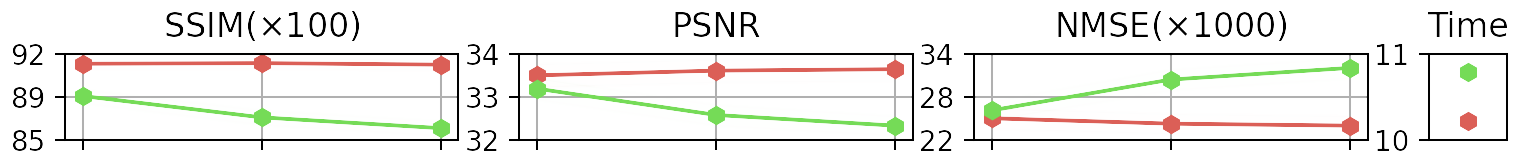}
    \end{subfigure}
    \hfill
    \begin{subfigure}{\columnwidth}
        \centering
        \includegraphics[width=\columnwidth]{\thischapter/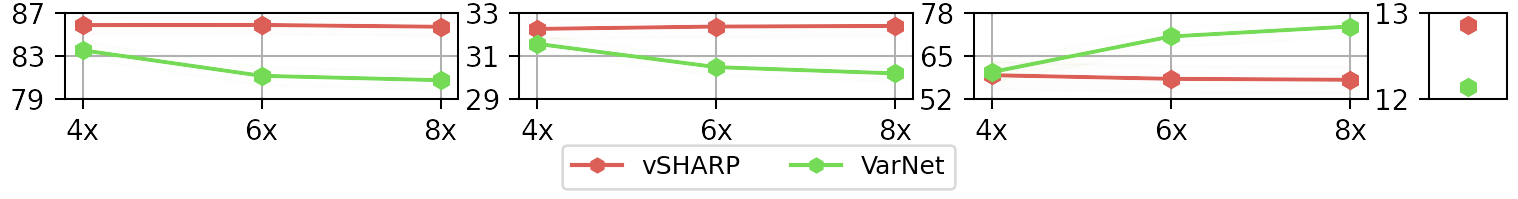}
    \end{subfigure}
    \caption{Impact of reconstruction model on registration results. Top: Cine. Bottom: Aorta (not seen during training).}
    \label{fig:chapter8:recon_comparison}
\end{figure}

\begin{figure}[!ht]
    \centering
    \begin{subfigure}{\columnwidth}
        \centering
        \includegraphics[width=\columnwidth]{\thischapter/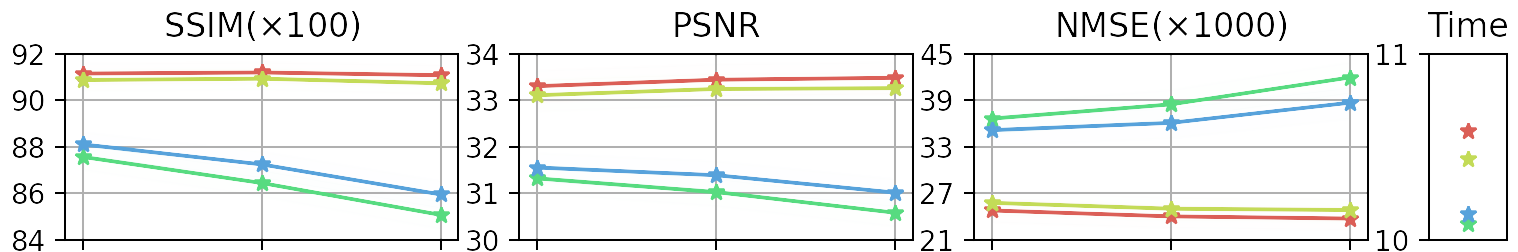}
    \end{subfigure}
    \hfill
    \begin{subfigure}{\columnwidth}
        \centering
        \includegraphics[width=\columnwidth]{\thischapter/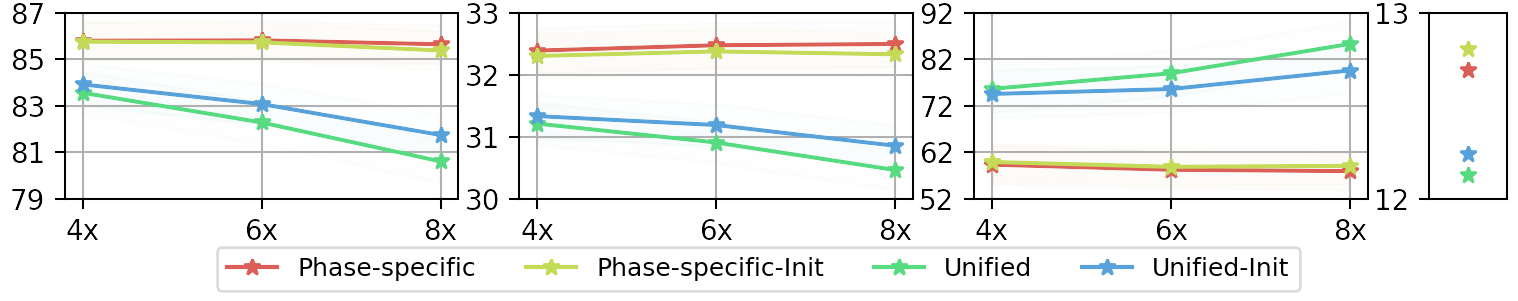}
    \end{subfigure}
    \caption{Impact of phase-specific vs unified adaptive sampling with ACS and equispaced initialization on registration results. Top: Cine. Bottom: Aorta (not seen during training).}
    \label{fig:chapter8:metrics_ads_dim}
\end{figure}

\begin{figure}[!ht]
    \centering
    \begin{subfigure}{\columnwidth}
        \centering
        \includegraphics[width=\columnwidth]{\thischapter/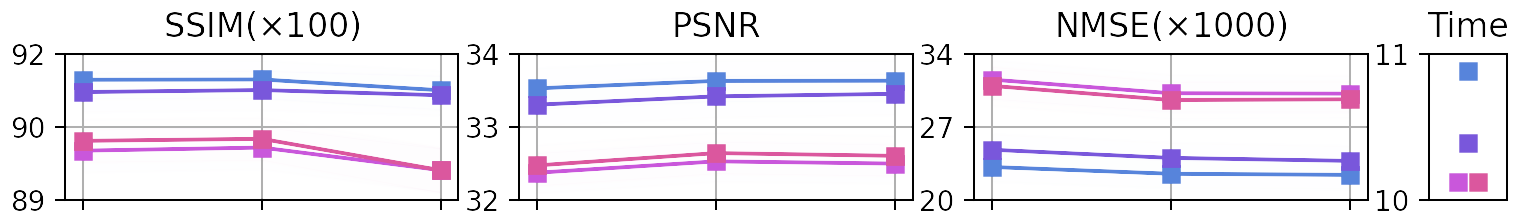}
    \end{subfigure}
    \hfill
    \begin{subfigure}{\columnwidth}
        \centering
        \includegraphics[width=\columnwidth]{\thischapter/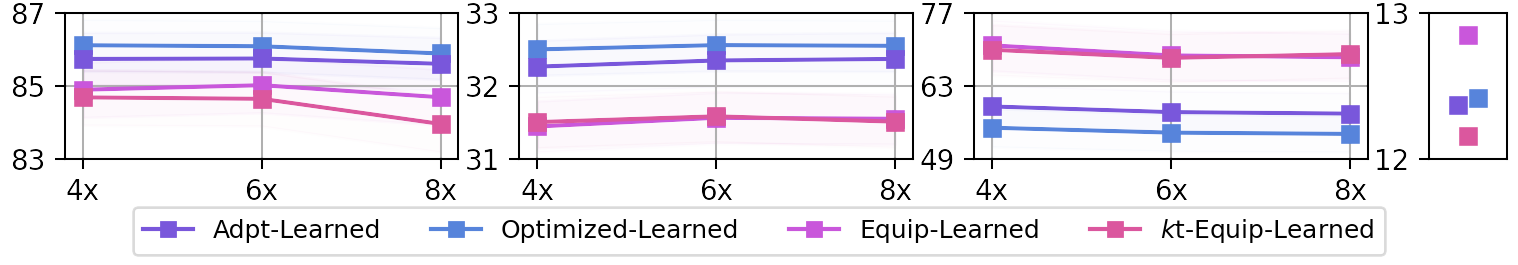}
    \end{subfigure}
    \caption{Impact of learned vs fixed non-adaptive sampling schemes on registration results. Top: Cine. Bottom: Aorta (not seen during training).}
    \label{fig:chapter8:metrics_ads}
\end{figure}

\begin{figure}[!ht]
    \centering
    \begin{subfigure}{\columnwidth}
        \centering
        \includegraphics[width=\columnwidth]{\thischapter/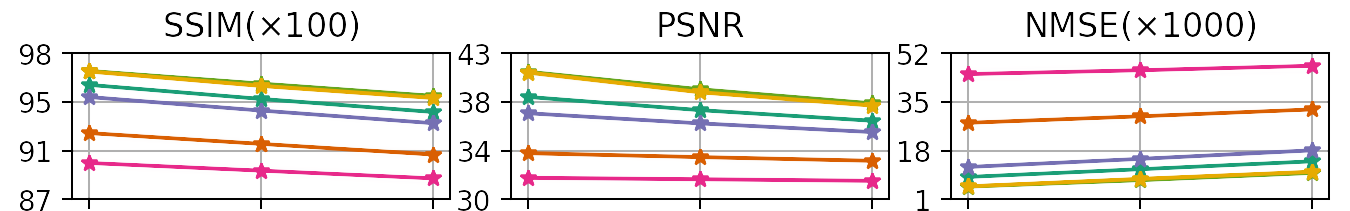}
    \end{subfigure}
    \hfill
    \begin{subfigure}{\columnwidth}
        \centering
        \includegraphics[width=\columnwidth]{\thischapter/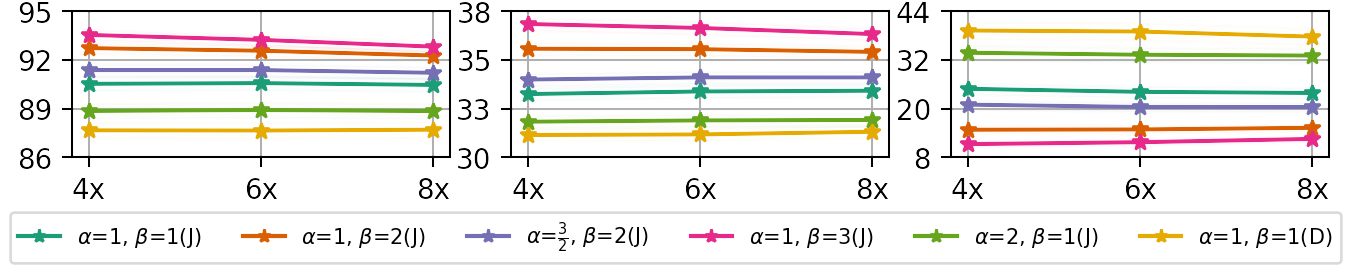}
    \end{subfigure}
    \caption{Impact of joint (J) vs decoupled (D) training, and varying loss weights (cine dataset). Top: Reconstruction performance. Bottom: Registration performance.}
    \label{fig:chapter8:loss_comparison_combined}
\end{figure}
\begin{figure}[!ht]
    \centering
    \begin{subfigure}{\columnwidth}
        \centering
        \includegraphics[width=\columnwidth]{\thischapter/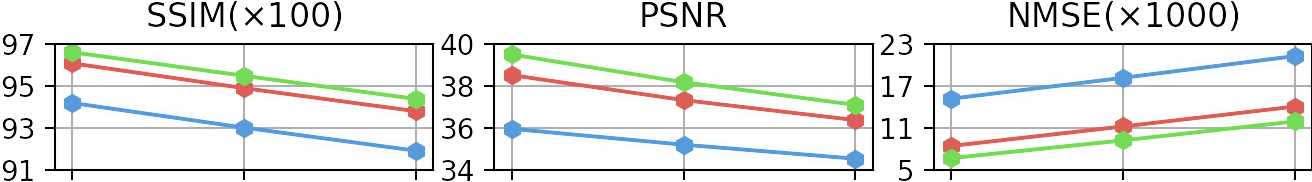}
    \end{subfigure}
    \hfill
    \begin{subfigure}{\columnwidth}
        \centering
        \includegraphics[width=\columnwidth]{\thischapter/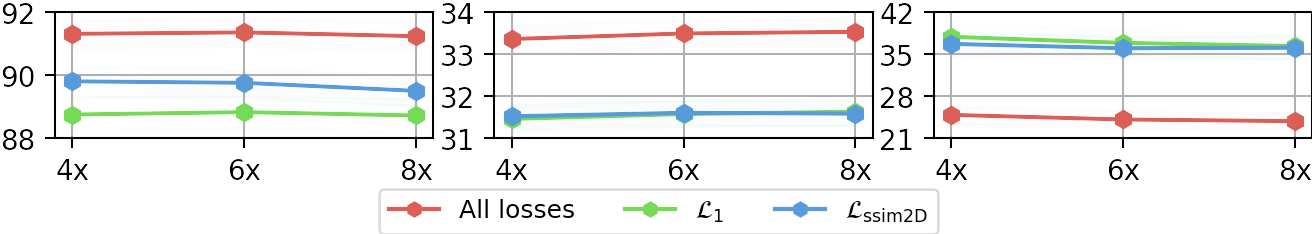}
    \end{subfigure}
    \caption{Impact of  loss choice (cine dataset). Top: Reconstruction performance. Bottom: Registration performance.}
    \label{fig:chapter8:loss_function_comparison}
\end{figure}

\paragraph{Registration Component}
\Figure{chapter8:metrics_registration} presents the registration performance of our pipeline across various registration modules, evaluated on both datasets. Across all acceleration factors (4$\times$, 6$\times$, 8$\times$), our proposed lightweight model consistently outperforms traditional methods (OptFlow ILK, TV-$L_1$, DEMONS) and the two considered DL-based approaches (Voxelmorph, Transmorph) in all metrics on both datasets, indicating more accurate motion estimation. Additionally, our pipeline with learned registration methods (including ours) achieves significantly lower inference times than non-DL-based methods. 

\paragraph{Reconstruction Component} \Figure{chapter8:recon_comparison} shows that replacing vSHARP with VarNet yields lower registration metrics, suggesting vSHARP’s optimization-based design better supports accurate motion estimation in our pipeline.

\paragraph{Adaptive Sampling Component}  In \Figure{chapter8:metrics_ads_dim} we compare the effects of phase-specific versus unified adaptive sampling with both ACS initialization and equispaced-fused ACS initialization. Phase-specific sampling consistently outperforms unified sampling across all metrics, particularly at higher accelerations, indicating better registration quality with unique trajectories per phase.  In contrast to the reconstruction quality findings in \cite{yiasemis2024endtoendadaptivedynamicsubsampling}, equispaced initialization did not notably benefit registration quality. \Figure{chapter8:metrics_ads} further shows that learned sampling (adaptive, dataset-optimized) outperforms fixed schemes across all metrics with minimal inference time impact, with dataset-optimized sampling showing a slight advantage over adaptive.

\paragraph{Weighting Parameters for Joint Loss}  \Figure{chapter8:loss_comparison_combined} illustrates the impact of varying loss weights on each task. Increasing the registration weight (\(\alpha < \beta\)) improves registration but degrades reconstruction. However, analysis of motion estimates and images (Appendix \ref{sec:chapter8:appendix:appendix3}, Figures \ref{fig:chapter8:appendix:example} and \ref{fig:chapter8:appendix:example2}) indicates that when \(\alpha < \beta\), the network may approximate the reference rather than reconstructing the moving image accurately. Conversely, when \(\alpha > \beta\), reconstruction quality improves at the cost of motion estimation. Among all weighting schemes, equal weighting preserves motion field accuracy while maintaining reconstruction quality.

\paragraph{Joint vs. Decoupled Training} \Figure{chapter8:loss_comparison_combined} further contrasts joint vs. decoupled training. While joint optimization slightly reduces reconstruction performance, it significantly improves registration quality. 

\paragraph{Choice of Loss Function} Results are shown in \Figure{chapter8:loss_function_comparison}. Both \(\mathcal{L}_{\text{ssim}}\) and \(\mathcal{L}_1\) are important for maintaining higher registration accuracy.

\begin{figure}[!ht]
    \centering
    \begin{subfigure}[b]{0.7\textwidth}
        \centering
        \parbox[b]{0.2\textwidth}{%
            \centering
            \includegraphics[height=0.11\textheight]{\thischapter/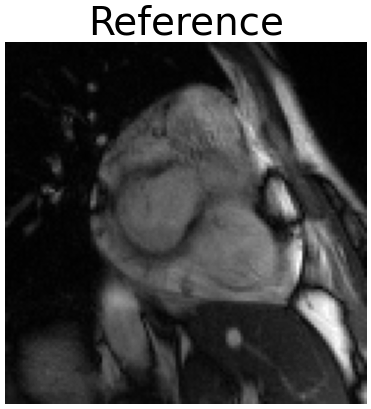}
            \vfill
            \vspace{0pt} 
            \includegraphics[height=0.2\textheight]{\thischapter/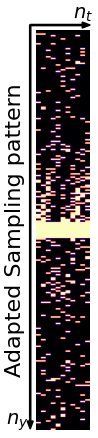}
        }%
        \hfill
        \raisebox{0pt}[0pt][0pt]{%
            \includegraphics[height=0.31\textheight]{\thischapter/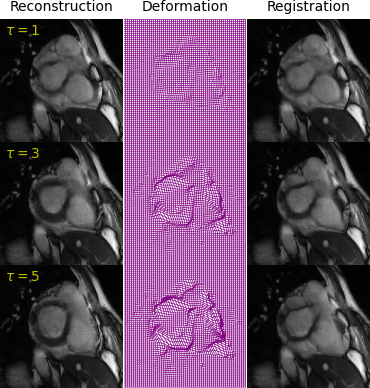}
        }
        \caption{Case I, \( R = 6 \).}
        \vspace{5pt}
        \label{fig:chapter8:example_6x}
    \end{subfigure}
    \hfill
    \begin{subfigure}[b]{0.7\textwidth}
        \centering
        \parbox[b]{0.2\textwidth}{%
            \centering
            \includegraphics[height=0.11\textheight]{\thischapter/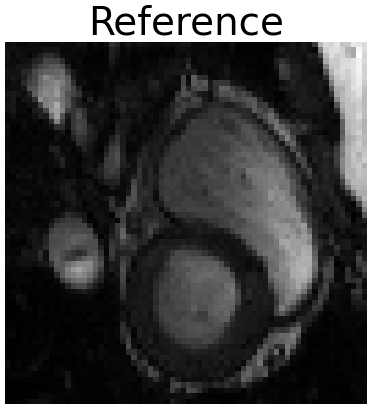}
            \vfill
            \vspace{0pt} 
            \includegraphics[height=0.2\textheight]{\thischapter/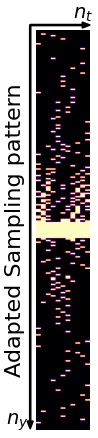}
        }%
        \hfill
        \raisebox{0pt}[0pt][0pt]{%
            \includegraphics[height=0.31\textheight]{\thischapter/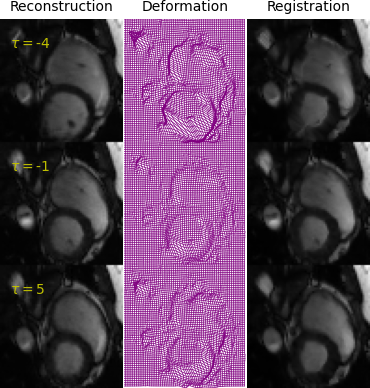}
        }
        \caption{Case II, \( R = 8 \).}
        \label{fig:chapter8:example_8x}
    \end{subfigure}
    \caption{Example results for two cases, shown at various temporal frames (\( \tau \)) relative to the reference image.}
    \vspace{-10pt}
    \label{fig:chapter8:example}
\end{figure}

\section{Discussion and Conclusion}
\label{sec:chapter8:sec5}

In this work we present a novel pipeline integrating adaptive (under)sampling, reconstruction, and registration in an end-to-end framework for dynamic MRI. Our experiments provide a thorough evaluation of each component, on both in and out-of-distribution datasets, highlighting the flexibility and performance of this approach.

Our results show that our motion estimation module achieves strong registration performance, measured by registered moving image's similarity to the reference, and provides efficient inference speeds compared to other considered deep learning and traditional methods.

Similarly, when testing an additional reconstruction network (VarNet), results indicated that our pipeline performed better with vSHARP as the reconstruction model, as evidenced by the registration outcomes.

While alternative hyperparameters for the registration component or utilizing different reconstruction algorithms might yield different outcomes, our aim is not to identify the optimal models but to underscore the modularity of our pipeline, allowing for interchangeable reconstruction or registration modules in a plug-and-play configuration.

Our evaluation of learned versus fixed sampling methods indicates that using an adaptive or optimized learned sampler leads to better registration field estimation than equidistant sampling schemes. This aligns with \cite{yiasemis2024endtoendadaptivedynamicsubsampling}, where adaptive sampling also improved reconstruction quality over fixed schemes, although the difference between adaptive and optimized sampling remained minor here. Additionally, phase-specific sampling patterns, rather than unified, improved registration, while equispaced initialization did not enhance motion estimation, contrasting with its positive impact on reconstruction found in \cite{yiasemis2024endtoendadaptivedynamicsubsampling}.

In addition, we compared joint versus decoupled training schemes. In the decoupled setup, task-specific losses were optimized by freezing the registration model weights during reconstruction loss computation, and vice versa, with joint optimization yielding significant registration improvements over decoupled training, with minimal reconstruction sacrifice. This suggests that joint optimization fosters better motion estimation by aligning reconstruction with registration objectives, whereas decoupled training isolates the tasks at the expense of registration quality. However, when prioritizing registration weight, this led to improved registration as the reconstruction model was tricked into producing the reference image (as a result of the joint optimization). Conversely, prioritizing reconstruction weight, can hinder registration performance. These suggest that balanced weighting yields more reliable motion estimation.

Finally, we analyzed the impact of different components in $\mathcal{L}_{\text{reg}}$ and $ \mathcal{L}_{\text{sim}}$, finding that incorporating all components led to the best overall performance.

Since we compared registered reconstructed moving images ($\hat{\vec{x}}_{\text{mov}}$) to the reference, our results reflect both registration and reconstruction quality. Evaluating motion estimation independently by comparing registered ground truth moving images ($\vec{x}_{\text{mov}}$)-using the predicted deformation fields-to the reference would provide a more objective assessment. Future research should also directly assess registration accuracy with relevant methods \cite{vandemeulebroucke2007popi,saleh2014distance}.

A limitation of our study is the reliance on fully sampled dynamic/moving $k$-space as ground truths, which are often unavailable due to motion constraints. Although self-supervised learning (SSL) methods exist for reconstruction \cite{9098514,yiasemis2023jssl,yan2023dc}, the adaptive sampling component requires access to a ground truth $k$-space. For the registration component, however, since it is trained in an unsupervised manner and only requires a fully sampled reference image, this issue is less problematic. Future work could explore SSL adaptations for the adaptive sampling component.

A further limitation is that our approach is two-dimensional, estimating motion between dynamic 2D (moving) slices and a reference 2D slice. A 3D approach could capture inter-slice correlations, though it would require a 4D (3D + time) reconstruction network, which would be computationally intensive for GPU memory during training (e.g., for back-propagation). Future work should consider exploring such a 3D approach.

This study marks the first application of an end-to-end, DL-based pipeline that jointly optimizes coil sensitivity estimation, adaptive sampling, reconstruction, and registration to enhance motion estimation between moving undersampled and reference images. The results suggest that each component contributes distinct benefits. Further research should investigate approaches that bypass reconstruction altogether, aiming to estimate motion directly from adaptively undersampled measurements for a more efficient workflow.

\newpage
\section*{Acknowledgments}
{This work was supported by institutional grants of the Dutch Cancer Society and of the Dutch Ministry of Health, Welfare and Sport. The authors would like to acknowledge the Research High Performance Computing (RHPC) facility of the Netherlands Cancer Institute (NKI).}
%
%
\bibliographystyle{splncs04}
\bibliography{references}

\newpage
\appendix
\renewcommand \thesection {\Alph{section}}

\noindent \Large{\textbf{Supplementary Material}}

\normalsize
\section{Methods - Additional Information}
\label{sec:chapter8:appendix:appendix1}
\renewcommand{\theequation}{A\arabic{equation}}
\setcounter{equation}{0} 

\renewcommand{\thefigure}{A\arabic{figure}}
\setcounter{figure}{0}

\renewcommand{\thetable}{A\arabic{table}}
\setcounter{table}{0}

\subsection{Sensitivity Profile Network}
As detailed in the main text and following \cite{Sriram2020}, the sensitivity profiles for the multi-coil data, denoted as \(\tilde{\vec{S}} \in \mathbb{C}^{n^2 \times n_c \times n_t}\), are initially estimated using the fully sampled central region of the $k$-space, \(\tilde{\vec{y}}_{\text{mov}}^{\mat{M}^\text{acs}}\), corresponding to the low-frequency autocalibration signal. This estimation is computed as:

\footnotesize
\begin{equation}
    \tilde{\mat{S}}_{\tau}^k\, = \,\mathcal{F}^{-1} \big( (\tilde{\vec{y}}_{\text{mov}}^{\mat{M}^\text{acs}})_{\cdot, k, \tau}\big) \oslash \text{RSS} \Big( \mathcal{F}^{-1} \big( (\tilde{\vec{y}}_{\text{mov}}^{\mat{M}^\text{acs}})_{\cdot, 1, \tau}\big), \cdots, \mathcal{F}^{-1} \big( (\tilde{\vec{y}}_{\text{mov}}^{\mat{M}^\text{acs}})_{\cdot, n_c, \tau}\big)\Big),
    \label{eq:chapter8:appendix1:sensemaps}
\end{equation}
\normalsize

\noindent
where \(\oslash\) represents element-wise (Hadamard) division, and \(\text{RSS}(\cdot)\) computes the root sum of squares.

These initial sensitivity estimates, consistent with the methodology in \cite{Sriram2020}, are further refined using a deep learning-based model. Specifically, a 2D U-Net \cite{Ronneberger2015}, denoted as \(\mathcal{S}_{\boldsymbol{\sigma}}\), is employed to enhance the sensitivity profiles:

\begin{equation}
    \mat{S}_{\tau}^{k} :=  \mathcal{S}_{\boldsymbol{\sigma}}(\tilde{\mat{S}}_{\tau}^{k}), \quad \tau = 1, \cdots, n_t, \quad k = 1, \cdots, n_c.
\end{equation}

\noindent
The refined sensitivity profiles are subsequently normalized to satisfy:
\begin{equation}
    \sum_{k=1}^{n_c} (\mat{S}_{\tau}^{k})^* \mat{S}_{\tau}^{k} = \mathbf{I}_{n} \in \real^{n\times n}.    
\end{equation}
 
\subsection{Reconstruction Network - vSHARP}

In our original pipeline, we utilized the variable Splitting Half-quadratic ADMM algorithm for Reconstruction of inverse Problems (vSHARP) as the reconstruction network. vSHARP is an unrolled, physics-guided deep learning method \cite{yiasemis2025vsharp} that has been successfully applied in accelerated dynamic cardiac MRI reconstruction, including among the winning solutions of the CMRxRecon 2023 and 2024 challenges \cite{cmrxrecon,wang2025universallearningbasedmodelcardiac}. The vSHARP algorithm leverages the half-quadratic variable splitting technique for the optimization problem defined in (\textcolor{red}{6}), introducing an auxiliary variable $\vec{z}$ as follows:
\begin{equation}
    \min_{\vec{x}^{'}, \, \vec{z} \, \in  \,\complex^{n \times n_t}} \frac{1}{2} \sum_{\tau=1}^{n_t} \left|\left| \mathcal{T}_{\vec{M}_{\tau}, \mat{C}_{\tau}}(\vec{x}_{\cdot, \tau}^{'}) - \Tilde{\vec{y}}_{\cdot, \cdot, \tau}^{\mat{M}}\right|\right|_2^2 +  \mathcal{H}(\vec{z}) \quad \text{subject to } \vec{x}^{'} = \vec{z}.
    \label{eq:chapter8:appendix1:vsharp_var}
\end{equation}

\noindent
Equation \eqref{eq:chapter8:appendix1:vsharp_var} is then unrolled over $T$ iterations using the Alternating Direction Method of Multipliers (ADMM), comprising three main steps at each iteration: \textbf{(i)} denoising to refine the auxiliary variable $\vec{z}$, \textbf{(ii)} data consistency for the target image $\vec{x}$, and \textbf{(iii)} updating the Lagrange multipliers $\vec{m}$ introduced by ADMM:

\footnotesize
\begin{subequations}
    \begin{gather}
        \vec{z}^{t+1} = \argmin_{\vec{z} \, \in  \,\complex^{n \times n_t}} \mathcal{H}(\vec{z})  + \frac{\lambda^{t+1}}{2} \left|\left|\vec{x}^{t} - \vec{z} + \frac{\vec{m}^{t}}{{\lambda}^{t+1}} \right|\right|_2^2 := \mathcal{D}_{\boldsymbol{\theta}^{t+1}}\left( \vec{z}^{t}, \vec{x}^{t}, \frac{\vec{m}^{t}}{{\lambda}^{t+1}} \right),\label{eq:chapter8:appendix1:z_step}\\
        \vec{x}^{t+1}  =  \argmin_{\vec{x}^{'} \, \in  \,\complex^{n \times n_t}}  \frac{1}{2} \sum_{\tau=1}^{n_t} \left|\left| \mathcal{T}_{\vec{M}_{\tau}, \mat{C}_{\tau}}(\vec{x}_{\cdot, \tau}^{'}) - \Tilde{\vec{y}}_{\cdot, \cdot, \tau}^{\mat{M}}\right|\right|_2^2  \notag \\
        +  \lambda^{t+1}\left|\left|\vec{x}^{'} - \vec{z}^{t+1} + \frac{\vec{m}^{t}}{{\lambda}^{t+1}} \right|\right|_2^2,
        \label{eq:chapter8:appendix1:x_step}\\
        \vec{m}^{t+1} = \vec{m}^{t} + \lambda^{t+1} \left(\vec{x}^{t+1} - \vec{z}^{t+1}\right).
        \label{eq:chapter8:appendix1:u_step}
    \end{gather}
\end{subequations}
\normalsize

\noindent
In \eqref{eq:chapter8:appendix1:z_step}, $\mathcal{D}_{\boldsymbol{\theta}^{t+1}}$ denotes a convolutional DL-based image denoiser with trainable parameters $\boldsymbol{\theta}^{t+1}$, and $\lambda^{t+1}$ is a trainable learning rate. At each iteration, $\mathcal{D}_{\boldsymbol{\theta}^{t+1}}$ processes the previous estimates of the three variables, yielding an updated estimate for $\vec{z}$. Equation \ref{eq:chapter8:appendix1:x_step} is numerically optimized using a differentiable gradient descent scheme unrolled over $T_{\vec{x}}$ iterations. The update in \eqref{eq:chapter8:appendix1:u_step} involves a straightforward calculation. The initial estimates for $\vec{x}$ and $\vec{z}$ are defined as follows:

\begin{equation}
    \vec{x}_{\tau}^{0},\, \vec{z}_{\tau}^{0} = \mathcal{R}_{\mat{S}_{\tau}} \circ \mathcal{F}^{-1} \left(\Tilde{\vec{y}}_{\tau}^{\mat{M}}\right).
\end{equation} 

\noindent
For $\vec{m}^{0}$, vSHARP employs a trainable network $\mathcal{I}^{\boldsymbol{\theta}^{\mathcal{I}}}$ initialized using $\vec{x}_{\tau}^{0}$:

\begin{equation}
    \mat{m}^{0} := \mathcal{I}^{\boldsymbol{\theta}^{\mathcal{I}}}(\vec{x}^{0}).
\end{equation}

\noindent
Accordingly, the trainable parameters of the vSHARP reconstruction model are defined as:

\begin{equation}
    \boldsymbol{\theta} = \Big\{\boldsymbol{\theta}^{1}, \cdots, \boldsymbol{\theta}^{T}, \boldsymbol{\theta}^{\mathcal{I}}, \lambda^{1}, \cdots, \lambda^{T} \Big\}.
\end{equation}

\section{Experiments - Additional Information}
\label{sec:chapter8:appendix:appendix2}

\renewcommand{\theequation}{B\arabic{equation}}
\setcounter{equation}{0} 

\renewcommand{\thefigure}{B\arabic{figure}}
\setcounter{figure}{0}

\renewcommand{\thetable}{B\arabic{table}}
\setcounter{table}{0}



\subsection{Datasets}
As outlined in the main text, the cardiac cine dataset from the CMRxRecon 2023 challenge \cite{cmrxrecon2023,Wang2021} was used. This dataset includes 471 4D cardiac $k$-space scans, resulting in 3,185 2D dynamic sequences. It features short- and long-axis views across two-, three-, and four-chamber configurations. Each scan is fully sampled, ECG-triggered, and acquired using multi-coil ($n_c=10$) setups. The dataset spans 3–12 dynamic slices per case, with each cardiac cycle divided into 12 temporal frames. The dataset was split into 251, 100, and 100 4D volumes for training, validation, and testing, respectively (comprising 1,710, 731, and 744 dynamic slices).

For external validation, the aorta dataset from the CMRxRecon 2024 challenge \cite{wang2024cmrxrecon2024multimodalitymultiviewkspace,wang2025universallearningbasedmodelcardiac} was employed. This dataset also contains fully sampled, ECG-triggered multi-coil acquisitions segmented into 12 temporal frames. Transverse and sagittal views of the aorta were provided, with the subset used for inference consisting of 111 scans and 883 dynamic slices.

\begin{figure}[!htb]
    \centering
    \begin{subfigure}{0.24\textwidth}
        \includegraphics[width=\textwidth]{\thischapter/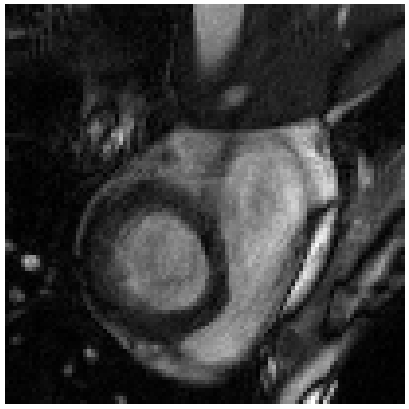}
        \caption{Cine SAX.}
        \label{fig:chapter8:appendix:sub1}
    \end{subfigure}
    \hfill
    \begin{subfigure}{0.24\textwidth}
        \includegraphics[width=\textwidth]{\thischapter/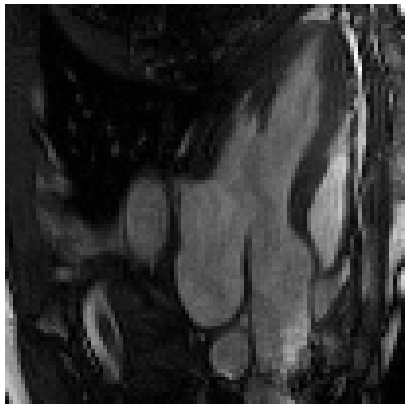}
        \caption{Cine LAX.}
        \label{fig:chapter8:appendix:sub2}
    \end{subfigure}
    \hfill
    \begin{subfigure}{0.24\textwidth}
        \includegraphics[width=\textwidth]{\thischapter/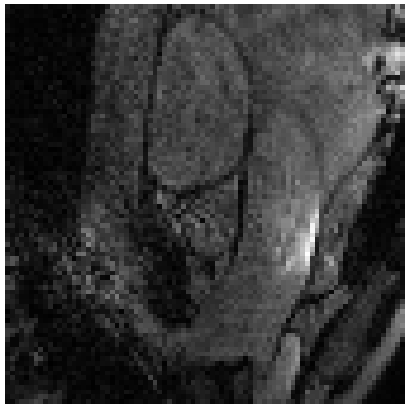}
        \caption{Aorta SAG.}
        \label{fig:chapter8:appendix:sub3}
    \end{subfigure}
    \hfill
    \begin{subfigure}{0.24\textwidth}
        \includegraphics[width=\textwidth]{\thischapter/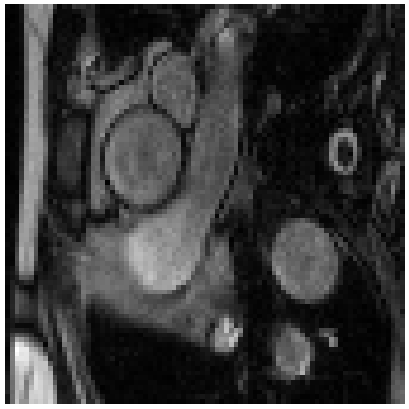}
        \caption{Aorta TRA.}
        \label{fig:chapter8:appendix:sub4}
    \end{subfigure}
    \caption{Representative fully sampled images across cardiac phases and views from each dataset.}
    \label{fig:chapter8:appendix:datasets}
\end{figure}

\Figure{chapter8:appendix:datasets} illustrates representative fully sampled reconstructed images from each dataset, showcasing the diversity of anatomical views.

\subsection{Undersampling}

\subsubsection{Learned Sampling}
Experiments utilized 1D line (rectilinear) sampling patterns, retaining certain lines in the phase-encoding direction of the Cartesian grid. For learned sampling patterns (adaptive or optimized), data were subsampled using the following strategies:

\begin{enumerate}
    \item Autocalibration Signal (ACS): The central $k$-space region was retained, covering $4\%$ of the fully sampled data.
    \item Equispaced Sampling: Applied with a target acceleration factor of $(R-4)\times$, where $R$ is the target acceleration. ACS measurements (4\% of the fully sampled data) were included in this sampling pattern.
\end{enumerate}

\noindent
For adaptive sampling, the initial undersampled data informed the acquisition of additional data at the target acceleration factor. In optimized sampling experiments, initialization was not strictly required but ensured inclusion of ACS data.

\subsubsection{Fixed Sampling}
For fixed (non-learned) sampling, we applied two distinct strategies:

\begin{figure}[!htb]
    \centering
    \begin{subfigure}{0.3\textwidth}
        \centering
        \includegraphics[width=0.3\textwidth]{\thischapter/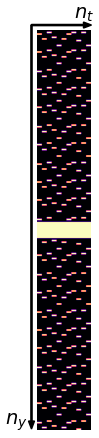}
        \caption{Phase-specific equispaced sampling pattern. Each time-step represents a scheme per each phase. Each point represents a line along the $n_x$ direction.}
        \label{fig:chapter8:appendix:phase_specific_mask}
    \end{subfigure}
    \hfill
    \begin{subfigure}{0.38\textwidth}
        \centering
        \includegraphics[width=0.55\textwidth]{\thischapter/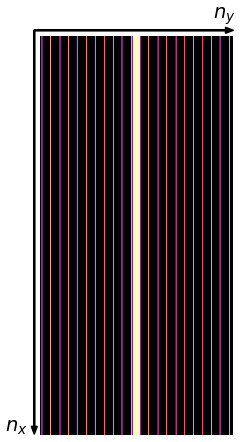}
        \caption{Unified equispaced sampling pattern.}
        \vspace{54pt}
        \label{fig:chapter8:appendix:unified_mask}
    \end{subfigure}
    \begin{subfigure}{0.3\textwidth}
        \centering
        \includegraphics[width=0.3\textwidth]{\thischapter/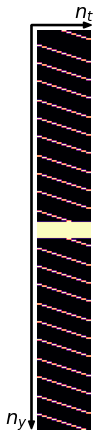}
        \caption{$k$t-Equispaced sampling pattern. Each time-step represents a scheme per each phase. Each point represents a line along the $n_x$ direction.}
        \vspace{13pt}
        \label{fig:chapter8:appendix:kt_mask}
    \end{subfigure}
    \caption{Representative examples of (fixed) sampling patterns.}
    \label{fig:chapter8:appendix:masks_equip}
\end{figure}

\begin{enumerate}
    \item Equispaced: Lines were evenly spaced to achieve the target acceleration, with the central offset randomized during training and fixed during inference. Separate patterns were used for phase-specific experiments, while a unified pattern was applied across all phases in unified experiments. Examples are shown in \Figure{chapter8:appendix:phase_specific_mask} and \Figure{chapter8:appendix:unified_mask}.
    \item $k$t-Equispaced: Used exclusively in phase-specific experiments, this strategy followed the method in \cite{tsao2003k}, where equispaced sampling was performed per phase with temporal interleaving along the time dimension. An example of this pattern is presented in \Figure{chapter8:appendix:kt_mask}.
\end{enumerate}

\subsection{Evaluation Metrics}

Assume $\vec{f}, \vec{d} \in \real^{n}$, where $\vec{f}$ denotes a ground truth image and $\vec{d}$ a prediction. Then the evaluation metrics we used are defined as follows:

  \begin{itemize}
    \item Structural Similarity Index Measure (SSIM)
    
        \begin{equation}
            \text{SSIM}(\vec{f},\,\vec{d}) =
            \frac{1}{N}\sum_{i=1}^{N} \frac{(2\mu_{\vec{f}_i}\mu_{\vec{d}_i} + c_1)(2\sigma_{\vec{f}_i\vec{d}_i} + c_2)}{({\mu^2_{\vec{f}_i}} +{\mu^2_{\vec{d}_i}} + c_1)({\sigma^2_{\vec{f}_i}} + {\sigma^2_{\vec{d}_i}} + c_2)},
            \label{eq:chapter8:appendix:ssim_metric} 
        \end{equation}
    
        where $\vec{f}_i, \vec{d}_i, i=1,...,N$ represent $7\times 7$ square windows of  $\vec{f}, \vec{d}$, respectively, and  $c_1 = 0.01$, $c_1 = 0.03$. Additionally, $\mu_{\vec{f}_i}$, $\mu_{\vec{d}_i}$ denote the means of each window, $\sigma_{\vec{f}_i}$ and $\sigma_{\vec{d}_i}$ represent the corresponding standard deviations. Lastly, $\sigma_{\vec{f}_i\vec{d}_i}$ represents the covariance between $\vec{f}_i$ and $\vec{d}_i$.

    \item Peak signal-to-noise ratio (PSNR)
    \begin{equation}
        \text{PSNR}(\vec{f},\, \vec{d})\, = \, 20\log_{10}\bigg(\frac{\max(\vec{f})}{{\,\sqrt{\frac{1}{n}\sum_{i}^n(\vec{f}_{i} - {\vec{d}}_{i})^2}}}\bigg).
        \label{eq:chapter8:appendix:psnr}
    \end{equation}
    
    \item Normalized Mean Squared Error (NMSE)
        \begin{equation}
            \text{NMSE}(\vec{f},\, \vec{d})\,= \, \frac{||\vec{f}\,-\,\vec{d}||_2^2}{||\vec{f}||_2^2}\,= \, \frac{\sum_{i=1}^n(\vec{f}_{i} - \vec{d}_{i})^2}{\sum_{i=1}^n \vec{f}_{i}^{2}}.
            \label{eq:chapter8:appendix:nmse}
        \end{equation}

    \end{itemize}

\subsection{Data Preprocessing and Postprocessing}
\subsubsection{Zero-padding}

Each 2D sequence comprised $n_t + 1 = 12$ cardiac cycle segments, $n_c = 10$ coils, and spatial dimensions varying between \( n_x' = \{448, 512\} \) and \( n_y' = \{132, 162, 168, 204, 246\} \) for the cine dataset, and \( n_x' = 416 \) and \( n_y' = 168 \) for the aorta dataset. Here, \( n_x', n_y' \) represent the pre-padding shapes.

To accommodate the fixed input size required by the MLP component of the adaptive sampling module \cite{yiasemis2024endtoendadaptivedynamicsubsampling}, data were zero-padded to the largest spatial dimensions, \( (n_x, n_y) = (512, 246) \). Specifically, fully sampled $k$-space data were transformed to the image domain using the inverse FFT, zero-padded in the image domain, and projected back to $k$-space using the FFT.

\subsubsection{Normalization}
Data were normalized using the 99.5\textsuperscript{th} percentile of the flattened magnitude of the autocalibration signal $k$-space for each moving sequence:

\begin{equation}
    s = \text{quantile}_{99.5}\Big(\text{flatten}(|\vec{y}_{\text{mov}}^{\mat{M}^{\text{acs}}}|)\Big).
\end{equation}

\subsubsection{Postprocessing Cropping}
Registered data were evaluated on a center-cropped region (in the image domain) of size \( (n_x' / 3, n_y' / 2) \), focusing on the region of interest (cardiac or aorta) as specified by the CMRxRecon challenge organizers.

\subsection{Hyperparameter Choices}

\subsubsection{Warping Transform}
The warping transform is implemented in PyTorch \cite{paszke2019pytorch} to ensure differentiability, following the spatial transformer framework described in the literature \cite{jaderberg2015spatial,balakrishnan2019voxelmorph}. More specifically, the aim is to deform an input image \( I \) based on a displacement field \( \mathbf{v} \), yielding a warped image \( I' \). The process begins by integrating the displacement field \( \mathbf{v} \) over two steps using the scaling and squaring method. Mathematically, this computes \( \mathbf{v}_{\text{integrated}} = \mathbf{v} + \mathbf{v} \circ \mathbf{v} + \dots \), where \( \circ \) denotes the composition of displacement fields. The integrated field defines a smooth transformation that is used to update spatial coordinates, ensuring the motion is captured accurately. The warped image is computed as \( I'(\mathbf{x}) = I\left(\mathbf{x} + \mathbf{v}_{\text{integrated}}(\mathbf{x})\right) \), where \( \mathbf{x} \) represents pixel coordinates. The method also normalizes the transformed grid to the range \([-1, 1]\) to match the input tensor's dimensions, applies bilinear sampling for smooth interpolation, and uses a binary mask to account for invalid or out-of-bound regions, ensuring the transformation is robust and accurate.

\subsubsection{Registration Comparisons}
We evaluated multiple registration approaches, each configured with specific hyperparameters:

\begin{enumerate}
    \item Voxelmorph with a 2D U-Net architecture comprising 4 scales with filter sizes of 16, 32, 64, and 128 was employed. For each moving (dynamic) image, we performed a forward pass for each temporal phase along with the reference image, consistent with the original implementation in \cite{balakrishnan2019voxelmorph}.image.
    \item Transmorph with a vision transformer model (ViT) inspired by the Swin Transformer \cite{liu2021swin}, configured with a patch size of \(10 \times 10\), an embedding dimension of 64, 8 layers, and 9 attention heads. The setup also includes Global Position-Sensitive Attention (GPSA) intervals and locality strength to focus on spatial regions. Each moving image phase is processed with the reference image during the forward pass, adhering to the approach described in \cite{chen2022transmorph}.
    \item  Iterative Lucas-Kanade (ILK) optical flow method from \verb|scikit-image|, configured with a radius of 5, 3 warp iterations, prefiltering enabled, and without Gaussian smoothing. All other parameters are based on the default settings of the \verb|scikit-image| in \url{https://scikit-image.org/docs/stable/api/skimage.registration.html#skimage.registration.optical_flow_ilk}.
    \item Total Variation $L_1$ optical flow method from \verb|scikit-image|, configured with an attachment weight of 15, tightness parameter of 0.3, 3 warp iterations, 5 main iterations, and a tolerance of \(1 \times 10^{-2}\). All other parameters are based on the default settings of the \verb|scikit-image| in \url{https://scikit-image.org/docs/stable/api/skimage.registration.html#skimage.registration.optical_flow_tvl1}.
    \item Demons registration method configured with 10 iterations, a Gaussian smoothing standard deviation of 1.0, and smoothing applied to the displacement field, following the standard settings of the \verb|SimpleITK| Demons algorithm. See \url{https://simpleitk.org/doxygen/latest/html/classitk_1_1simple_1_1DemonsRegistrationFilter.html}. 
\end{enumerate}
\section{Additional Results}
\label{sec:chapter8:appendix:appendix3}

\renewcommand{\theequation}{C\arabic{equation}}
\setcounter{equation}{0} 

\renewcommand{\thefigure}{C\arabic{figure}}
\setcounter{figure}{0}

\renewcommand{\thetable}{C\arabic{table}}
\setcounter{table}{0}

\subsection{Additional Experiments}

Due to space constraints in the main paper, we present additional experiments here to further validate our proposed pipeline across various settings, analyzing different aspects of its performance. We report all average results, including those from the main paper, in Tables \ref{tab:chapter8:appendix:metrics-phase-specific} to \ref{tab:chapter8:appendix:metrics-unified-aorta-recon}. These results include evaluations on both the cine test set—sharing the same data distribution as the training set—and an out-of-distribution aorta inference set.

First, we extend our registration component experiments by considering the methods described in the main paper while testing additional configurations. Specifically, we assess performance using non-adaptive sampling strategies, such as equispaced and $k$t-equispaced trajectories, alongside adaptive sampling with initialization. These experiments are conducted for both phase-specific and unified sampling approaches. Additionally, we expand our reconstruction model comparisons to include the unified sampling setting, complementing the phase-specific evaluations performed in the original experiments.

\begin{figure}[ht]

    \centering

    \begin{subfigure}[b]{0.48\textwidth}
        \centering
        \parbox[b]{0.35\columnwidth}{%
            \centering
            \includegraphics[height=0.075\textheight]{\thischapter/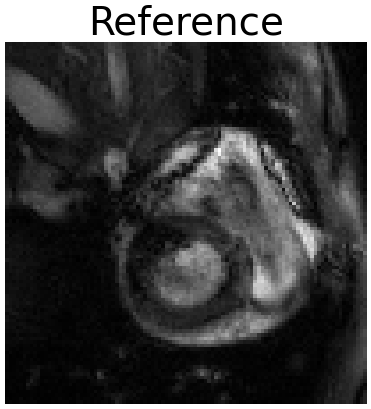}
            \vfill
            \includegraphics[height=0.12\textheight]{\thischapter/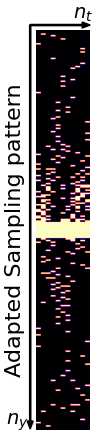}
        }%
        \hfill
        \raisebox{0pt}[0pt][0pt]{%
            \includegraphics[height=0.2\textheight]{\thischapter/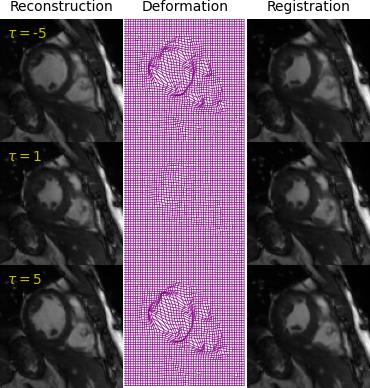}
        }
        \caption{$\alpha=1$, $\beta=1$ (original setup)}
        \label{fig:chapter8:appendix:example_original_setup}
    \end{subfigure}
    \hfill
    \begin{subfigure}[b]{0.48\textwidth}
        \centering
        \parbox[b]{0.35\columnwidth}{%
            \centering
            \includegraphics[height=0.075\textheight]{\thischapter/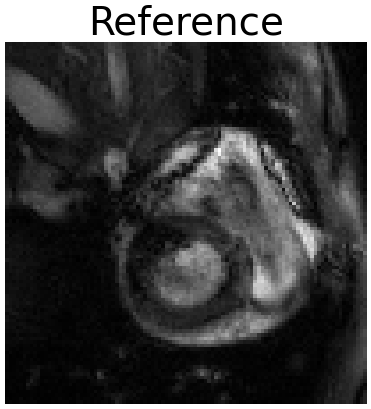}
            \vfill
            \includegraphics[height=0.12\textheight]{\thischapter/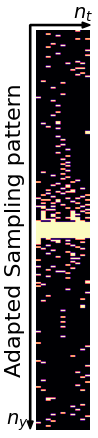}
        }%
        \hfill
        \raisebox{0pt}[0pt][0pt]{%
            \includegraphics[height=0.2\textheight]{\thischapter/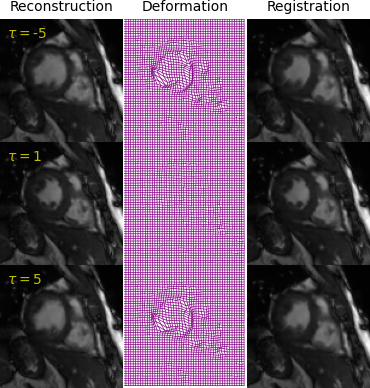}
        }
        \caption{$\alpha=1$, $\beta=2$}
        \label{fig:chapter8:appendix:example_a1_b2}
    \end{subfigure}
    \vspace{10pt} 

        \begin{subfigure}[b]{0.48\textwidth}
        \centering
        \parbox[b]{0.35\columnwidth}{%
            \centering
            \includegraphics[height=0.075\textheight]{\thischapter/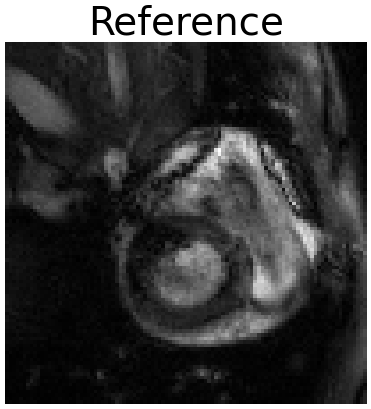}
            \vfill
            \includegraphics[height=0.12\textheight]{\thischapter/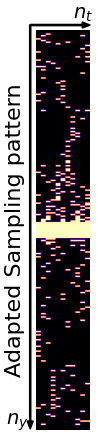}
        }%
        \hfill
        \raisebox{0pt}[0pt][0pt]{%
            \includegraphics[height=0.2\textheight]{\thischapter/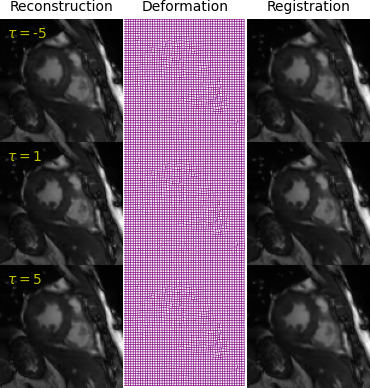}
        }
        \caption{$\alpha=1$, $\beta=3$}
        \label{fig:chapter8:appendix:example_a1_b3}
    \end{subfigure}
    \hfill
    \begin{subfigure}[b]{0.48\textwidth}
        \centering
        \parbox[b]{0.35\columnwidth}{%
            \centering
            \includegraphics[height=0.075\textheight]{\thischapter/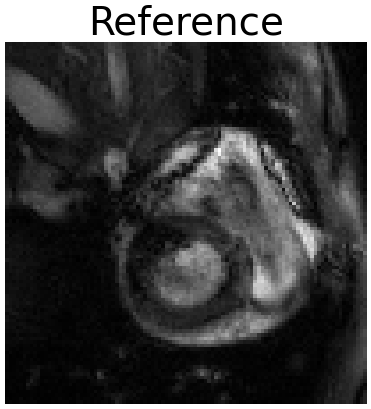}
            \vfill
            \includegraphics[height=0.12\textheight]{\thischapter/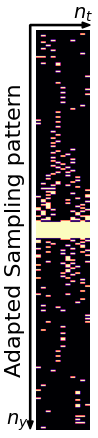}
        }%
        \hfill
        \raisebox{0pt}[0pt][0pt]{%
            \includegraphics[height=0.2\textheight]{\thischapter/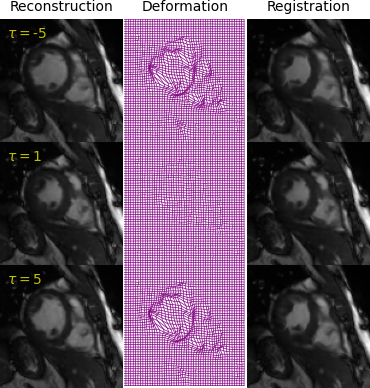}
        }
        \caption{$\alpha=\frac{3}{2}$, $\beta=2$}
        \label{fig:chapter8:appendix:example_a15_b2}
    \end{subfigure}
    \vspace{10pt} 

    \begin{subfigure}[b]{0.48\textwidth}
        \centering
        \parbox[b]{0.35\columnwidth}{%
            \centering
            \includegraphics[height=0.075\textheight]{\thischapter/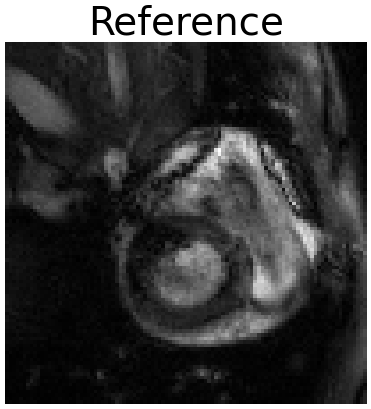}
            \vfill
            \includegraphics[height=0.12\textheight]{\thischapter/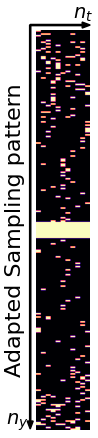}
        }%
        \hfill
        \raisebox{0pt}[0pt][0pt]{%
            \includegraphics[height=0.2\textheight]{\thischapter/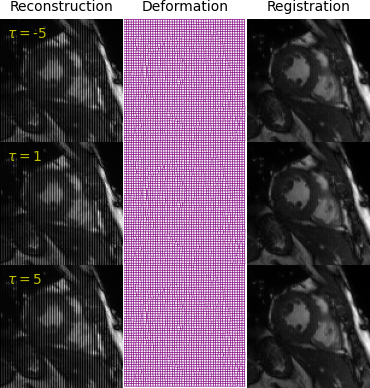}
        }
        \caption{$\alpha=0$, $\beta=1$}
        \label{fig:chapter8:appendix:example_a0}
    \end{subfigure}
    \hfill
    \begin{subfigure}[b]{0.48\textwidth}
        \centering
        \parbox[b]{0.35\columnwidth}{%
            \centering
            \includegraphics[height=0.075\textheight]{\thischapter/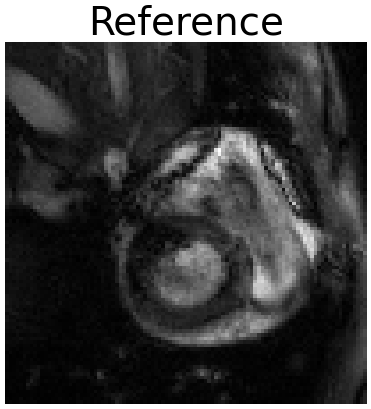}
            \vfill
            \includegraphics[height=0.12\textheight]{\thischapter/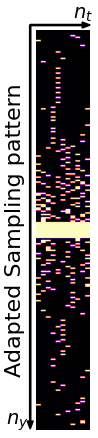}
        }%
        \hfill
        \raisebox{0pt}[0pt][0pt]{%
            \includegraphics[height=0.2\textheight]{\thischapter/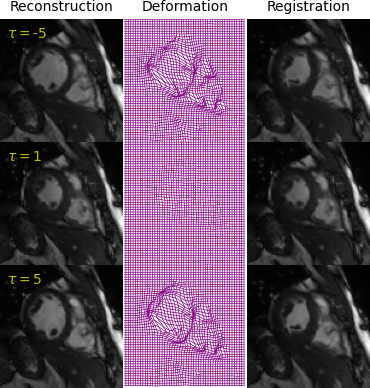}
        }
        \caption{$\alpha=2$, $\beta=1$}
        \label{fig:chapter8:appendix:example_a2_b1}
    \end{subfigure}
    
    \caption{Example I results for a case, shown at various temporal frames (\( \tau \)) relative to the reference image, for different joint loss weighting parameters at $R = 8$.}
    \label{fig:chapter8:appendix:example}
\end{figure}

\begin{figure}[ht]

    \centering

    \begin{subfigure}[b]{0.48\textwidth}
        \centering
        \parbox[b]{0.35\columnwidth}{%
            \centering
            \includegraphics[height=0.075\textheight]{\thischapter/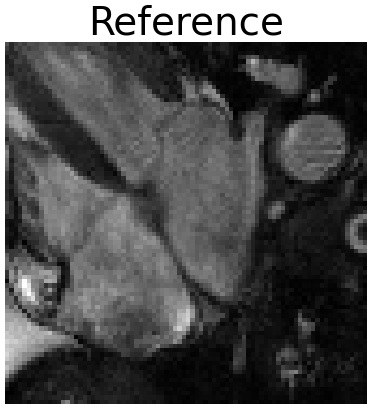}
            \vfill
            \includegraphics[height=0.12\textheight]{\thischapter/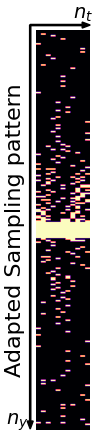}
        }%
        \hfill
        \raisebox{0pt}[0pt][0pt]{%
            \includegraphics[height=0.2\textheight]{\thischapter/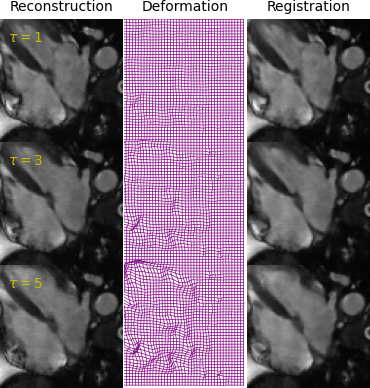}
        }
        \caption{$\alpha=1$, $\beta=1$ (original setup)}
        \label{fig:chapter8:appendix:example2_original_setup}
    \end{subfigure}
    \hfill
    \begin{subfigure}[b]{0.48\textwidth}
        \centering
        \parbox[b]{0.35\columnwidth}{%
            \centering
            \includegraphics[height=0.075\textheight]{\thischapter/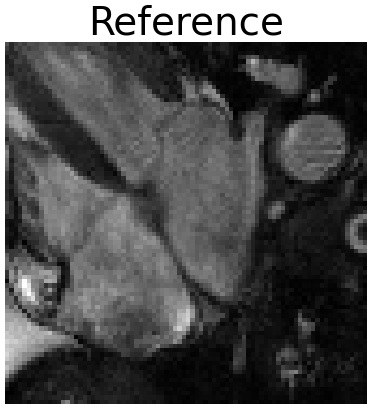}
            \vfill
            \includegraphics[height=0.12\textheight]{\thischapter/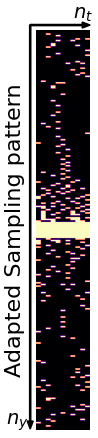}
        }%
        \hfill
        \raisebox{0pt}[0pt][0pt]{%
            \includegraphics[height=0.2\textheight]{\thischapter/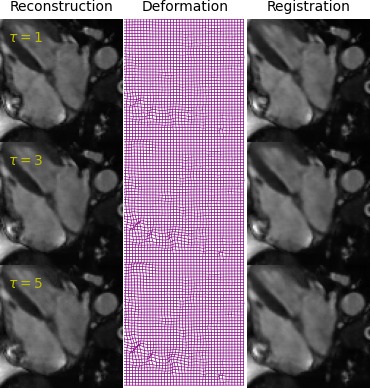}
        }
        \caption{$\alpha=1$, $\beta=2$}
        \label{fig:chapter8:appendix:example2_a1_b2}
    \end{subfigure}
    \vspace{10pt} 

        \begin{subfigure}[b]{0.48\textwidth}
        \centering
        \parbox[b]{0.35\columnwidth}{%
            \centering
            \includegraphics[height=0.075\textheight]{\thischapter/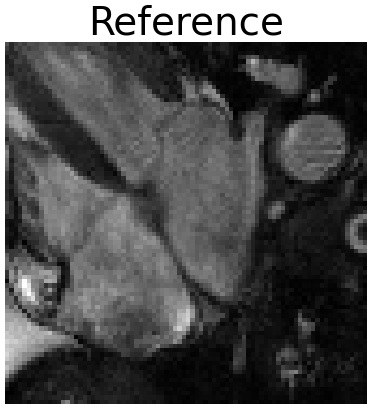}
            \vfill
            \includegraphics[height=0.12\textheight]{\thischapter/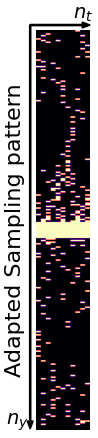}
        }%
        \hfill
        \raisebox{0pt}[0pt][0pt]{%
            \includegraphics[height=0.2\textheight]{\thischapter/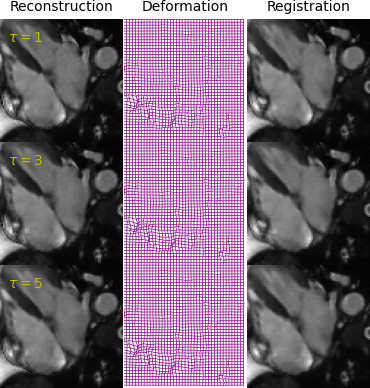}
        }
        \caption{$\alpha=1$, $\beta=3$}
        \label{fig:chapter8:appendix:example2_a1_b3}
    \end{subfigure}
    \hfill
    \begin{subfigure}[b]{0.48\textwidth}
        \centering
        \parbox[b]{0.35\columnwidth}{%
            \centering
            \includegraphics[height=0.075\textheight]{\thischapter/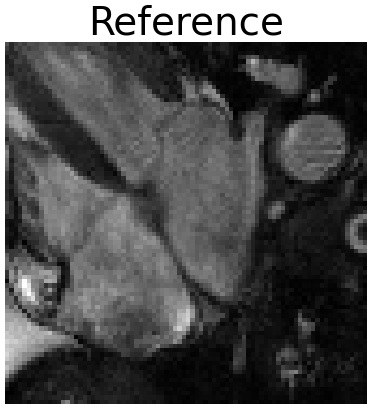}
            \vfill
            \includegraphics[height=0.12\textheight]{\thischapter/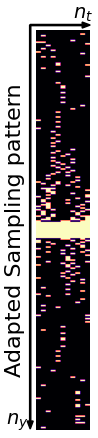}
        }%
        \hfill
        \raisebox{0pt}[0pt][0pt]{%
            \includegraphics[height=0.2\textheight]{\thischapter/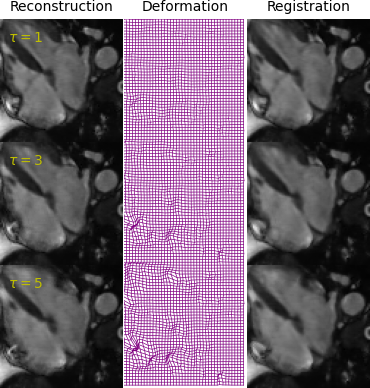}
        }
        \caption{$\alpha=\frac{3}{2}$, $\beta=2$}
        \label{fig:chapter8:appendix:example2_a15_b2}
    \end{subfigure}
    \vspace{10pt} 

    \begin{subfigure}[b]{0.48\textwidth}
        \centering
        \parbox[b]{0.35\columnwidth}{%
            \centering
            \includegraphics[height=0.075\textheight]{\thischapter/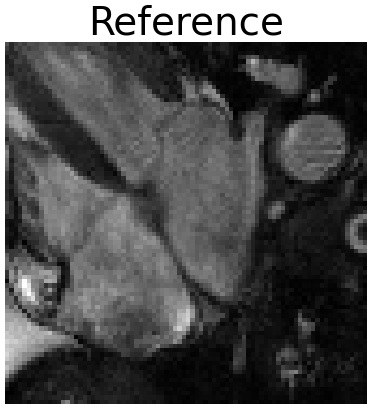}
            \vfill
            \includegraphics[height=0.12\textheight]{\thischapter/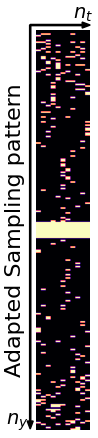}
        }%
        \hfill
        \raisebox{0pt}[0pt][0pt]{%
            \includegraphics[height=0.2\textheight]{\thischapter/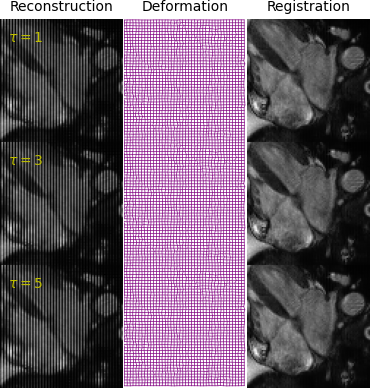}
        }
        \caption{$\alpha=0$, $\beta=1$}
        \label{fig:chapter8:appendix:example2_a0}
    \end{subfigure}
    \hfill
    \begin{subfigure}[b]{0.48\textwidth}
        \centering
        \parbox[b]{0.35\columnwidth}{%
            \centering
            \includegraphics[height=0.075\textheight]{\thischapter/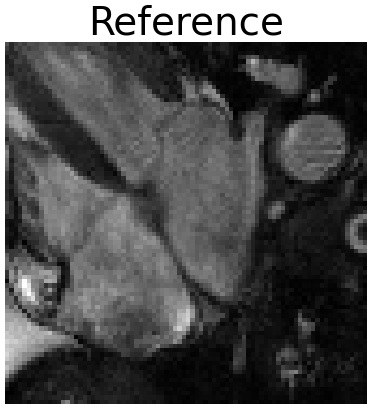}
            \vfill
            \includegraphics[height=0.12\textheight]{\thischapter/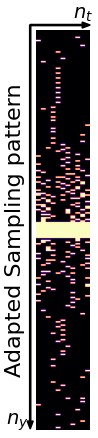}
        }%
        \hfill
        \raisebox{0pt}[0pt][0pt]{%
            \includegraphics[height=0.2\textheight]{\thischapter/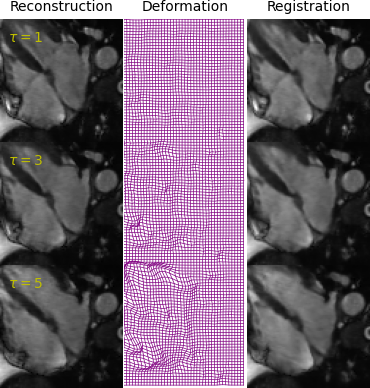}
        }
        \caption{$\alpha=2$, $\beta=1$}
        \label{fig:chapter8:appendix:example2_a2_b1}
    \end{subfigure}
    
    \caption{Example II results for a case, shown at various temporal frames (\( \tau \)) relative to the reference image, for different joint loss weighting parameters at $R = 8$.}
    \label{fig:chapter8:appendix:example2}
\end{figure}

Furthermore, we investigate the limiting case where $\alpha=0$, effectively setting the reconstruction loss weight in the joint loss calculation to zero, thereby omitting reconstruction loss optimization. We also provide qualitative visualizations of this scenario in Figures \ref{fig:chapter8:appendix:example} and \ref{fig:chapter8:appendix:example2}.

Finally, to analyze the impact of the smoothness/regularization component $\mathcal{L}_{\text{smooth}}$ within the registration loss, we conduct phase-specific experiments by removing this term. We also explore different combinations of $\alpha$ and $\beta$ in the absence of this component to assess its contribution to the overall performance.

\clearpage
\subsection{Additional Tables}

\setlength{\tabcolsep}{2.pt}
{\renewcommand{\arraystretch}{2}

\begin{table}[!hbt]
\rotatebox{90}{
\centering
\resizebox{0.85\textheight}{!}{%
\begin{tabular}{ccccccccccccccccccc}
\hline
\multirow{2}{*}{\textbf{\begin{tabular}[c]{@{}c@{}}Registration\\ Model\end{tabular}}} &
  \multirow{2}{*}{\textbf{\begin{tabular}[c]{@{}c@{}}Reconstruction\\ Model\end{tabular}}} &
  \multirow{2}{*}{\textbf{\begin{tabular}[c]{@{}c@{}}Sampling\\ Type\end{tabular}}} &
  \multirow{2}{*}{\textbf{\begin{tabular}[c]{@{}c@{}}Sampling\\ Initialization\end{tabular}}} &
  \multicolumn{5}{c}{\textbf{Loss Details}} &
  \multicolumn{3}{c}{\textbf{$4\times$}} &
  \multicolumn{3}{c}{\textbf{$6\times$}} &
  \multicolumn{3}{c}{\textbf{$8\times$}} &
  \multirow{2}{*}{\textbf{\begin{tabular}[c]{@{}c@{}}Inference\\ Time\end{tabular}}} \\
 &
   &
   &
   &
  \textbf{$\mathcal{L}_{\mathrm{sim}}, \mathcal{L}_{\mathrm{reg}}$} &
  \textbf{$\alpha$} &
  \textbf{$\beta$} &
  \textbf{$\mathcal{L}_{\mathrm{smooth}}$} &
  \textbf{E2E Loss} &
  \textbf{SSIM ($\uparrow$)} &
  \textbf{PSNR ($\uparrow$)} &
  \textbf{NMSE ($\downarrow$)} &
  \textbf{SSIM ($\uparrow$)} &
  \textbf{PSNR ($\uparrow$)} &
  \textbf{NMSE ($\downarrow$)} &
  \textbf{SSIM ($\uparrow$)} &
  \textbf{PSNR ($\uparrow$)} &
  \textbf{NMSE ($\downarrow$)} &
   \\ \hline
Learned &
  vSHARP &
  Adaptive &
  \crossmark &
  $\mathcal{L}_{\mathrm{ssim}} + \mathcal{L}_{1}$ &
  1.0 &
  1.0 &
  \checkmark &
  \checkmark &
  $0.909 \pm 0.041$ &
  $33.51 \pm 3.19$ &
  $0.025 \pm 0.015$ &
  $0.910 \pm 0.041$ &
  $33.63 \pm 3.18$ &
  $0.024 \pm 0.014$ &
  $0.909 \pm 0.042$ &
  $33.67 \pm 3.16$ &
  $0.024 \pm 0.014$ &
  10.16 \\
Voxelmorph &
  vSHARP &
  Adaptive &
  \crossmark &
  $\mathcal{L}_{\mathrm{ssim}} + \mathcal{L}_{1}$ &
  1.0 &
  1.0 &
  \checkmark &
  \checkmark &
  $0.889 \pm 0.046$ &
  $32.16 \pm 2.92$ &
  $0.033 \pm 0.017$ &
  $0.886 \pm 0.047$ &
  $32.13 \pm 2.91$ &
  $0.033 \pm 0.017$ &
  $0.882 \pm 0.048$ &
  $32.04 \pm 2.89$ &
  $0.034 \pm 0.017$ &
  10.74 \\
Transfmorph &
  vSHARP &
  Adaptive &
  \crossmark &
  $\mathcal{L}_{\mathrm{ssim}} + \mathcal{L}_{1}$ &
  1.0 &
  1.0 &
  \checkmark &
  \checkmark &
  $0.886 \pm 0.046$ &
  $31.37 \pm 3.02$ &
  $0.040 \pm 0.022$ &
  $0.886 \pm 0.046$ &
  $31.47 \pm 3.00$ &
  $0.039 \pm 0.022$ &
  $0.886 \pm 0.046$ &
  $31.50 \pm 2.95$ &
  $0.039 \pm 0.021$ &
  10.11 \\
Learned &
  vSHARP &
  Adaptive &
  \crossmark &
  $\mathcal{L}_{\mathrm{ssim}} + \mathcal{L}_{1}$ &
  1.0 &
  3.0 &
  \checkmark &
  \checkmark &
  $0.939 \pm 0.031$ &
  $36.99 \pm 3.26$ &
  $0.011 \pm 0.007$ &
  $0.936 \pm 0.032$ &
  $36.79 \pm 3.22$ &
  $0.012 \pm 0.008$ &
  $0.932 \pm 0.034$ &
  $36.49 \pm 3.19$ &
  $0.013 \pm 0.008$ &
  10.90 \\
Learned &
  vSHARP &
  Adaptive &
  \crossmark &
  $\mathcal{L}_{\mathrm{ssim}} + \mathcal{L}_{1}$ &
  1.0 &
  2.0 &
  \checkmark &
  \checkmark &
  $0.931 \pm 0.033$ &
  $35.76 \pm 3.17$ &
  $0.015 \pm 0.009$ &
  $0.930 \pm 0.034$ &
  $35.74 \pm 3.16$ &
  $0.015 \pm 0.009$ &
  $0.927 \pm 0.034$ &
  $35.60 \pm 3.12$ &
  $0.015 \pm 0.009$ &
  10.40 \\
Learned &
  vSHARP &
  Adaptive &
  \crossmark &
  $\mathcal{L}_{\mathrm{ssim}} + \mathcal{L}_{1}$ &
  1.5 &
  2.0 &
  \checkmark &
  \checkmark &
  $0.918 \pm 0.038$ &
  $34.23 \pm 3.14$ &
  $0.021 \pm 0.012$ &
  $0.918 \pm 0.038$ &
  $34.33 \pm 3.12$ &
  $0.021 \pm 0.012$ &
  $0.916 \pm 0.038$ &
  $34.33 \pm 3.10$ &
  $0.021 \pm 0.012$ &
  11.78 \\
Learned &
  vSHARP &
  Adaptive &
  \crossmark &
  $\mathcal{L}_{\mathrm{ssim}} + \mathcal{L}_{1}$ &
  0.0 &
  1.0 &
  \checkmark &
  \checkmark &
  $0.987 \pm 0.006$ &
  $44.15 \pm 2.53$ &
  $0.002 \pm 0.001$ &
  $0.985 \pm 0.007$ &
  $43.78 \pm 2.45$ &
  $0.002 \pm 0.001$ &
  $0.984 \pm 0.008$ &
  $43.22 \pm 2.41$ &
  $0.003 \pm 0.001$ &
  10.69 \\
Learned &
  vSHARP &
  Adaptive &
  \crossmark &
  $\mathcal{L}_{\mathrm{ssim}} + \mathcal{L}_{1}$ &
  2.0 &
  1.0 &
  \checkmark &
  \checkmark &
  $0.893 \pm 0.046$ &
  $32.13 \pm 3.07$ &
  $0.034 \pm 0.019$ &
  $0.893 \pm 0.046$ &
  $32.19 \pm 3.07$ &
  $0.033 \pm 0.019$ &
  $0.893 \pm 0.047$ &
  $32.22 \pm 3.06$ &
  $0.033 \pm 0.019$ &
  10.87 \\
Learned &
  vSHARP &
  Adaptive &
  \crossmark &
  $\mathcal{L}_{\mathrm{ssim}} + \mathcal{L}_{1}$ &
  1.0 &
  1.0 &
  \checkmark &
  \crossmark &
  $0.881 \pm 0.051$ &
  $31.47 \pm 3.06$ &
  $0.039 \pm 0.022$ &
  $0.881 \pm 0.051$ &
  $31.50 \pm 3.06$ &
  $0.039 \pm 0.022$ &
  $0.881 \pm 0.051$ &
  $31.63 \pm 3.04$ &
  $0.038 \pm 0.021$ &
  10.89 \\
Learned &
  vSHARP &
  Adaptive &
  \crossmark &
  $\mathcal{L}_{\mathrm{ssim}} + \mathcal{L}_{1}$ &
  1.0 &
  1.0 &
  \crossmark &
  \crossmark &
  $0.925 \pm 0.033$ &
  $33.66 \pm 3.05$ &
  $0.024 \pm 0.014$ &
  $0.926 \pm 0.033$ &
  $33.72 \pm 3.04$ &
  $0.024 \pm 0.014$ &
  $0.927 \pm 0.032$ &
  $33.82 \pm 3.04$ &
  $0.023 \pm 0.014$ &
  10.99 \\
Learned &
  vSHARP &
  Adaptive &
  \crossmark &
  $\mathcal{L}_{1}$ &
  1.0 &
  1.0 &
  \checkmark &
  \checkmark &
  $0.886 \pm 0.049$ &
  $31.69 \pm 3.10$ &
  $0.038 \pm 0.021$ &
  $0.887 \pm 0.048$ &
  $31.80 \pm 3.09$ &
  $0.037 \pm 0.021$ &
  $0.886 \pm 0.049$ &
  $31.85 \pm 3.08$ &
  $0.036 \pm 0.020$ &
  11.91 \\
Learned &
  vSHARP &
  Adaptive &
  \crossmark &
  $\mathcal{L}_{\mathrm{ssim2D}}$ &
  1.0 &
  1.0 &
  \checkmark &
  \checkmark &
  $0.896 \pm 0.043$ &
  $31.75 \pm 2.92$ &
  $0.037 \pm 0.020$ &
  $0.895 \pm 0.043$ &
  $31.83 \pm 2.90$ &
  $0.036 \pm 0.019$ &
  $0.893 \pm 0.044$ &
  $31.81 \pm 2.88$ &
  $0.036 \pm 0.019$ &
  10.47 \\
Learned &
  vSHARP &
  Adaptive &
  \crossmark &
  $\mathcal{L}_{\mathrm{ssim}} + \mathcal{L}_{1}$ &
  1.0 &
  1.0 &
  \crossmark &
  \checkmark &
  $0.948 \pm 0.024$ &
  $35.54 \pm 3.18$ &
  $0.016 \pm 0.010$ &
  $0.949 \pm 0.023$ &
  $35.74 \pm 3.18$ &
  $0.015 \pm 0.009$ &
  $0.950 \pm 0.023$ &
  $35.88 \pm 3.17$ &
  $0.015 \pm 0.009$ &
  10.66 \\
Learned &
  vSHARP &
  Adaptive &
  \crossmark &
  $\mathcal{L}_{\mathrm{ssim}} + \mathcal{L}_{1}$ &
  1.0 &
  2.0 &
  \crossmark &
  \checkmark &
  $0.967 \pm 0.015$ &
  $38.02 \pm 3.23$ &
  $0.009 \pm 0.005$ &
  $0.967 \pm 0.016$ &
  $38.08 \pm 3.21$ &
  $0.009 \pm 0.005$ &
  $0.967 \pm 0.016$ &
  $38.08 \pm 3.19$ &
  $0.009 \pm 0.005$ &
  10.40 \\
Learned &
  vSHARP &
  Adaptive &
  \crossmark &
  $\mathcal{L}_{\mathrm{ssim}} + \mathcal{L}_{1}$ &
  2.0 &
  1.0 &
  \crossmark &
  \checkmark &
  $0.933 \pm 0.030$ &
  $34.18 \pm 3.11$ &
  $0.022 \pm 0.013$ &
  $0.935 \pm 0.029$ &
  $34.34 \pm 3.10$ &
  $0.021 \pm 0.013$ &
  $0.937 \pm 0.028$ &
  $34.47 \pm 3.10$ &
  $0.020 \pm 0.013$ &
  10.95 \\
Learned &
  vSHARP &
  Adaptive &
  \checkmark &
  $\mathcal{L}_{\mathrm{ssim}} + \mathcal{L}_{1}$ &
  1.0 &
  1.0 &
  \checkmark &
  \checkmark &
  $0.907 \pm 0.042$ &
  $33.33 \pm 3.15$ &
  $0.026 \pm 0.015$ &
  $0.907 \pm 0.042$ &
  $33.45 \pm 3.15$ &
  $0.025 \pm 0.015$ &
  $0.906 \pm 0.042$ &
  $33.47 \pm 3.12$ &
  $0.025 \pm 0.015$ &
  10.18 \\
Learned &
  vSHARP &
  Equispaced &
  NA &
  $\mathcal{L}_{\mathrm{ssim}} + \mathcal{L}_{1}$ &
  1.0 &
  1.0 &
  \checkmark &
  \checkmark &
  $0.897 \pm 0.045$ &
  $32.47 \pm 3.08$ &
  $0.031 \pm 0.018$ &
  $0.898 \pm 0.045$ &
  $32.65 \pm 3.08$ &
  $0.030 \pm 0.017$ &
  $0.893 \pm 0.046$ &
  $32.61 \pm 3.00$ &
  $0.030 \pm 0.017$ &
  9.93 \\
Learned &
  vSHARP &
  kt-Equispaced &
  NA &
  $\mathcal{L}_{\mathrm{ssim}} + \mathcal{L}_{1}$ &
  1.0 &
  1.0 &
  \checkmark &
  \checkmark &
  $0.899 \pm 0.045$ &
  $32.59 \pm 3.14$ &
  $0.031 \pm 0.018$ &
  $0.900 \pm 0.045$ &
  $32.77 \pm 3.14$ &
  $0.030 \pm 0.017$ &
  $0.893 \pm 0.047$ &
  $32.73 \pm 3.09$ &
  $0.030 \pm 0.017$ &
  9.91 \\
Learned &
  vSHARP &
  Optimized &
  \crossmark &
  $\mathcal{L}_{\mathrm{ssim}} + \mathcal{L}_{1}$ &
  1.0 &
  1.0 &
  \checkmark &
  \checkmark &
  $0.912 \pm 0.040$ &
  $33.76 \pm 3.14$ &
  $0.023 \pm 0.013$ &
  $0.912 \pm 0.040$ &
  $33.87 \pm 3.14$ &
  $0.023 \pm 0.013$ &
  $0.910 \pm 0.041$ &
  $33.87 \pm 3.11$ &
  $0.023 \pm 0.013$ &
  10.85 \\
Learned &
  VarNet &
  Adaptive &
  \crossmark &
  $\mathcal{L}_{\mathrm{ssim}} + \mathcal{L}_{1}$ &
  1.0 &
  1.0 &
  \checkmark &
  \checkmark &
  $0.886 \pm 0.052$ &
  $33.15 \pm 2.97$ &
  $0.026 \pm 0.013$ &
  $0.870 \pm 0.056$ &
  $32.44 \pm 2.87$ &
  $0.030 \pm 0.015$ &
  $0.863 \pm 0.055$ &
  $32.15 \pm 2.73$ &
  $0.032 \pm 0.015$ &
  10.59 \\
OptFlowILK &
  vSHARP &
  Adaptive &
  \crossmark &
  $\mathcal{L}_{\mathrm{ssim}} + \mathcal{L}_{1}$ &
  1.0 &
  1.0 &
  \checkmark &
  NA &
  $0.890 \pm 0.044$ &
  $32.20 \pm 2.89$ &
  $0.033 \pm 0.018$ &
  $0.889 \pm 0.044$ &
  $32.21 \pm 2.89$ &
  $0.033 \pm 0.018$ &
  $0.887 \pm 0.045$ &
  $32.17 \pm 2.87$ &
  $0.033 \pm 0.018$ &
  27.27 \\
OptFlowTVL1 &
  vSHARP &
  Adaptive &
  \crossmark &
  $\mathcal{L}_{\mathrm{ssim}} + \mathcal{L}_{1}$ &
  1.0 &
  1.0 &
  \checkmark &
  NA &
  $0.893 \pm 0.042$ &
  $30.69 \pm 3.08$ &
  $0.049 \pm 0.030$ &
  $0.892 \pm 0.042$ &
  $30.66 \pm 3.06$ &
  $0.049 \pm 0.030$ &
  $0.891 \pm 0.042$ &
  $30.63 \pm 3.05$ &
  $0.049 \pm 0.030$ &
  30.20 \\
DEMONS &
  vSHARP &
  Adaptive &
  \crossmark &
  $\mathcal{L}_{\mathrm{ssim}} + \mathcal{L}_{1}$ &
  1.0 &
  1.0 &
  \checkmark &
  NA &
  $0.892 \pm 0.044$ &
  $30.77 \pm 2.87$ &
  $0.045 \pm 0.023$ &
  $0.891 \pm 0.045$ &
  $30.80 \pm 2.86$ &
  $0.045 \pm 0.023$ &
  $0.890 \pm 0.045$ &
  $30.82 \pm 2.85$ &
  $0.045 \pm 0.023$ &
  29.31 \\
OptFlowILK &
  vSHARP &
  Adaptive &
  \checkmark &
  $\mathcal{L}_{\mathrm{ssim}} + \mathcal{L}_{1}$ &
  1.0 &
  1.0 &
  \checkmark &
  NA &
  $0.891 \pm 0.044$ &
  $32.19 \pm 2.90$ &
  $0.033 \pm 0.018$ &
  $0.890 \pm 0.044$ &
  $32.21 \pm 2.89$ &
  $0.033 \pm 0.018$ &
  $0.888 \pm 0.045$ &
  $32.19 \pm 2.88$ &
  $0.033 \pm 0.018$ &
  27.18 \\
OptFlowTVL1 &
  vSHARP &
  Adaptive &
  \checkmark &
  $\mathcal{L}_{\mathrm{ssim}} + \mathcal{L}_{1}$ &
  1.0 &
  1.0 &
  \checkmark &
  NA &
  $0.895 \pm 0.041$ &
  $30.79 \pm 3.12$ &
  $0.048 \pm 0.030$ &
  $0.894 \pm 0.041$ &
  $30.76 \pm 3.12$ &
  $0.048 \pm 0.030$ &
  $0.892 \pm 0.042$ &
  $30.72 \pm 3.11$ &
  $0.048 \pm 0.030$ &
  30.16 \\
DEMONS &
  vSHARP &
  Adaptive &
  \checkmark &
  $\mathcal{L}_{\mathrm{ssim}} + \mathcal{L}_{1}$ &
  1.0 &
  1.0 &
  \checkmark &
  NA &
  $0.892 \pm 0.044$ &
  $30.75 \pm 2.89$ &
  $0.046 \pm 0.023$ &
  $0.892 \pm 0.045$ &
  $30.80 \pm 2.88$ &
  $0.045 \pm 0.023$ &
  $0.891 \pm 0.045$ &
  $30.83 \pm 2.87$ &
  $0.045 \pm 0.023$ &
  28.92 \\
OptFlowILK &
  vSHARP &
  Equispaced &
  NA &
  $\mathcal{L}_{\mathrm{ssim}} + \mathcal{L}_{1}$ &
  1.0 &
  1.0 &
  \checkmark &
  NA &
  $0.888 \pm 0.045$ &
  $32.14 \pm 2.89$ &
  $0.033 \pm 0.018$ &
  $0.886 \pm 0.045$ &
  $32.14 \pm 2.88$ &
  $0.033 \pm 0.018$ &
  $0.880 \pm 0.048$ &
  $32.01 \pm 2.84$ &
  $0.034 \pm 0.018$ &
  26.58 \\
OptFlowTVL1 &
  vSHARP &
  Equispaced &
  NA &
  $\mathcal{L}_{\mathrm{ssim}} + \mathcal{L}_{1}$ &
  1.0 &
  1.0 &
  \checkmark &
  NA &
  $0.891 \pm 0.042$ &
  $30.58 \pm 3.09$ &
  $0.050 \pm 0.031$ &
  $0.889 \pm 0.043$ &
  $30.55 \pm 3.07$ &
  $0.050 \pm 0.031$ &
  $0.883 \pm 0.045$ &
  $30.37 \pm 3.01$ &
  $0.052 \pm 0.032$ &
  30.08 \\
DEMONS &
  vSHARP &
  Equispaced &
  NA &
  $\mathcal{L}_{\mathrm{ssim}} + \mathcal{L}_{1}$ &
  1.0 &
  1.0 &
  \checkmark &
  NA &
  $0.890 \pm 0.045$ &
  $30.74 \pm 2.88$ &
  $0.046 \pm 0.023$ &
  $0.889 \pm 0.046$ &
  $30.78 \pm 2.87$ &
  $0.045 \pm 0.023$ &
  $0.884 \pm 0.047$ &
  $30.79 \pm 2.85$ &
  $0.045 \pm 0.023$ &
  29.00 \\
OptFlowILK &
  vSHARP &
  kt-Equispaced &
  NA &
  $\mathcal{L}_{\mathrm{ssim}} + \mathcal{L}_{1}$ &
  1.0 &
  1.0 &
  \checkmark &
  NA &
  $0.889 \pm 0.045$ &
  $32.18 \pm 2.91$ &
  $0.033 \pm 0.018$ &
  $0.886 \pm 0.046$ &
  $32.16 \pm 2.91$ &
  $0.033 \pm 0.018$ &
  $0.879 \pm 0.049$ &
  $32.03 \pm 2.88$ &
  $0.034 \pm 0.018$ &
  26.46 \\
OptFlowTVL1 &
  vSHARP &
  kt-Equispaced &
  NA &
  $\mathcal{L}_{\mathrm{ssim}} + \mathcal{L}_{1}$ &
  1.0 &
  1.0 &
  \checkmark &
  NA &
  $0.891 \pm 0.042$ &
  $30.62 \pm 3.09$ &
  $0.049 \pm 0.030$ &
  $0.888 \pm 0.044$ &
  $30.51 \pm 3.08$ &
  $0.051 \pm 0.031$ &
  $0.881 \pm 0.046$ &
  $30.33 \pm 3.04$ &
  $0.053 \pm 0.032$ &
  29.74 \\
DEMONS &
  vSHARP &
  kt-Equispaced &
  NA &
  $\mathcal{L}_{\mathrm{ssim}} + \mathcal{L}_{1}$ &
  1.0 &
  1.0 &
  \checkmark &
  NA &
  $0.891 \pm 0.045$ &
  $30.77 \pm 2.89$ &
  $0.045 \pm 0.023$ &
  $0.889 \pm 0.046$ &
  $30.81 \pm 2.90$ &
  $0.045 \pm 0.023$ &
  $0.884 \pm 0.048$ &
  $30.84 \pm 2.87$ &
  $0.045 \pm 0.023$ &
  28.78 \\ \hline
\end{tabular}
}
}
\caption{Registration quantitative results for various configurations on the cardiac cine test set (under phase-specific settings).}
\label{tab:chapter8:appendix:metrics-phase-specific}
\end{table}

\begin{table}[!hbt]
\rotatebox{90}{
\centering

\label{tab:chapter8:appendix:metrics-phase-specific-recon}
\resizebox{0.85\textheight}{!}{%
\begin{tabular}{ccccccccccccccccccc}
\hline
\multirow{2}{*}{\textbf{\begin{tabular}[c]{@{}c@{}}Registration\\ Model\end{tabular}}} &
  \multirow{2}{*}{\textbf{\begin{tabular}[c]{@{}c@{}}Reconstruction\\ Model\end{tabular}}} &
  \multirow{2}{*}{\textbf{\begin{tabular}[c]{@{}c@{}}Sampling\\ Type\end{tabular}}} &
  \multirow{2}{*}{\textbf{\begin{tabular}[c]{@{}c@{}}Sampling\\ Initialization\end{tabular}}} &
  \multicolumn{5}{c}{\textbf{Loss Details}} &
  \multicolumn{3}{c}{\textbf{$4\times$}} &
  \multicolumn{3}{c}{\textbf{$6\times$}} &
  \multicolumn{3}{c}{\textbf{$8\times$}} &
  \multirow{2}{*}{\textbf{\begin{tabular}[c]{@{}c@{}}Inference\\ Time\end{tabular}}} \\
 &
   &
   &
   &
  \textbf{$\mathcal{L}_{\mathrm{sim}}, \mathcal{L}_{\mathrm{reg}}$} &
  \textbf{$\alpha$} &
  \textbf{$\beta$} &
  \textbf{$\mathcal{L}_{\mathrm{smooth}}$} &
  \textbf{E2E Loss} &
  \textbf{SSIM ($\uparrow$)} &
  \textbf{PSNR ($\uparrow$)} &
  \textbf{NMSE ($\downarrow$)} &
  \textbf{SSIM ($\uparrow$)} &
  \textbf{PSNR ($\uparrow$)} &
  \textbf{NMSE ($\downarrow$)} &
  \textbf{SSIM ($\uparrow$)} &
  \textbf{PSNR ($\uparrow$)} &
  \textbf{NMSE ($\downarrow$)} &
   \\ \hline
Learned &
  vSHARP &
  Adaptive &
  \crossmark &
  $\mathcal{L}_{\mathrm{ssim}} + \mathcal{L}_{1}$ &
  1.0 &
  1.0 &
  $\checkmark$ &
  $\checkmark$ &
  $0.960 \pm 0.016$ &
  $38.70 \pm 2.43$ &
  $0.009 \pm 0.004$ &
  $0.949 \pm 0.020$ &
  $37.52 \pm 2.46$ &
  $0.011 \pm 0.006$ &
  $0.939 \pm 0.024$ &
  $36.57 \pm 2.47$ &
  $0.014 \pm 0.007$ &
  10.16 \\
Voxelmorph &
  vSHARP &
  Adaptive &
  \crossmark &
  $\mathcal{L}_{\mathrm{ssim}} + \mathcal{L}_{1}$ &
  1.0 &
  1.0 &
  $\checkmark$ &
  $\checkmark$ &
  $0.948 \pm 0.020$ &
  $37.65 \pm 2.48$ &
  $0.011 \pm 0.006$ &
  $0.934 \pm 0.026$ &
  $36.28 \pm 2.50$ &
  $0.015 \pm 0.008$ &
  $0.920 \pm 0.030$ &
  $35.16 \pm 2.53$ &
  $0.020 \pm 0.011$ &
  10.74 \\
Transfmorph &
  vSHARP &
  Adaptive &
  \crossmark &
  $\mathcal{L}_{\mathrm{ssim}} + \mathcal{L}_{1}$ &
  1.0 &
  1.0 &
  $\checkmark$ &
  $\checkmark$ &
  $0.953 \pm 0.019$ &
  $37.81 \pm 2.55$ &
  $0.010 \pm 0.005$ &
  $0.942 \pm 0.023$ &
  $36.79 \pm 2.59$ &
  $0.013 \pm 0.007$ &
  $0.931 \pm 0.028$ &
  $35.90 \pm 2.62$ &
  $0.016 \pm 0.008$ &
  10.11 \\
Learned &
  vSHARP &
  Adaptive &
  \crossmark &
  $\mathcal{L}_{\mathrm{ssim}} + \mathcal{L}_{1}$ &
  1.0 &
  3.0 &
  $\checkmark$ &
  $\checkmark$ &
  $0.900 \pm 0.034$ &
  $31.48 \pm 2.47$ &
  $0.045 \pm 0.020$ &
  $0.894 \pm 0.036$ &
  $31.35 \pm 2.47$ &
  $0.046 \pm 0.021$ &
  $0.888 \pm 0.039$ &
  $31.21 \pm 2.47$ &
  $0.048 \pm 0.022$ &
  10.90 \\
Learned &
  vSHARP &
  Adaptive &
  \crossmark &
  $\mathcal{L}_{\mathrm{ssim}} + \mathcal{L}_{1}$ &
  1.0 &
  2.0 &
  $\checkmark$ &
  $\checkmark$ &
  $0.923 \pm 0.029$ &
  $33.67 \pm 2.69$ &
  $0.028 \pm 0.014$ &
  $0.914 \pm 0.033$ &
  $33.33 \pm 2.68$ &
  $0.030 \pm 0.015$ &
  $0.906 \pm 0.035$ &
  $33.00 \pm 2.69$ &
  $0.032 \pm 0.016$ &
  10.40 \\
Learned &
  vSHARP &
  Adaptive &
  \crossmark &
  $\mathcal{L}_{\mathrm{ssim}} + \mathcal{L}_{1}$ &
  1.5 &
  2.0 &
  $\checkmark$ &
  $\checkmark$ &
  $0.950 \pm 0.019$ &
  $37.24 \pm 2.55$ &
  $0.012 \pm 0.006$ &
  $0.940 \pm 0.024$ &
  $36.34 \pm 2.56$ &
  $0.015 \pm 0.008$ &
  $0.930 \pm 0.027$ &
  $35.56 \pm 2.59$ &
  $0.018 \pm 0.009$ &
  11.78 \\
Learned &
  vSHARP &
  Adaptive &
  \crossmark &
  $\mathcal{L}_{\mathrm{ssim}} + \mathcal{L}_{1}$ &
  0.0 &
  1.0 &
  $\checkmark$ &
  $\checkmark$ &
  $0.479 \pm 0.085$ &
  $22.30 \pm 2.32$ &
  $0.353 \pm 0.119$ &
  $0.475 \pm 0.085$ &
  $22.28 \pm 2.32$ &
  $0.354 \pm 0.117$ &
  $0.471 \pm 0.086$ &
  $22.26 \pm 2.30$ &
  $0.355 \pm 0.116$ &
  10.69 \\
Learned &
  vSHARP &
  Adaptive &
  \crossmark &
  $\mathcal{L}_{\mathrm{ssim}} + \mathcal{L}_{1}$ &
  2.0 &
  1.0 &
  $\checkmark$ &
  $\checkmark$ &
  $0.971 \pm 0.012$ &
  $40.95 \pm 2.45$ &
  $0.005 \pm 0.003$ &
  $0.961 \pm 0.016$ &
  $39.38 \pm 2.48$ &
  $0.007 \pm 0.004$ &
  $0.951 \pm 0.020$ &
  $38.13 \pm 2.55$ &
  $0.010 \pm 0.005$ &
  10.87 \\
Learned &
  vSHARP &
  Adaptive &
  \crossmark &
  $\mathcal{L}_{\mathrm{ssim}} + \mathcal{L}_{1}$ &
  1.0 &
  1.0 &
  $\checkmark$ &
  \crossmark &
  $0.970 \pm 0.012$ &
  $40.87 \pm 2.44$ &
  $0.005 \pm 0.003$ &
  $0.959 \pm 0.016$ &
  $39.11 \pm 2.47$ &
  $0.008 \pm 0.004$ &
  $0.950 \pm 0.020$ &
  $37.93 \pm 2.48$ &
  $0.010 \pm 0.006$ &
  10.89 \\
Learned &
  vSHARP &
  Adaptive &
  \crossmark &
  $\mathcal{L}_{\mathrm{ssim}} + \mathcal{L}_{1}$ &
  1.0 &
  1.0 &
  \crossmark &
  \crossmark &
  $0.972 \pm 0.011$ &
  $41.26 \pm 2.44$ &
  $0.005 \pm 0.002$ &
  $0.962 \pm 0.015$ &
  $39.54 \pm 2.48$ &
  $0.007 \pm 0.004$ &
  $0.952 \pm 0.019$ &
  $38.19 \pm 2.53$ &
  $0.010 \pm 0.005$ &
  10.99 \\
Learned &
  vSHARP &
  Adaptive &
  \crossmark &
  $\mathcal{L}_{1}$ &
  1.0 &
  1.0 &
  $\checkmark$ &
  $\checkmark$ &
  $0.965 \pm 0.014$ &
  $39.68 \pm 2.46$ &
  $0.007 \pm 0.003$ &
  $0.954 \pm 0.018$ &
  $38.37 \pm 2.50$ &
  $0.009 \pm 0.005$ &
  $0.944 \pm 0.022$ &
  $37.29 \pm 2.52$ &
  $0.012 \pm 0.006$ &
  11.91 \\
Learned &
  vSHARP &
  Adaptive &
  \crossmark &
  $\mathcal{L}_{\mathrm{ssim2D}}$ &
  1.0 &
  1.0 &
  $\checkmark$ &
  $\checkmark$ &
  $0.942 \pm 0.022$ &
  $36.16 \pm 2.49$ &
  $0.015 \pm 0.007$ &
  $0.932 \pm 0.026$ &
  $35.41 \pm 2.49$ &
  $0.018 \pm 0.008$ &
  $0.922 \pm 0.029$ &
  $34.75 \pm 2.48$ &
  $0.021 \pm 0.010$ &
  10.47 \\
Learned &
  vSHARP &
  Adaptive &
  \crossmark &
  $\mathcal{L}_{\mathrm{ssim}} + \mathcal{L}_{1}$ &
  1.0 &
  1.0 &
  \crossmark &
  $\checkmark$ &
  $0.961 \pm 0.015$ &
  $39.23 \pm 2.39$ &
  $0.008 \pm 0.004$ &
  $0.950 \pm 0.020$ &
  $37.82 \pm 2.47$ &
  $0.011 \pm 0.006$ &
  $0.939 \pm 0.024$ &
  $36.70 \pm 2.51$ &
  $0.014 \pm 0.008$ &
  10.66 \\
Learned &
  vSHARP &
  Adaptive &
  \crossmark &
  $\mathcal{L}_{\mathrm{ssim}} + \mathcal{L}_{1}$ &
  1.0 &
  2.0 &
  \crossmark &
  $\checkmark$ &
  $0.931 \pm 0.027$ &
  $35.36 \pm 2.57$ &
  $0.019 \pm 0.010$ &
  $0.922 \pm 0.030$ &
  $34.84 \pm 2.58$ &
  $0.021 \pm 0.012$ &
  $0.913 \pm 0.034$ &
  $34.35 \pm 2.57$ &
  $0.024 \pm 0.013$ &
  10.40 \\
Learned &
  vSHARP &
  Adaptive &
  \crossmark &
  $\mathcal{L}_{\mathrm{ssim}} + \mathcal{L}_{1}$ &
  2.0 &
  1.0 &
  \crossmark &
  $\checkmark$ &
  $0.971 \pm 0.012$ &
  $41.04 \pm 2.43$ &
  $0.005 \pm 0.002$ &
  $0.961 \pm 0.016$ &
  $39.44 \pm 2.49$ &
  $0.007 \pm 0.004$ &
  $0.951 \pm 0.020$ &
  $38.15 \pm 2.53$ &
  $0.010 \pm 0.005$ &
  10.95 \\
Learned &
  vSHARP &
  Adaptive &
  $\checkmark$ &
  $\mathcal{L}_{\mathrm{ssim}} + \mathcal{L}_{1}$ &
  1.0 &
  1.0 &
  $\checkmark$ &
  $\checkmark$ &
  $0.964 \pm 0.014$ &
  $39.59 \pm 2.43$ &
  $0.007 \pm 0.003$ &
  $0.953 \pm 0.018$ &
  $38.29 \pm 2.47$ &
  $0.009 \pm 0.005$ &
  $0.943 \pm 0.023$ &
  $37.19 \pm 2.51$ &
  $0.012 \pm 0.006$ &
  10.18 \\
Learned &
  vSHARP &
  Equispaced &
  NA &
  $\mathcal{L}_{\mathrm{ssim}} + \mathcal{L}_{1}$ &
  1.0 &
  1.0 &
  $\checkmark$ &
  $\checkmark$ &
  $0.970 \pm 0.012$ &
  $40.80 \pm 2.41$ &
  $0.005 \pm 0.002$ &
  $0.954 \pm 0.019$ &
  $38.52 \pm 2.51$ &
  $0.009 \pm 0.004$ &
  $0.934 \pm 0.026$ &
  $36.54 \pm 2.45$ &
  $0.014 \pm 0.007$ &
  9.93 \\
Learned &
  vSHARP &
  kt-Equispaced &
  NA &
  $\mathcal{L}_{\mathrm{ssim}} + \mathcal{L}_{1}$ &
  1.0 &
  1.0 &
  $\checkmark$ &
  $\checkmark$ &
  $0.971 \pm 0.011$ &
  $40.73 \pm 2.40$ &
  $0.005 \pm 0.002$ &
  $0.952 \pm 0.019$ &
  $38.14 \pm 2.44$ &
  $0.010 \pm 0.005$ &
  $0.928 \pm 0.028$ &
  $35.92 \pm 2.43$ &
  $0.016 \pm 0.008$ &
  9.91 \\
Learned &
  vSHARP &
  Optimized &
  \crossmark &
  $\mathcal{L}_{\mathrm{ssim}} + \mathcal{L}_{1}$ &
  1.0 &
  1.0 &
  $\checkmark$ &
  $\checkmark$ &
  $0.959 \pm 0.017$ &
  $38.60 \pm 2.52$ &
  $0.009 \pm 0.005$ &
  $0.947 \pm 0.021$ &
  $37.29 \pm 2.57$ &
  $0.012 \pm 0.006$ &
  $0.936 \pm 0.026$ &
  $36.28 \pm 2.57$ &
  $0.015 \pm 0.008$ &
  10.85 \\
Learned &
  VarNet &
  Adaptive &
  \crossmark &
  $\mathcal{L}_{\mathrm{ssim}} + \mathcal{L}_{1}$ &
  1.0 &
  1.0 &
  $\checkmark$ &
  $\checkmark$ &
  $0.862 \pm 0.051$ &
  $31.39 \pm 2.56$ &
  $0.046 \pm 0.022$ &
  $0.834 \pm 0.055$ &
  $30.35 \pm 2.47$ &
  $0.057 \pm 0.025$ &
  $0.814 \pm 0.056$ &
  $29.65 \pm 2.32$ &
  $0.066 \pm 0.026$ &
  10.59 \\
OptFlowILK &
  vSHARP &
  Adaptive &
  \crossmark &
  $\mathcal{L}_{\mathrm{ssim}} + \mathcal{L}_{1}$ &
  1.0 &
  1.0 &
  $\checkmark$ &
  NA &
  $0.971 \pm 0.011$ &
  $41.05 \pm 2.42$ &
  $0.005 \pm 0.002$ &
  $0.961 \pm 0.016$ &
  $39.35 \pm 2.46$ &
  $0.007 \pm 0.004$ &
  $0.951 \pm 0.020$ &
  $38.09 \pm 2.49$ &
  $0.010 \pm 0.005$ &
  27.27 \\
OptFlowTVL1 &
  vSHARP &
  Adaptive &
  \crossmark &
  $\mathcal{L}_{\mathrm{ssim}} + \mathcal{L}_{1}$ &
  1.0 &
  1.0 &
  $\checkmark$ &
  NA &
  $0.971 \pm 0.011$ &
  $41.04 \pm 2.43$ &
  $0.005 \pm 0.002$ &
  $0.961 \pm 0.016$ &
  $39.34 \pm 2.47$ &
  $0.007 \pm 0.004$ &
  $0.951 \pm 0.020$ &
  $38.09 \pm 2.50$ &
  $0.010 \pm 0.005$ &
  30.20 \\
DEMONS &
  vSHARP &
  Adaptive &
  \crossmark &
  $\mathcal{L}_{\mathrm{ssim}} + \mathcal{L}_{1}$ &
  1.0 &
  1.0 &
  $\checkmark$ &
  NA &
  $0.971 \pm 0.011$ &
  $41.06 \pm 2.42$ &
  $0.005 \pm 0.002$ &
  $0.960 \pm 0.016$ &
  $39.33 \pm 2.47$ &
  $0.007 \pm 0.004$ &
  $0.951 \pm 0.019$ &
  $38.10 \pm 2.49$ &
  $0.010 \pm 0.005$ &
  29.31 \\
OptFlowILK &
  vSHARP &
  Adaptive &
  $\checkmark$ &
  $\mathcal{L}_{\mathrm{ssim}} + \mathcal{L}_{1}$ &
  1.0 &
  1.0 &
  $\checkmark$ &
  NA &
  $0.974 \pm 0.010$ &
  $41.73 \pm 2.41$ &
  $0.004 \pm 0.002$ &
  $0.963 \pm 0.015$ &
  $39.90 \pm 2.48$ &
  $0.006 \pm 0.003$ &
  $0.953 \pm 0.019$ &
  $38.50 \pm 2.49$ &
  $0.009 \pm 0.005$ &
  27.18 \\
OptFlowTVL1 &
  vSHARP &
  Adaptive &
  $\checkmark$ &
  $\mathcal{L}_{\mathrm{ssim}} + \mathcal{L}_{1}$ &
  1.0 &
  1.0 &
  $\checkmark$ &
  NA &
  $0.974 \pm 0.010$ &
  $41.73 \pm 2.43$ &
  $0.004 \pm 0.002$ &
  $0.963 \pm 0.015$ &
  $39.90 \pm 2.47$ &
  $0.006 \pm 0.003$ &
  $0.952 \pm 0.019$ &
  $38.50 \pm 2.49$ &
  $0.009 \pm 0.005$ &
  30.16 \\
DEMONS &
  vSHARP &
  Adaptive &
  $\checkmark$ &
  $\mathcal{L}_{\mathrm{ssim}} + \mathcal{L}_{1}$ &
  1.0 &
  1.0 &
  $\checkmark$ &
  NA &
  $0.974 \pm 0.010$ &
  $41.76 \pm 2.43$ &
  $0.004 \pm 0.002$ &
  $0.963 \pm 0.015$ &
  $39.89 \pm 2.48$ &
  $0.006 \pm 0.003$ &
  $0.952 \pm 0.019$ &
  $38.50 \pm 2.49$ &
  $0.009 \pm 0.005$ &
  28.92 \\
OptFlowILK &
  vSHARP &
  Equispaced &
  NA &
  $\mathcal{L}_{\mathrm{ssim}} + \mathcal{L}_{1}$ &
  1.0 &
  1.0 &
  $\checkmark$ &
  NA &
  $0.978 \pm 0.009$ &
  $42.78 \pm 2.45$ &
  $0.003 \pm 0.001$ &
  $0.962 \pm 0.016$ &
  $39.84 \pm 2.51$ &
  $0.006 \pm 0.003$ &
  $0.942 \pm 0.023$ &
  $37.41 \pm 2.46$ &
  $0.011 \pm 0.006$ &
  26.58 \\
OptFlowTVL1 &
  vSHARP &
  Equispaced &
  NA &
  $\mathcal{L}_{\mathrm{ssim}} + \mathcal{L}_{1}$ &
  1.0 &
  1.0 &
  $\checkmark$ &
  NA &
  $0.978 \pm 0.009$ &
  $42.78 \pm 2.45$ &
  $0.003 \pm 0.001$ &
  $0.962 \pm 0.016$ &
  $39.84 \pm 2.51$ &
  $0.006 \pm 0.003$ &
  $0.942 \pm 0.023$ &
  $37.41 \pm 2.46$ &
  $0.011 \pm 0.006$ &
  30.08 \\
DEMONS &
  vSHARP &
  Equispaced &
  NA &
  $\mathcal{L}_{\mathrm{ssim}} + \mathcal{L}_{1}$ &
  1.0 &
  1.0 &
  $\checkmark$ &
  NA &
  $0.978 \pm 0.009$ &
  $42.78 \pm 2.45$ &
  $0.003 \pm 0.001$ &
  $0.962 \pm 0.016$ &
  $39.84 \pm 2.51$ &
  $0.006 \pm 0.003$ &
  $0.942 \pm 0.023$ &
  $37.41 \pm 2.46$ &
  $0.011 \pm 0.006$ &
  29.00 \\
OptFlowILK &
  vSHARP &
  kt-Equispaced &
  NA &
  $\mathcal{L}_{\mathrm{ssim}} + \mathcal{L}_{1}$ &
  1.0 &
  1.0 &
  $\checkmark$ &
  NA &
  $0.979 \pm 0.008$ &
  $42.92 \pm 2.42$ &
  $0.003 \pm 0.001$ &
  $0.960 \pm 0.015$ &
  $39.53 \pm 2.45$ &
  $0.007 \pm 0.003$ &
  $0.938 \pm 0.024$ &
  $36.91 \pm 2.40$ &
  $0.013 \pm 0.006$ &
  26.46 \\
OptFlowTVL1 &
  vSHARP &
  kt-Equispaced &
  NA &
  $\mathcal{L}_{\mathrm{ssim}} + \mathcal{L}_{1}$ &
  1.0 &
  1.0 &
  $\checkmark$ &
  NA &
  $0.979 \pm 0.008$ &
  $42.92 \pm 2.42$ &
  $0.003 \pm 0.001$ &
  $0.960 \pm 0.015$ &
  $39.53 \pm 2.45$ &
  $0.007 \pm 0.003$ &
  $0.938 \pm 0.024$ &
  $36.91 \pm 2.40$ &
  $0.013 \pm 0.006$ &
  29.74 \\
DEMONS &
  vSHARP &
  kt-Equispaced &
  NA &
  $\mathcal{L}_{\mathrm{ssim}} + \mathcal{L}_{1}$ &
  1.0 &
  1.0 &
  $\checkmark$ &
  NA &
  $0.979 \pm 0.008$ &
  $42.92 \pm 2.42$ &
  $0.003 \pm 0.001$ &
  $0.960 \pm 0.015$ &
  $39.53 \pm 2.45$ &
  $0.007 \pm 0.003$ &
  $0.938 \pm 0.024$ &
  $36.91 \pm 2.40$ &
  $0.013 \pm 0.006$ &
  28.78 \\ \hline
\end{tabular}%
}
}
\caption{Reconstruction quantitative results for various configurations on the cardiac cine test set (under phase-specific settings).}
\end{table}
}

\setlength{\tabcolsep}{2.pt}
{\renewcommand{\arraystretch}{3.5}

\begin{table}[!hbt]
\rotatebox{90}{
\centering

\label{tab:chapter8:appendix:metrics-unified}
\resizebox{0.9\textheight}{!}{%
\begin{tabular}{ccccccccccccccccccc}
\hline
\multirow{2}{*}{\textbf{\begin{tabular}[c]{@{}c@{}}Registration\\ Model\end{tabular}}} & \multirow{2}{*}{\textbf{\begin{tabular}[c]{@{}c@{}}Reconstruction\\ Model\end{tabular}}} & \multirow{2}{*}{\textbf{\begin{tabular}[c]{@{}c@{}}Sampling\\ Type\end{tabular}}} & \multirow{2}{*}{\textbf{\begin{tabular}[c]{@{}c@{}}Sampling\\ Initialization\end{tabular}}} & \multicolumn{5}{c}{\textbf{Loss Details}}                                                                                                                               & \multicolumn{3}{c}{\textbf{$4\times$}}                                                 & \multicolumn{3}{c}{\textbf{$6\times$}}                                                 & \multicolumn{3}{c}{\textbf{$8\times$}}                                                 & \multirow{2}{*}{\textbf{\begin{tabular}[c]{@{}c@{}}Inference\\ Time\end{tabular}}} \\
                                                                                       &                                                                                          &                                                                                   &                                                                                             & \textbf{$\mathcal{L}_{\mathrm{sim}}, \mathcal{L}_{\mathrm{reg}}$} & \textbf{$\alpha$} & \textbf{$\beta$} & \textbf{$\mathcal{L}_{\mathrm{smooth}}$} & \textbf{E2E Loss} & \textbf{SSIM ($\uparrow$)} & \textbf{PSNR ($\uparrow$)} & \textbf{NMSE ($\downarrow$)} & \textbf{SSIM ($\uparrow$)} & \textbf{PSNR ($\uparrow$)} & \textbf{NMSE ($\downarrow$)} & \textbf{SSIM ($\uparrow$)} & \textbf{PSNR ($\uparrow$)} & \textbf{NMSE ($\downarrow$)} &                                                                                    \\ \hline
Learned                                                                                & vSHARP                                                                                   & Adaptive                                                                          & \crossmark                                                                                & $\mathcal{L}_{\mathrm{ssim}} + \mathcal{L}_{1}$                   & 1                 & 1                & \checkmark                             & \checkmark      & $0.877 \pm 0.050$          & $31.68 \pm 2.86$           & $0.037 \pm 0.019$            & $0.867 \pm 0.052$          & $31.41 \pm 2.76$           & $0.038 \pm 0.019$            & $0.854 \pm 0.054$          & $31.01 \pm 2.69$           & $0.042 \pm 0.020$            & 10.06                                                                              \\
Learned                                                                                & vSHARP                                                                                   & Adaptive                                                                          & \checkmark                                                                                & $\mathcal{L}_{\mathrm{ssim}} + \mathcal{L}_{1}$                   & 1                 & 1                & \checkmark                             & \checkmark      & $0.882 \pm 0.049$          & $31.90 \pm 2.92$           & $0.035 \pm 0.019$            & $0.874 \pm 0.051$          & $31.75 \pm 2.84$           & $0.036 \pm 0.019$            & $0.862 \pm 0.054$          & $31.40 \pm 2.75$           & $0.039 \pm 0.020$            & 10.08                                                                              \\
Learned                                                                                & vSHARP                                                                                   & Equispaced                                                                        & NA                                                                                          & $\mathcal{L}_{\mathrm{ssim}} + \mathcal{L}_{1}$                   & 1                 & 1                & \checkmark                             & \checkmark      & $0.879 \pm 0.050$          & $31.62 \pm 2.98$           & $0.038 \pm 0.021$            & $0.868 \pm 0.053$          & $31.51 \pm 2.88$           & $0.038 \pm 0.020$            & $0.848 \pm 0.057$          & $30.93 \pm 2.72$           & $0.043 \pm 0.021$            & 9.96                                                                               \\
Learned                                                                                & VarNet                                                                                   & Adaptive                                                                          & \crossmark                                                                                & $\mathcal{L}_{\mathrm{ssim}} + \mathcal{L}_{1}$                   & 1                 & 1                & \checkmark                             & \checkmark      & $0.837 \pm 0.056$          & $30.92 \pm 2.40$           & $0.041 \pm 0.015$            & $0.832 \pm 0.057$          & $30.77 \pm 2.40$           & $0.043 \pm 0.016$            & $0.829 \pm 0.058$          & $30.66 \pm 2.38$           & $0.044 \pm 0.016$            & 10.54                                                                              \\
DEMONS                                                                                 & vSHARP                                                                                   & Adaptive                                                                          & \crossmark                                                                                & $\mathcal{L}_{\mathrm{ssim}} + \mathcal{L}_{1}$                   & 1                 & 1                & \checkmark                             & NA                & $0.868 \pm 0.049$          & $31.53 \pm 2.68$           & $0.037 \pm 0.019$            & $0.856 \pm 0.052$          & $31.15 \pm 2.65$           & $0.041 \pm 0.019$            & $0.842 \pm 0.056$          & $30.70 \pm 2.62$           & $0.045 \pm 0.021$            & 24.93                                                                              \\
DEMONS                                                                                 & vSHARP                                                                                   & Adaptive                                                                          & \crossmark                                                                                & $\mathcal{L}_{\mathrm{ssim}} + \mathcal{L}_{1}$                   & 1                 & 1                & \checkmark                             & NA                & $0.873 \pm 0.045$          & $30.01 \pm 2.78$           & $0.055 \pm 0.031$            & $0.861 \pm 0.049$          & $29.55 \pm 2.76$           & $0.061 \pm 0.034$            & $0.848 \pm 0.053$          & $29.01 \pm 2.77$           & $0.069 \pm 0.038$            & 29.98                                                                              \\
DEMONS                                                                                 & vSHARP                                                                                   & Adaptive                                                                          & \crossmark                                                                                & $\mathcal{L}_{\mathrm{ssim}} + \mathcal{L}_{1}$                   & 1                 & 1                & \checkmark                             & NA                & $0.873 \pm 0.050$          & $30.36 \pm 2.80$           & $0.049 \pm 0.024$            & $0.863 \pm 0.053$          & $30.19 \pm 2.78$           & $0.051 \pm 0.025$            & $0.852 \pm 0.056$          & $29.93 \pm 2.73$           & $0.054 \pm 0.026$            & 29.11                                                                              \\
DEMONS                                                                                 & vSHARP                                                                                   & Adaptive                                                                          & \checkmark                                                                                & $\mathcal{L}_{\mathrm{ssim}} + \mathcal{L}_{1}$                   & 1                 & 1                & \checkmark                             & NA                & $0.874 \pm 0.047$          & $31.69 \pm 2.69$           & $0.036 \pm 0.018$            & $0.862 \pm 0.050$          & $31.34 \pm 2.61$           & $0.039 \pm 0.019$            & $0.849 \pm 0.053$          & $30.94 \pm 2.57$           & $0.042 \pm 0.020$            & 25.51                                                                              \\
DEMONS                                                                                 & vSHARP                                                                                   & Adaptive                                                                          & \checkmark                                                                                & $\mathcal{L}_{\mathrm{ssim}} + \mathcal{L}_{1}$                   & 1                 & 1                & \checkmark                             & NA                & $0.877 \pm 0.044$          & $30.18 \pm 2.86$           & $0.053 \pm 0.030$            & $0.867 \pm 0.047$          & $29.76 \pm 2.73$           & $0.058 \pm 0.033$            & $0.855 \pm 0.050$          & $29.31 \pm 2.70$           & $0.064 \pm 0.035$            & 29.90                                                                              \\
DEMONS                                                                                 & vSHARP                                                                                   & Adaptive                                                                          & \checkmark                                                                                & $\mathcal{L}_{\mathrm{ssim}} + \mathcal{L}_{1}$                   & 1                 & 1                & \checkmark                             & NA                & $0.877 \pm 0.048$          & $30.50 \pm 2.79$           & $0.048 \pm 0.024$            & $0.868 \pm 0.052$          & $30.34 \pm 2.76$           & $0.049 \pm 0.024$            & $0.857 \pm 0.054$          & $30.14 \pm 2.71$           & $0.051 \pm 0.025$            & 28.91                                                                              \\
DEMONS                                                                                 & vSHARP                                                                                   & Equispaced                                                                        & NA                                                                                          & $\mathcal{L}_{\mathrm{ssim}} + \mathcal{L}_{1}$                   & 1                 & 1                & \checkmark                             & NA                & $0.877 \pm 0.046$          & $31.82 \pm 2.74$           & $0.035 \pm 0.018$            & $0.861 \pm 0.050$          & $31.43 \pm 2.65$           & $0.038 \pm 0.018$            & $0.838 \pm 0.055$          & $30.72 \pm 2.50$           & $0.044 \pm 0.019$            & 26.66                                                                              \\
DEMONS                                                                                 & vSHARP                                                                                   & Equispaced                                                                        & NA                                                                                          & $\mathcal{L}_{\mathrm{ssim}} + \mathcal{L}_{1}$                   & 1                 & 1                & \checkmark                             & NA                & $0.882 \pm 0.044$          & $30.40 \pm 2.99$           & $0.051 \pm 0.031$            & $0.868 \pm 0.047$          & $30.06 \pm 2.87$           & $0.054 \pm 0.031$            & $0.847 \pm 0.052$          & $29.32 \pm 2.72$           & $0.064 \pm 0.035$            & 30.04                                                                              \\
DEMONS                                                                                 & vSHARP                                                                                   & Equispaced                                                                        & NA                                                                                          & $\mathcal{L}_{\mathrm{ssim}} + \mathcal{L}_{1}$                   & 1                 & 1                & \checkmark                             & NA                & $0.880 \pm 0.048$          & $30.55 \pm 2.83$           & $0.047 \pm 0.024$            & $0.865 \pm 0.052$          & $30.32 \pm 2.77$           & $0.050 \pm 0.024$            & $0.845 \pm 0.056$          & $29.89 \pm 2.64$           & $0.054 \pm 0.025$            & 29.29                                                                              \\
Learned                                                                                & vSHARP                                                                                   & Optimized                                                                         & \crossmark                                                                                & $\mathcal{L}_{\mathrm{ssim}} + \mathcal{L}_{1}$                   & 1                 & 1                & \checkmark                             & \checkmark      & $0.886 \pm 0.047$          & $32.17 \pm 2.87$           & $0.033 \pm 0.018$            & $0.879 \pm 0.048$          & $31.96 \pm 2.80$           & $0.034 \pm 0.018$            & $0.870 \pm 0.050$          & $31.71 \pm 2.72$           & $0.036 \pm 0.018$            & 10.83                                                                              \\ \hline
\end{tabular}%
}
}
\caption{Registration quantitative results for various configurations on the cardiac cine test set (under unified sampling settings).}
\end{table}

\begin{table}[!hbt]
\rotatebox{90}{
\centering

\label{tab:chapter8:appendix:metrics-unified-recon}
\resizebox{0.9\textheight}{!}{%

\begin{tabular}{ccccccccccccccccccc}
\hline
\multirow{2}{*}{\textbf{\begin{tabular}[c]{@{}c@{}}Registration\\ Model\end{tabular}}} &
  \multirow{2}{*}{\textbf{\begin{tabular}[c]{@{}c@{}}Reconstruction\\ Model\end{tabular}}} &
  \multirow{2}{*}{\textbf{\begin{tabular}[c]{@{}c@{}}Sampling\\ Type\end{tabular}}} &
  \multirow{2}{*}{\textbf{\begin{tabular}[c]{@{}c@{}}Sampling\\ Initialization\end{tabular}}} &
  \multicolumn{5}{c}{\textbf{Loss Details}} &
  \multicolumn{3}{c}{\textbf{$4\times$}} &
  \multicolumn{3}{c}{\textbf{$6\times$}} &
  \multicolumn{3}{c}{\textbf{$8\times$}} &
  \multirow{2}{*}{\textbf{\begin{tabular}[c]{@{}c@{}}Inference\\ Time\end{tabular}}} \\
 &
   &
   &
   &
  \textbf{$\mathcal{L}_{\mathrm{sim}}, \mathcal{L}_{\mathrm{reg}}$} &
  \textbf{$\alpha$} &
  \textbf{$\beta$} &
  \textbf{$\mathcal{L}_{\mathrm{smooth}}$} &
  \textbf{E2E Loss} &
  \textbf{SSIM ($\uparrow$)} &
  \textbf{PSNR ($\uparrow$)} &
  \textbf{NMSE ($\downarrow$)} &
  \textbf{SSIM ($\uparrow$)} &
  \textbf{PSNR ($\uparrow$)} &
  \textbf{NMSE ($\downarrow$)} &
  \textbf{SSIM ($\uparrow$)} &
  \textbf{PSNR ($\uparrow$)} &
  \textbf{NMSE ($\downarrow$)} &
   \\ \hline
Learned &
  vSHARP &
  Adaptive &
  \crossmark &
  $\mathcal{L}_{\mathrm{ssim}} + \mathcal{L}_{1}$ &
  1 &
  1 &
  $\checkmark$ &
  $\checkmark$ &
  $0.934 \pm 0.023$ &
  $36.07 \pm 2.26$ &
  $0.015 \pm 0.007$ &
  $0.908 \pm 0.031$ &
  $34.03 \pm 2.30$ &
  $0.024 \pm 0.010$ &
  $0.883 \pm 0.037$ &
  $32.61 \pm 2.28$ &
  $0.033 \pm 0.013$ &
  10.06 \\
Learned &
  vSHARP &
  Adaptive &
  $\checkmark$ &
  $\mathcal{L}_{\mathrm{ssim}} + \mathcal{L}_{1}$ &
  1 &
  1 &
  $\checkmark$ &
  $\checkmark$ &
  $0.942 \pm 0.021$ &
  $37.01 \pm 2.32$ &
  $0.012 \pm 0.005$ &
  $0.919 \pm 0.029$ &
  $35.10 \pm 2.29$ &
  $0.019 \pm 0.008$ &
  $0.894 \pm 0.034$ &
  $33.53 \pm 2.26$ &
  $0.027 \pm 0.013$ &
  10.08 \\
Learned &
  vSHARP &
  Equispaced &
  NA &
  $\mathcal{L}_{\mathrm{ssim}} + \mathcal{L}_{1}$ &
  1 &
  1 &
  $\checkmark$ &
  $\checkmark$ &
  $0.960 \pm 0.016$ &
  $39.46 \pm 2.43$ &
  $0.007 \pm 0.003$ &
  $0.923 \pm 0.028$ &
  $35.92 \pm 2.35$ &
  $0.016 \pm 0.007$ &
  $0.882 \pm 0.037$ &
  $33.23 \pm 2.25$ &
  $0.029 \pm 0.011$ &
  9.96 \\
Learned &
  VarNet &
  Adaptive &
  \crossmark &
  $\mathcal{L}_{\mathrm{ssim}} + \mathcal{L}_{1}$ &
  1 &
  1 &
  $\checkmark$ &
  $\checkmark$ &
  $0.768 \pm 0.052$ &
  $28.00 \pm 1.94$ &
  $0.093 \pm 0.027$ &
  $0.755 \pm 0.056$ &
  $27.69 \pm 1.99$ &
  $0.101 \pm 0.030$ &
  $0.749 \pm 0.058$ &
  $27.60 \pm 2.00$ &
  $0.103 \pm 0.031$ &
  10.54 \\
DEMONS &
  vSHARP &
  Adaptive &
  \crossmark &
  $\mathcal{L}_{\mathrm{ssim}} + \mathcal{L}_{1}$ &
  1 &
  1 &
  $\checkmark$ &
  NA &
  $0.946 \pm 0.020$ &
  $37.25 \pm 2.24$ &
  $0.012 \pm 0.005$ &
  $0.923 \pm 0.027$ &
  $35.18 \pm 2.36$ &
  $0.019 \pm 0.009$ &
  $0.900 \pm 0.034$ &
  $33.66 \pm 2.31$ &
  $0.027 \pm 0.012$ &
  24.93 \\
DEMONS &
  vSHARP &
  Adaptive &
  \crossmark &
  $\mathcal{L}_{\mathrm{ssim}} + \mathcal{L}_{1}$ &
  1 &
  1 &
  $\checkmark$ &
  NA &
  $0.947 \pm 0.020$ &
  $37.37 \pm 2.30$ &
  $0.011 \pm 0.005$ &
  $0.923 \pm 0.027$ &
  $35.21 \pm 2.30$ &
  $0.018 \pm 0.008$ &
  $0.900 \pm 0.034$ &
  $33.62 \pm 2.36$ &
  $0.027 \pm 0.012$ &
  29.98 \\
DEMONS &
  vSHARP &
  Adaptive &
  \crossmark &
  $\mathcal{L}_{\mathrm{ssim}} + \mathcal{L}_{1}$ &
  1 &
  1 &
  $\checkmark$ &
  NA &
  $0.946 \pm 0.021$ &
  $37.31 \pm 2.39$ &
  $0.012 \pm 0.006$ &
  $0.923 \pm 0.028$ &
  $35.22 \pm 2.35$ &
  $0.019 \pm 0.009$ &
  $0.900 \pm 0.033$ &
  $33.66 \pm 2.30$ &
  $0.027 \pm 0.012$ &
  29.11 \\
DEMONS &
  vSHARP &
  Adaptive &
  $\checkmark$ &
  $\mathcal{L}_{\mathrm{ssim}} + \mathcal{L}_{1}$ &
  1 &
  1 &
  $\checkmark$ &
  NA &
  $0.954 \pm 0.017$ &
  $38.37 \pm 2.28$ &
  $0.009 \pm 0.004$ &
  $0.931 \pm 0.024$ &
  $36.11 \pm 2.24$ &
  $0.015 \pm 0.007$ &
  $0.909 \pm 0.030$ &
  $34.44 \pm 2.21$ &
  $0.022 \pm 0.010$ &
  25.51 \\
DEMONS &
  vSHARP &
  Adaptive &
  $\checkmark$ &
  $\mathcal{L}_{\mathrm{ssim}} + \mathcal{L}_{1}$ &
  1 &
  1 &
  $\checkmark$ &
  NA &
  $0.954 \pm 0.017$ &
  $38.33 \pm 2.33$ &
  $0.009 \pm 0.004$ &
  $0.932 \pm 0.024$ &
  $36.14 \pm 2.24$ &
  $0.015 \pm 0.007$ &
  $0.909 \pm 0.030$ &
  $34.45 \pm 2.18$ &
  $0.022 \pm 0.010$ &
  29.90 \\
DEMONS &
  vSHARP &
  Adaptive &
  $\checkmark$ &
  $\mathcal{L}_{\mathrm{ssim}} + \mathcal{L}_{1}$ &
  1 &
  1 &
  $\checkmark$ &
  NA &
  $0.953 \pm 0.017$ &
  $38.31 \pm 2.24$ &
  $0.009 \pm 0.004$ &
  $0.931 \pm 0.025$ &
  $36.10 \pm 2.26$ &
  $0.015 \pm 0.007$ &
  $0.909 \pm 0.030$ &
  $34.48 \pm 2.19$ &
  $0.022 \pm 0.010$ &
  28.91 \\
DEMONS &
  vSHARP &
  Equispaced &
  NA &
  $\mathcal{L}_{\mathrm{ssim}} + \mathcal{L}_{1}$ &
  1 &
  1 &
  $\checkmark$ &
  NA &
  $0.967 \pm 0.013$ &
  $40.67 \pm 2.48$ &
  $0.005 \pm 0.002$ &
  $0.934 \pm 0.024$ &
  $36.94 \pm 2.40$ &
  $0.012 \pm 0.006$ &
  $0.896 \pm 0.033$ &
  $34.05 \pm 2.26$ &
  $0.024 \pm 0.010$ &
  26.66 \\
DEMONS &
  vSHARP &
  Equispaced &
  NA &
  $\mathcal{L}_{\mathrm{ssim}} + \mathcal{L}_{1}$ &
  1 &
  1 &
  $\checkmark$ &
  NA &
  $0.967 \pm 0.013$ &
  $40.67 \pm 2.48$ &
  $0.005 \pm 0.002$ &
  $0.934 \pm 0.024$ &
  $36.94 \pm 2.40$ &
  $0.012 \pm 0.006$ &
  $0.896 \pm 0.033$ &
  $34.05 \pm 2.26$ &
  $0.024 \pm 0.010$ &
  30.04 \\
DEMONS &
  vSHARP &
  Equispaced &
  NA &
  $\mathcal{L}_{\mathrm{ssim}} + \mathcal{L}_{1}$ &
  1 &
  1 &
  $\checkmark$ &
  NA &
  $0.967 \pm 0.013$ &
  $40.67 \pm 2.48$ &
  $0.005 \pm 0.002$ &
  $0.934 \pm 0.024$ &
  $36.94 \pm 2.40$ &
  $0.012 \pm 0.006$ &
  $0.896 \pm 0.033$ &
  $34.05 \pm 2.26$ &
  $0.024 \pm 0.010$ &
  29.29 \\
Learned &
  vSHARP &
  Optimized &
  \crossmark &
  $\mathcal{L}_{\mathrm{ssim}} + \mathcal{L}_{1}$ &
  1 &
  1 &
  $\checkmark$ &
  $\checkmark$ &
  $0.941 \pm 0.021$ &
  $36.59 \pm 2.23$ &
  $0.013 \pm 0.006$ &
  $0.920 \pm 0.026$ &
  $34.78 \pm 2.19$ &
  $0.020 \pm 0.008$ &
  $0.899 \pm 0.031$ &
  $33.54 \pm 2.16$ &
  $0.027 \pm 0.010$ &
  10.83 \\ \hline
\end{tabular}%
}
}
\caption{Reconstruction quantitative results for various configurations on the cardiac cine test set (under unified sampling settings).}
\end{table}
}

\setlength{\tabcolsep}{2.5pt}
{\renewcommand{\arraystretch}{3}

\begin{table}[!hbt]
\rotatebox{90}{
\centering

\label{tab:chapter8:appendix:metrics-phase-specific-aorta}
\resizebox{0.85\textheight}{!}{%
\begin{tabular}{ccccccccccccccccccc}
\hline
\multicolumn{1}{c}{\multirow{2}{*}{\textbf{\begin{tabular}[c]{@{}c@{}}Registration\\ Model\end{tabular}}}} &
  \multicolumn{1}{c}{\multirow{2}{*}{\textbf{\begin{tabular}[c]{@{}c@{}}Reconstruction\\ Model\end{tabular}}}} &
  \multicolumn{1}{c}{\multirow{2}{*}{\textbf{\begin{tabular}[c]{@{}c@{}}Sampling\\ Type\end{tabular}}}} &
  \multicolumn{1}{c}{\multirow{2}{*}{\textbf{\begin{tabular}[c]{@{}c@{}}Sampling\\ Initialization\end{tabular}}}} &
  \multicolumn{5}{c}{\textbf{Loss Details}} &
  \multicolumn{3}{c}{\textbf{$4\times$}} &
  \multicolumn{3}{c}{\textbf{$6\times$}} &
  \multicolumn{3}{c}{\textbf{$8\times$}} &
  \multicolumn{1}{c}{\multirow{2}{*}{\textbf{\begin{tabular}[c]{@{}c@{}}Inference\\ Time\end{tabular}}}} \\
\multicolumn{1}{c}{} &
  \multicolumn{1}{c}{} &
  \multicolumn{1}{c}{} &
  \multicolumn{1}{c}{} &
  \multicolumn{1}{c}{\textbf{$\mathcal{L}_{\mathrm{sim}}, \mathcal{L}_{\mathrm{reg}}$}} &
  \multicolumn{1}{c}{\textbf{$\alpha$}} &
  \multicolumn{1}{c}{\textbf{$\beta$}} &
  \multicolumn{1}{c}{\textbf{$\mathcal{L}_{\mathrm{smooth}}$}} &
  \multicolumn{1}{c}{\textbf{E2E Loss}} &
  \multicolumn{1}{c}{\textbf{SSIM ($\uparrow$)}} &
  \multicolumn{1}{c}{\textbf{PSNR ($\uparrow$)}} &
  \multicolumn{1}{c}{\textbf{NMSE ($\downarrow$)}} &
  \multicolumn{1}{c}{\textbf{SSIM ($\uparrow$)}} &
  \multicolumn{1}{c}{\textbf{PSNR ($\uparrow$)}} &
  \multicolumn{1}{c}{\textbf{NMSE ($\downarrow$)}} &
  \multicolumn{1}{c}{\textbf{SSIM ($\uparrow$)}} &
  \multicolumn{1}{c}{\textbf{PSNR ($\uparrow$)}} &
  \multicolumn{1}{c}{\textbf{NMSE ($\downarrow$)}} &
  \multicolumn{1}{c}{} \\ \hline
Learned &
  vSHARP &
  Adaptive &
  \crossmark &
  $\mathcal{L}_{\mathrm{ssim}} + \mathcal{L}_{1}$ &
  1 &
  1.0 &
  \checkmark &
  \checkmark &
  $0.859 \pm 0.082$ &
  $31.96 \pm 3.65$ &
  $0.059 \pm 0.042$ &
  $0.859 \pm 0.082$ &
  $32.04 \pm 3.67$ &
  $0.058 \pm 0.041$ &
  $0.857 \pm 0.082$ &
  $32.07 \pm 3.68$ &
  $0.058 \pm 0.041$ &
  12.38 \\
Voxelmorph &
  vSHARP &
  Adaptive &
  \crossmark &
  $\mathcal{L}_{\mathrm{ssim}} + \mathcal{L}_{1}$ &
  1 &
  1.0 &
  \checkmark &
  \checkmark &
  $0.839 \pm 0.085$ &
  $31.10 \pm 3.40$ &
  $0.069 \pm 0.044$ &
  $0.835 \pm 0.086$ &
  $31.05 \pm 3.42$ &
  $0.070 \pm 0.044$ &
  $0.830 \pm 0.086$ &
  $30.94 \pm 3.42$ &
  $0.071 \pm 0.044$ &
  12.05 \\
Transfmorph &
  vSHARP &
  Adaptive &
  \crossmark &
  $\mathcal{L}_{\mathrm{ssim}} + \mathcal{L}_{1}$ &
  1 &
  1.0 &
  \checkmark &
  \checkmark &
  $0.835 \pm 0.092$ &
  $30.46 \pm 3.48$ &
  $0.083 \pm 0.060$ &
  $0.834 \pm 0.091$ &
  $30.53 \pm 3.48$ &
  $0.082 \pm 0.059$ &
  $0.835 \pm 0.091$ &
  $30.65 \pm 3.51$ &
  $0.080 \pm 0.058$ &
  12.10 \\
Learned &
  vSHARP &
  Adaptive &
  \crossmark &
  $\mathcal{L}_{\mathrm{ssim}} + \mathcal{L}_{1}$ &
  1 &
  3.0 &
  \checkmark &
  \checkmark &
  $0.891 \pm 0.067$ &
  $34.36 \pm 3.82$ &
  $0.034 \pm 0.023$ &
  $0.887 \pm 0.069$ &
  $34.18 \pm 3.83$ &
  $0.035 \pm 0.023$ &
  $0.882 \pm 0.070$ &
  $33.93 \pm 3.80$ &
  $0.037 \pm 0.024$ &
  12.05 \\
Learned &
  vSHARP &
  Adaptive &
  \checkmark &
  $\mathcal{L}_{\mathrm{ssim}} + \mathcal{L}_{1}$ &
  1 &
  1.0 &
  \checkmark &
  \checkmark &
  $0.858 \pm 0.083$ &
  $31.87 \pm 3.61$ &
  $0.060 \pm 0.041$ &
  $0.858 \pm 0.082$ &
  $31.95 \pm 3.62$ &
  $0.059 \pm 0.040$ &
  $0.855 \pm 0.082$ &
  $31.90 \pm 3.58$ &
  $0.059 \pm 0.040$ &
  12.40 \\
Learned &
  vSHARP &
  Equispaced &
  NA &
  $\mathcal{L}_{\mathrm{ssim}} + \mathcal{L}_{1}$ &
  1 &
  1.0 &
  \checkmark &
  \checkmark &
  $0.849 \pm 0.086$ &
  $31.13 \pm 3.53$ &
  $0.071 \pm 0.049$ &
  $0.851 \pm 0.086$ &
  $31.25 \pm 3.53$ &
  $0.069 \pm 0.048$ &
  $0.847 \pm 0.087$ &
  $31.24 \pm 3.52$ &
  $0.069 \pm 0.047$ &
  11.97 \\
Learned &
  vSHARP &
  kt-Equispaced &
  NA &
  $\mathcal{L}_{\mathrm{ssim}} + \mathcal{L}_{1}$ &
  1 &
  1.0 &
  \checkmark &
  \checkmark &
  $0.847 \pm 0.086$ &
  $31.19 \pm 3.56$ &
  $0.070 \pm 0.049$ &
  $0.847 \pm 0.085$ &
  $31.27 \pm 3.53$ &
  $0.069 \pm 0.047$ &
  $0.839 \pm 0.087$ &
  $31.20 \pm 3.55$ &
  $0.069 \pm 0.047$ &
  12.19 \\
Learned &
  vSHARP &
  Optimized &
  \crossmark &
  $\mathcal{L}_{\mathrm{ssim}} + \mathcal{L}_{1}$ &
  1 &
  1.0 &
  \checkmark &
  \checkmark &
  $0.863 \pm 0.081$ &
  $32.20 \pm 3.60$ &
  $0.055 \pm 0.037$ &
  $0.863 \pm 0.080$ &
  $32.26 \pm 3.58$ &
  $0.054 \pm 0.036$ &
  $0.861 \pm 0.080$ &
  $32.25 \pm 3.56$ &
  $0.054 \pm 0.035$ &
  12.43 \\
Learned &
  VarNet &
  Adaptive &
  \crossmark &
  $\mathcal{L}_{\mathrm{ssim}} + \mathcal{L}_{1}$ &
  1 &
  1.0 &
  \checkmark &
  \checkmark &
  $0.835 \pm 0.083$ &
  $31.42 \pm 3.25$ &
  $0.060 \pm 0.031$ &
  $0.811 \pm 0.087$ &
  $30.57 \pm 3.07$ &
  $0.071 \pm 0.031$ &
  $0.807 \pm 0.089$ &
  $30.34 \pm 2.99$ &
  $0.074 \pm 0.031$ &
  11.65 \\
OptFlowILK &
  vSHARP &
  Adaptive &
  \crossmark &
  $\mathcal{L}_{\mathrm{ssim}} + \mathcal{L}_{1}$ &
  1 &
  1.0 &
  \checkmark &
  NA &
  $0.829 \pm 0.089$ &
  $30.41 \pm 3.25$ &
  $0.081 \pm 0.053$ &
  $0.827 \pm 0.089$ &
  $30.42 \pm 3.23$ &
  $0.081 \pm 0.052$ &
  $0.825 \pm 0.089$ &
  $30.41 \pm 3.20$ &
  $0.081 \pm 0.051$ &
  31.02 \\
OptFlowTVL1 &
  vSHARP &
  Adaptive &
  \crossmark &
  $\mathcal{L}_{\mathrm{ssim}} + \mathcal{L}_{1}$ &
  1 &
  1.0 &
  \checkmark &
  NA &
  $0.817 \pm 0.071$ &
  $28.11 \pm 2.32$ &
  $0.129 \pm 0.064$ &
  $0.814 \pm 0.071$ &
  $28.02 \pm 2.28$ &
  $0.132 \pm 0.065$ &
  $0.811 \pm 0.070$ &
  $27.93 \pm 2.27$ &
  $0.134 \pm 0.065$ &
  35.28 \\
DEMONS &
  vSHARP &
  Adaptive &
  \crossmark &
  $\mathcal{L}_{\mathrm{ssim}} + \mathcal{L}_{1}$ &
  1 &
  1.0 &
  \checkmark &
  NA &
  $0.848 \pm 0.087$ &
  $30.58 \pm 3.74$ &
  $0.085 \pm 0.072$ &
  $0.847 \pm 0.087$ &
  $30.57 \pm 3.72$ &
  $0.085 \pm 0.070$ &
  $0.844 \pm 0.087$ &
  $30.55 \pm 3.71$ &
  $0.085 \pm 0.069$ &
  32.77 \\
OptFlowILK &
  vSHARP &
  Adaptive &
  \checkmark &
  $\mathcal{L}_{\mathrm{ssim}} + \mathcal{L}_{1}$ &
  1 &
  1.0 &
  \checkmark &
  NA &
  $0.830 \pm 0.090$ &
  $30.43 \pm 3.29$ &
  $0.081 \pm 0.053$ &
  $0.830 \pm 0.090$ &
  $30.46 \pm 3.26$ &
  $0.080 \pm 0.052$ &
  $0.827 \pm 0.090$ &
  $30.46 \pm 3.24$ &
  $0.080 \pm 0.051$ &
  31.39 \\
OptFlowTVL1 &
  vSHARP &
  Adaptive &
  \checkmark &
  $\mathcal{L}_{\mathrm{ssim}} + \mathcal{L}_{1}$ &
  1 &
  1.0 &
  \checkmark &
  NA &
  $0.823 \pm 0.073$ &
  $28.29 \pm 2.40$ &
  $0.125 \pm 0.064$ &
  $0.822 \pm 0.073$ &
  $28.29 \pm 2.39$ &
  $0.125 \pm 0.064$ &
  $0.819 \pm 0.073$ &
  $28.21 \pm 2.40$ &
  $0.127 \pm 0.064$ &
  35.43 \\
DEMONS &
  vSHARP &
  Adaptive &
  \checkmark &
  $\mathcal{L}_{\mathrm{ssim}} + \mathcal{L}_{1}$ &
  1 &
  1.0 &
  \checkmark &
  NA &
  $0.850 \pm 0.088$ &
  $30.59 \pm 3.77$ &
  $0.085 \pm 0.072$ &
  $0.849 \pm 0.088$ &
  $30.63 \pm 3.76$ &
  $0.084 \pm 0.070$ &
  $0.846 \pm 0.088$ &
  $30.62 \pm 3.73$ &
  $0.083 \pm 0.068$ &
  33.39 \\
OptFlowILK &
  vSHARP &
  Equispaced &
  NA &
  $\mathcal{L}_{\mathrm{ssim}} + \mathcal{L}_{1}$ &
  1 &
  1.0 &
  \checkmark &
  NA &
  $0.827 \pm 0.091$ &
  $30.37 \pm 3.29$ &
  $0.082 \pm 0.054$ &
  $0.826 \pm 0.091$ &
  $30.40 \pm 3.26$ &
  $0.081 \pm 0.052$ &
  $0.821 \pm 0.092$ &
  $30.33 \pm 3.21$ &
  $0.082 \pm 0.050$ &
  29.50 \\
OptFlowTVL1 &
  vSHARP &
  Equispaced &
  NA &
  $\mathcal{L}_{\mathrm{ssim}} + \mathcal{L}_{1}$ &
  1 &
  1.0 &
  \checkmark &
  NA &
  $0.819 \pm 0.072$ &
  $28.12 \pm 2.32$ &
  $0.128 \pm 0.063$ &
  $0.817 \pm 0.072$ &
  $28.05 \pm 2.29$ &
  $0.130 \pm 0.063$ &
  $0.808 \pm 0.074$ &
  $27.81 \pm 2.27$ &
  $0.136 \pm 0.064$ &
  35.18 \\
DEMONS &
  vSHARP &
  Equispaced &
  NA &
  $\mathcal{L}_{\mathrm{ssim}} + \mathcal{L}_{1}$ &
  1 &
  1.0 &
  \checkmark &
  NA &
  $0.848 \pm 0.088$ &
  $30.54 \pm 3.76$ &
  $0.086 \pm 0.074$ &
  $0.847 \pm 0.089$ &
  $30.55 \pm 3.72$ &
  $0.085 \pm 0.071$ &
  $0.841 \pm 0.090$ &
  $30.49 \pm 3.65$ &
  $0.085 \pm 0.067$ &
  32.74 \\
OptFlowILK &
  vSHARP &
  kt-Equispaced &
  NA &
  $\mathcal{L}_{\mathrm{ssim}} + \mathcal{L}_{1}$ &
  1 &
  1.0 &
  \checkmark &
  NA &
  $0.826 \pm 0.091$ &
  $30.36 \pm 3.30$ &
  $0.082 \pm 0.054$ &
  $0.824 \pm 0.091$ &
  $30.37 \pm 3.28$ &
  $0.082 \pm 0.053$ &
  $0.816 \pm 0.093$ &
  $30.30 \pm 3.24$ &
  $0.082 \pm 0.051$ &
  30.42 \\
OptFlowTVL1 &
  vSHARP &
  kt-Equispaced &
  NA &
  $\mathcal{L}_{\mathrm{ssim}} + \mathcal{L}_{1}$ &
  1 &
  1.0 &
  \checkmark &
  NA &
  $0.815 \pm 0.073$ &
  $28.11 \pm 2.35$ &
  $0.129 \pm 0.064$ &
  $0.810 \pm 0.073$ &
  $27.98 \pm 2.33$ &
  $0.133 \pm 0.065$ &
  $0.798 \pm 0.076$ &
  $27.67 \pm 2.32$ &
  $0.142 \pm 0.067$ &
  34.99 \\
DEMONS &
  vSHARP &
  kt-Equispaced &
  NA &
  $\mathcal{L}_{\mathrm{ssim}} + \mathcal{L}_{1}$ &
  1 &
  1.0 &
  \checkmark &
  NA &
  $0.847 \pm 0.088$ &
  $30.54 \pm 3.77$ &
  $0.086 \pm 0.073$ &
  $0.845 \pm 0.088$ &
  $30.53 \pm 3.73$ &
  $0.086 \pm 0.071$ &
  $0.837 \pm 0.090$ &
  $30.48 \pm 3.68$ &
  $0.085 \pm 0.065$ &
  32.54 \\ \hline
\end{tabular}%
}
}
\caption{Registration uantitative results for various configurations on the aorta inference set (under phase-specific settings).}
\end{table}

\begin{table}[!hbt]
\rotatebox{90}{
\centering

\label{tab:chapter8:appendix:metrics-phase-specific-aorta-recon}
\resizebox{0.85\textheight}{!}{%
\begin{tabular}{lllllllllllllllllll}
\hline
\multicolumn{1}{c}{\multirow{2}{*}{\textbf{\begin{tabular}[c]{@{}c@{}}Registration\\ Model\end{tabular}}}} &
  \multicolumn{1}{c}{\multirow{2}{*}{\textbf{\begin{tabular}[c]{@{}c@{}}Reconstruction\\ Model\end{tabular}}}} &
  \multicolumn{1}{c}{\multirow{2}{*}{\textbf{\begin{tabular}[c]{@{}c@{}}Sampling\\ Type\end{tabular}}}} &
  \multicolumn{1}{c}{\multirow{2}{*}{\textbf{\begin{tabular}[c]{@{}c@{}}Sampling\\ Initialization\end{tabular}}}} &
  \multicolumn{5}{c}{\textbf{Loss Details}} &
  \multicolumn{3}{c}{\textbf{$4\times$}} &
  \multicolumn{3}{c}{\textbf{$6\times$}} &
  \multicolumn{3}{c}{\textbf{$8\times$}} &
  \multicolumn{1}{c}{\multirow{2}{*}{\textbf{\begin{tabular}[c]{@{}c@{}}Inference\\ Time\end{tabular}}}} \\
\multicolumn{1}{c}{} &
  \multicolumn{1}{c}{} &
  \multicolumn{1}{c}{} &
  \multicolumn{1}{c}{} &
  \multicolumn{1}{c}{\textbf{$\mathcal{L}_{\mathrm{sim}}, \mathcal{L}_{\mathrm{reg}}$}} &
  \multicolumn{1}{c}{\textbf{$\alpha$}} &
  \multicolumn{1}{c}{\textbf{$\beta$}} &
  \multicolumn{1}{c}{\textbf{$\mathcal{L}_{\mathrm{smooth}}$}} &
  \multicolumn{1}{c}{\textbf{E2E Loss}} &
  \multicolumn{1}{c}{\textbf{SSIM ($\uparrow$)}} &
  \multicolumn{1}{c}{\textbf{PSNR ($\uparrow$)}} &
  \multicolumn{1}{c}{\textbf{NMSE ($\downarrow$)}} &
  \multicolumn{1}{c}{\textbf{SSIM ($\uparrow$)}} &
  \multicolumn{1}{c}{\textbf{PSNR ($\uparrow$)}} &
  \multicolumn{1}{c}{\textbf{NMSE ($\downarrow$)}} &
  \multicolumn{1}{c}{\textbf{SSIM ($\uparrow$)}} &
  \multicolumn{1}{c}{\textbf{PSNR ($\uparrow$)}} &
  \multicolumn{1}{c}{\textbf{NMSE ($\downarrow$)}} &
  \multicolumn{1}{c}{} \\ \hline
Learned &
  vSHARP &
  Adaptive &
  \crossmark &
  $\mathcal{L}_{\mathrm{ssim}} + \mathcal{L}_{1}$ &
  1 &
  1.0 &
  $\checkmark$ &
  $\checkmark$ &
  $0.944 \pm 0.030$ &
  $38.63 \pm 3.61$ &
  $0.021 \pm 0.013$ &
  $0.929 \pm 0.039$ &
  $37.32 \pm 3.61$ &
  $0.028 \pm 0.017$ &
  $0.914 \pm 0.047$ &
  $36.26 \pm 3.57$ &
  $0.035 \pm 0.022$ &
  12.38 \\
Voxelmorph &
  vSHARP &
  Adaptive &
  \crossmark &
  $\mathcal{L}_{\mathrm{ssim}} + \mathcal{L}_{1}$ &
  1 &
  1.0 &
  $\checkmark$ &
  $\checkmark$ &
  $0.923 \pm 0.040$ &
  $37.02 \pm 3.66$ &
  $0.030 \pm 0.020$ &
  $0.906 \pm 0.049$ &
  $35.84 \pm 3.60$ &
  $0.039 \pm 0.025$ &
  $0.889 \pm 0.057$ &
  $34.80 \pm 3.50$ &
  $0.049 \pm 0.030$ &
  12.05 \\
Transfmorph &
  vSHARP &
  Adaptive &
  \crossmark &
  $\mathcal{L}_{\mathrm{ssim}} + \mathcal{L}_{1}$ &
  1 &
  1.0 &
  $\checkmark$ &
  $\checkmark$ &
  $0.939 \pm 0.034$ &
  $38.19 \pm 3.70$ &
  $0.023 \pm 0.014$ &
  $0.923 \pm 0.043$ &
  $36.99 \pm 3.68$ &
  $0.030 \pm 0.019$ &
  $0.909 \pm 0.052$ &
  $35.96 \pm 3.63$ &
  $0.038 \pm 0.023$ &
  12.10 \\
Learned &
  vSHARP &
  Adaptive &
  \crossmark &
  $\mathcal{L}_{\mathrm{ssim}} + \mathcal{L}_{1}$ &
  1 &
  3.0 &
  $\checkmark$ &
  $\checkmark$ &
  $0.893 \pm 0.057$ &
  $33.38 \pm 3.28$ &
  $0.070 \pm 0.049$ &
  $0.885 \pm 0.062$ &
  $33.17 \pm 3.28$ &
  $0.074 \pm 0.051$ &
  $0.876 \pm 0.067$ &
  $32.92 \pm 3.29$ &
  $0.078 \pm 0.053$ &
  12.05 \\
Learned &
  vSHARP &
  Adaptive &
  $\checkmark$ &
  $\mathcal{L}_{\mathrm{ssim}} + \mathcal{L}_{1}$ &
  1 &
  1.0 &
  $\checkmark$ &
  $\checkmark$ &
  $0.952 \pm 0.026$ &
  $39.48 \pm 3.62$ &
  $0.017 \pm 0.010$ &
  $0.936 \pm 0.035$ &
  $37.94 \pm 3.63$ &
  $0.024 \pm 0.015$ &
  $0.919 \pm 0.044$ &
  $36.70 \pm 3.57$ &
  $0.032 \pm 0.019$ &
  12.40 \\
Learned &
  vSHARP &
  Equispaced &
  NA &
  $\mathcal{L}_{\mathrm{ssim}} + \mathcal{L}_{1}$ &
  1 &
  1.0 &
  $\checkmark$ &
  $\checkmark$ &
  $0.964 \pm 0.019$ &
  $41.05 \pm 3.61$ &
  $0.012 \pm 0.007$ &
  $0.943 \pm 0.032$ &
  $38.75 \pm 3.66$ &
  $0.020 \pm 0.013$ &
  $0.918 \pm 0.046$ &
  $36.72 \pm 3.69$ &
  $0.032 \pm 0.021$ &
  11.97 \\
Learned &
  vSHARP &
  kt-Equispaced &
  NA &
  $\mathcal{L}_{\mathrm{ssim}} + \mathcal{L}_{1}$ &
  1 &
  1.0 &
  $\checkmark$ &
  $\checkmark$ &
  $0.962 \pm 0.019$ &
  $40.92 \pm 3.59$ &
  $0.012 \pm 0.008$ &
  $0.939 \pm 0.032$ &
  $38.54 \pm 3.67$ &
  $0.021 \pm 0.013$ &
  $0.909 \pm 0.047$ &
  $36.14 \pm 3.60$ &
  $0.036 \pm 0.022$ &
  12.19 \\
Learned &
  vSHARP &
  Optimized &
  \crossmark &
  $\mathcal{L}_{\mathrm{ssim}} + \mathcal{L}_{1}$ &
  1 &
  1.0 &
  $\checkmark$ &
  $\checkmark$ &
  $0.941 \pm 0.032$ &
  $38.13 \pm 3.55$ &
  $0.023 \pm 0.015$ &
  $0.923 \pm 0.043$ &
  $36.71 \pm 3.50$ &
  $0.032 \pm 0.019$ &
  $0.908 \pm 0.050$ &
  $35.70 \pm 3.43$ &
  $0.039 \pm 0.023$ &
  12.43 \\
Learned &
  VarNet &
  Adaptive &
  \crossmark &
  $\mathcal{L}_{\mathrm{ssim}} + \mathcal{L}_{1}$ &
  1 &
  1.0 &
  $\checkmark$ &
  $\checkmark$ &
  $0.834 \pm 0.071$ &
  $31.65 \pm 3.02$ &
  $0.093 \pm 0.042$ &
  $0.794 \pm 0.080$ &
  $30.49 \pm 2.96$ &
  $0.119 \pm 0.045$ &
  $0.777 \pm 0.085$ &
  $30.02 \pm 2.97$ &
  $0.132 \pm 0.048$ &
  11.65 \\
OptFlowILK &
  vSHARP &
  Adaptive &
  \crossmark &
  $\mathcal{L}_{\mathrm{ssim}} + \mathcal{L}_{1}$ &
  1 &
  1.0 &
  $\checkmark$ &
  NA &
  $0.957 \pm 0.022$ &
  $40.20 \pm 3.63$ &
  $0.014 \pm 0.009$ &
  $0.940 \pm 0.032$ &
  $38.38 \pm 3.64$ &
  $0.022 \pm 0.014$ &
  $0.925 \pm 0.040$ &
  $37.07 \pm 3.58$ &
  $0.029 \pm 0.018$ &
  31.02 \\
OptFlowTVL1 &
  vSHARP &
  Adaptive &
  \crossmark &
  $\mathcal{L}_{\mathrm{ssim}} + \mathcal{L}_{1}$ &
  1 &
  1.0 &
  $\checkmark$ &
  NA &
  $0.957 \pm 0.022$ &
  $40.20 \pm 3.62$ &
  $0.014 \pm 0.009$ &
  $0.940 \pm 0.032$ &
  $38.37 \pm 3.63$ &
  $0.022 \pm 0.014$ &
  $0.925 \pm 0.040$ &
  $37.08 \pm 3.58$ &
  $0.029 \pm 0.018$ &
  35.28 \\
DEMONS &
  vSHARP &
  Adaptive &
  \crossmark &
  $\mathcal{L}_{\mathrm{ssim}} + \mathcal{L}_{1}$ &
  1 &
  1.0 &
  $\checkmark$ &
  NA &
  $0.957 \pm 0.022$ &
  $40.19 \pm 3.62$ &
  $0.014 \pm 0.009$ &
  $0.940 \pm 0.032$ &
  $38.37 \pm 3.64$ &
  $0.022 \pm 0.014$ &
  $0.925 \pm 0.040$ &
  $37.07 \pm 3.58$ &
  $0.029 \pm 0.018$ &
  32.77 \\
OptFlowILK &
  vSHARP &
  Adaptive &
  $\checkmark$ &
  $\mathcal{L}_{\mathrm{ssim}} + \mathcal{L}_{1}$ &
  1 &
  1.0 &
  $\checkmark$ &
  NA &
  $0.963 \pm 0.020$ &
  $41.07 \pm 3.66$ &
  $0.012 \pm 0.007$ &
  $0.946 \pm 0.030$ &
  $39.03 \pm 3.71$ &
  $0.019 \pm 0.012$ &
  $0.929 \pm 0.039$ &
  $37.50 \pm 3.64$ &
  $0.027 \pm 0.017$ &
  31.39 \\
OptFlowTVL1 &
  vSHARP &
  Adaptive &
  $\checkmark$ &
  $\mathcal{L}_{\mathrm{ssim}} + \mathcal{L}_{1}$ &
  1 &
  1.0 &
  $\checkmark$ &
  NA &
  $0.963 \pm 0.020$ &
  $41.07 \pm 3.67$ &
  $0.012 \pm 0.007$ &
  $0.946 \pm 0.030$ &
  $39.04 \pm 3.72$ &
  $0.019 \pm 0.012$ &
  $0.929 \pm 0.039$ &
  $37.50 \pm 3.65$ &
  $0.027 \pm 0.017$ &
  35.43 \\
DEMONS &
  vSHARP &
  Adaptive &
  $\checkmark$ &
  $\mathcal{L}_{\mathrm{ssim}} + \mathcal{L}_{1}$ &
  1 &
  1.0 &
  $\checkmark$ &
  NA &
  $0.963 \pm 0.020$ &
  $41.07 \pm 3.67$ &
  $0.012 \pm 0.007$ &
  $0.946 \pm 0.030$ &
  $39.02 \pm 3.72$ &
  $0.019 \pm 0.012$ &
  $0.929 \pm 0.039$ &
  $37.50 \pm 3.64$ &
  $0.027 \pm 0.017$ &
  33.39 \\
OptFlowILK &
  vSHARP &
  Equispaced &
  NA &
  $\mathcal{L}_{\mathrm{ssim}} + \mathcal{L}_{1}$ &
  1 &
  1.0 &
  $\checkmark$ &
  NA &
  $0.976 \pm 0.013$ &
  $43.54 \pm 3.64$ &
  $0.006 \pm 0.004$ &
  $0.953 \pm 0.025$ &
  $39.96 \pm 3.63$ &
  $0.015 \pm 0.009$ &
  $0.925 \pm 0.041$ &
  $37.26 \pm 3.67$ &
  $0.028 \pm 0.018$ &
  29.50 \\
OptFlowTVL1 &
  vSHARP &
  Equispaced &
  NA &
  $\mathcal{L}_{\mathrm{ssim}} + \mathcal{L}_{1}$ &
  1 &
  1.0 &
  $\checkmark$ &
  NA &
  $0.976 \pm 0.013$ &
  $43.54 \pm 3.64$ &
  $0.006 \pm 0.004$ &
  $0.953 \pm 0.025$ &
  $39.96 \pm 3.63$ &
  $0.015 \pm 0.009$ &
  $0.925 \pm 0.041$ &
  $37.26 \pm 3.67$ &
  $0.028 \pm 0.018$ &
  35.18 \\
DEMONS &
  vSHARP &
  Equispaced &
  NA &
  $\mathcal{L}_{\mathrm{ssim}} + \mathcal{L}_{1}$ &
  1 &
  1.0 &
  $\checkmark$ &
  NA &
  $0.976 \pm 0.013$ &
  $43.54 \pm 3.64$ &
  $0.006 \pm 0.004$ &
  $0.953 \pm 0.025$ &
  $39.96 \pm 3.63$ &
  $0.015 \pm 0.009$ &
  $0.925 \pm 0.041$ &
  $37.26 \pm 3.67$ &
  $0.028 \pm 0.018$ &
  32.74 \\
OptFlowILK &
  vSHARP &
  kt-Equispaced &
  NA &
  $\mathcal{L}_{\mathrm{ssim}} + \mathcal{L}_{1}$ &
  1 &
  1.0 &
  $\checkmark$ &
  NA &
  $0.975 \pm 0.013$ &
  $43.41 \pm 3.59$ &
  $0.007 \pm 0.004$ &
  $0.951 \pm 0.025$ &
  $39.70 \pm 3.68$ &
  $0.016 \pm 0.009$ &
  $0.918 \pm 0.042$ &
  $36.64 \pm 3.61$ &
  $0.032 \pm 0.020$ &
  30.42 \\
OptFlowTVL1 &
  vSHARP &
  kt-Equispaced &
  NA &
  $\mathcal{L}_{\mathrm{ssim}} + \mathcal{L}_{1}$ &
  1 &
  1.0 &
  $\checkmark$ &
  NA &
  $0.975 \pm 0.013$ &
  $43.41 \pm 3.59$ &
  $0.007 \pm 0.004$ &
  $0.951 \pm 0.025$ &
  $39.70 \pm 3.68$ &
  $0.016 \pm 0.009$ &
  $0.918 \pm 0.042$ &
  $36.64 \pm 3.61$ &
  $0.032 \pm 0.020$ &
  34.99 \\
DEMONS &
  vSHARP &
  kt-Equispaced &
  NA &
  $\mathcal{L}_{\mathrm{ssim}} + \mathcal{L}_{1}$ &
  1 &
  1.0 &
  $\checkmark$ &
  NA &
  $0.975 \pm 0.013$ &
  $43.41 \pm 3.59$ &
  $0.007 \pm 0.004$ &
  $0.951 \pm 0.025$ &
  $39.70 \pm 3.68$ &
  $0.016 \pm 0.009$ &
  $0.918 \pm 0.042$ &
  $36.64 \pm 3.61$ &
  $0.032 \pm 0.020$ &
  32.54 \\ \hline
\end{tabular}%
}
}
\caption{Reconstruction quantitative results for various configurations on the aorta inference set (under phase-specific settings).}
\end{table}
}

\setlength{\tabcolsep}{2.pt}
{\renewcommand{\arraystretch}{3.5}

\begin{table}[!hbt]
\rotatebox{90}{
\centering

\label{tab:chapter8:appendix:metrics-unified-aorta}
\resizebox{0.9\textheight}{!}{%
\begin{tabular}{ccccccccccccccccccc}
\hline
\multirow{2}{*}{\textbf{\begin{tabular}[c]{@{}c@{}}Registration\\ Model\end{tabular}}} &
  \multirow{2}{*}{\textbf{\begin{tabular}[c]{@{}c@{}}Reconstruction\\ Model\end{tabular}}} &
  \multirow{2}{*}{\textbf{\begin{tabular}[c]{@{}c@{}}Sampling\\ Type\end{tabular}}} &
  \multirow{2}{*}{\textbf{\begin{tabular}[c]{@{}c@{}}Sampling\\ Initialization\end{tabular}}} &
  \multicolumn{5}{c}{\textbf{Loss Details}} &
  \multicolumn{3}{c}{\textbf{$4\times$}} &
  \multicolumn{3}{c}{\textbf{$6\times$}} &
  \multicolumn{3}{c}{\textbf{$8\times$}} &
  \multirow{2}{*}{\textbf{\begin{tabular}[c]{@{}c@{}}Inference\\ Time\end{tabular}}} \\
 &
   &
   &
   &
  \textbf{$\mathcal{L}_{\mathrm{sim}}, \mathcal{L}_{\mathrm{reg}}$} &
  \textbf{$\alpha$} &
  \textbf{$\beta$} &
  \textbf{$\mathcal{L}_{\mathrm{smooth}}$} &
  \textbf{E2E Loss} &
  \textbf{SSIM ($\uparrow$)} &
  \textbf{PSNR ($\uparrow$)} &
  \textbf{NMSE ($\downarrow$)} &
  \textbf{SSIM ($\uparrow$)} &
  \textbf{PSNR ($\uparrow$)} &
  \textbf{NMSE ($\downarrow$)} &
  \textbf{SSIM ($\uparrow$)} &
  \textbf{PSNR ($\uparrow$)} &
  \textbf{NMSE ($\downarrow$)} &
   \\ \hline
Learned &
  vSHARP &
  Adaptive &
  \crossmark &
  $\mathcal{L}_{\mathrm{ssim}} + \mathcal{L}_{1}$ &
  1 &
  1 &
  $\checkmark$ &
  $\checkmark$ &
  $0.837 \pm 0.088$ &
  $30.79 \pm 3.45$ &
  $0.075 \pm 0.051$ &
  $0.825 \pm 0.090$ &
  $30.49 \pm 3.33$ &
  $0.079 \pm 0.049$ &
  $0.809 \pm 0.092$ &
  $30.04 \pm 3.23$ &
  $0.085 \pm 0.049$ &
  12.25 \\
Learned &
  vSHARP &
  Adaptive &
  $\checkmark$ &
  $\mathcal{L}_{\mathrm{ssim}} + \mathcal{L}_{1}$ &
  1 &
  1 &
  $\checkmark$ &
  $\checkmark$ &
  $0.841 \pm 0.087$ &
  $30.91 \pm 3.49$ &
  $0.074 \pm 0.052$ &
  $0.832 \pm 0.086$ &
  $30.77 \pm 3.41$ &
  $0.075 \pm 0.050$ &
  $0.820 \pm 0.087$ &
  $30.43 \pm 3.30$ &
  $0.079 \pm 0.049$ &
  12.18 \\
Learned &
  vSHARP &
  Equispaced &
  NA &
  $\mathcal{L}_{\mathrm{ssim}} + \mathcal{L}_{1}$ &
  1 &
  1 &
  $\checkmark$ &
  $\checkmark$ &
  $0.837 \pm 0.090$ &
  $30.67 \pm 3.50$ &
  $0.078 \pm 0.054$ &
  $0.828 \pm 0.091$ &
  $30.59 \pm 3.40$ &
  $0.079 \pm 0.052$ &
  $0.803 \pm 0.093$ &
  $29.90 \pm 3.14$ &
  $0.088 \pm 0.051$ &
  12.10 \\
Learned &
  VarNet &
  Adaptive &
  \crossmark &
  $\mathcal{L}_{\mathrm{ssim}} + \mathcal{L}_{1}$ &
  1 &
  1 &
  $\checkmark$ &
  $\checkmark$ &
  $0.782 \pm 0.089$ &
  $29.56 \pm 2.80$ &
  $0.086 \pm 0.028$ &
  $0.769 \pm 0.090$ &
  $29.15 \pm 2.69$ &
  $0.094 \pm 0.029$ &
  $0.763 \pm 0.089$ &
  $28.97 \pm 2.66$ &
  $0.098 \pm 0.030$ &
  11.74 \\
DEMONS &
  vSHARP &
  Adaptive &
  \crossmark &
  $\mathcal{L}_{\mathrm{ssim}} + \mathcal{L}_{1}$ &
  1 &
  1 &
  $\checkmark$ &
  NA &
  $0.810 \pm 0.092$ &
  $29.95 \pm 3.09$ &
  $0.087 \pm 0.052$ &
  $0.795 \pm 0.092$ &
  $29.53 \pm 2.93$ &
  $0.094 \pm 0.050$ &
  $0.776 \pm 0.092$ &
  $29.04 \pm 2.80$ &
  $0.102 \pm 0.048$ &
  31.46 \\
DEMONS &
  vSHARP &
  Adaptive &
  \crossmark &
  $\mathcal{L}_{\mathrm{ssim}} + \mathcal{L}_{1}$ &
  1 &
  1 &
  $\checkmark$ &
  NA &
  $0.792 \pm 0.072$ &
  $27.52 \pm 2.20$ &
  $0.145 \pm 0.065$ &
  $0.771 \pm 0.072$ &
  $26.97 \pm 2.12$ &
  $0.162 \pm 0.069$ &
  $0.747 \pm 0.071$ &
  $26.44 \pm 2.07$ &
  $0.181 \pm 0.073$ &
  35.27 \\
DEMONS &
  vSHARP &
  Adaptive &
  \crossmark &
  $\mathcal{L}_{\mathrm{ssim}} + \mathcal{L}_{1}$ &
  1 &
  1 &
  $\checkmark$ &
  NA &
  $0.832 \pm 0.090$ &
  $30.13 \pm 3.57$ &
  $0.091 \pm 0.071$ &
  $0.817 \pm 0.091$ &
  $29.72 \pm 3.39$ &
  $0.096 \pm 0.069$ &
  $0.800 \pm 0.093$ &
  $29.28 \pm 3.23$ &
  $0.103 \pm 0.066$ &
  33.34 \\
DEMONS &
  vSHARP &
  Adaptive &
  $\checkmark$ &
  $\mathcal{L}_{\mathrm{ssim}} + \mathcal{L}_{1}$ &
  1 &
  1 &
  $\checkmark$ &
  NA &
  $0.817 \pm 0.091$ &
  $30.14 \pm 3.17$ &
  $0.084 \pm 0.052$ &
  $0.805 \pm 0.091$ &
  $29.84 \pm 3.03$ &
  $0.088 \pm 0.050$ &
  $0.787 \pm 0.090$ &
  $29.43 \pm 2.90$ &
  $0.095 \pm 0.049$ &
  31.41 \\
DEMONS &
  vSHARP &
  Adaptive &
  $\checkmark$ &
  $\mathcal{L}_{\mathrm{ssim}} + \mathcal{L}_{1}$ &
  1 &
  1 &
  $\checkmark$ &
  NA &
  $0.801 \pm 0.072$ &
  $27.70 \pm 2.24$ &
  $0.139 \pm 0.064$ &
  $0.784 \pm 0.071$ &
  $27.26 \pm 2.19$ &
  $0.152 \pm 0.066$ &
  $0.764 \pm 0.070$ &
  $26.79 \pm 2.15$ &
  $0.168 \pm 0.071$ &
  35.10 \\
DEMONS &
  vSHARP &
  Adaptive &
  $\checkmark$ &
  $\mathcal{L}_{\mathrm{ssim}} + \mathcal{L}_{1}$ &
  1 &
  1 &
  $\checkmark$ &
  NA &
  $0.837 \pm 0.089$ &
  $30.30 \pm 3.64$ &
  $0.088 \pm 0.071$ &
  $0.825 \pm 0.090$ &
  $30.03 \pm 3.50$ &
  $0.091 \pm 0.069$ &
  $0.809 \pm 0.091$ &
  $29.65 \pm 3.34$ &
  $0.096 \pm 0.066$ &
  33.46 \\
DEMONS &
  vSHARP &
  Equispaced &
  NA &
  $\mathcal{L}_{\mathrm{ssim}} + \mathcal{L}_{1}$ &
  1 &
  1 &
  $\checkmark$ &
  NA &
  $0.820 \pm 0.091$ &
  $30.22 \pm 3.27$ &
  $0.084 \pm 0.054$ &
  $0.806 \pm 0.092$ &
  $29.98 \pm 3.14$ &
  $0.088 \pm 0.053$ &
  $0.778 \pm 0.093$ &
  $29.38 \pm 2.92$ &
  $0.097 \pm 0.050$ &
  31.04 \\
DEMONS &
  vSHARP &
  Equispaced &
  NA &
  $\mathcal{L}_{\mathrm{ssim}} + \mathcal{L}_{1}$ &
  1 &
  1 &
  $\checkmark$ &
  NA &
  $0.809 \pm 0.073$ &
  $28.00 \pm 2.33$ &
  $0.132 \pm 0.064$ &
  $0.792 \pm 0.074$ &
  $27.64 \pm 2.27$ &
  $0.141 \pm 0.066$ &
  $0.758 \pm 0.074$ &
  $26.87 \pm 2.17$ &
  $0.165 \pm 0.068$ &
  35.10 \\
DEMONS &
  vSHARP &
  Equispaced &
  NA &
  $\mathcal{L}_{\mathrm{ssim}} + \mathcal{L}_{1}$ &
  1 &
  1 &
  $\checkmark$ &
  NA &
  $0.840 \pm 0.090$ &
  $30.40 \pm 3.73$ &
  $0.088 \pm 0.074$ &
  $0.826 \pm 0.092$ &
  $30.14 \pm 3.62$ &
  $0.091 \pm 0.072$ &
  $0.797 \pm 0.094$ &
  $29.50 \pm 3.34$ &
  $0.100 \pm 0.067$ &
  33.09 \\
Learned &
  vSHARP &
  Optimized &
  \crossmark &
  $\mathcal{L}_{\mathrm{ssim}} + \mathcal{L}_{1}$ &
  1 &
  1 &
  $\checkmark$ &
  $\checkmark$ &
  $0.846 \pm 0.085$ &
  $31.15 \pm 3.51$ &
  $0.070 \pm 0.049$ &
  $0.838 \pm 0.086$ &
  $30.85 \pm 3.41$ &
  $0.073 \pm 0.048$ &
  $0.828 \pm 0.086$ &
  $30.59 \pm 3.31$ &
  $0.076 \pm 0.047$ &
  12.23 \\ \hline
\end{tabular}%
}
}
\caption{Registration quantitative results for various configurations on the aorta inference set (under unified sampling settings).}
\end{table}

\begin{table}[!hbt]
\rotatebox{90}{
\centering
\resizebox{0.9\textheight}{!}{%
\begin{tabular}{ccccccccccccccccccc}
\hline
\multirow{2}{*}{\textbf{\begin{tabular}[c]{@{}c@{}}Registration\\ Model\end{tabular}}} &
  \multirow{2}{*}{\textbf{\begin{tabular}[c]{@{}c@{}}Reconstruction\\ Model\end{tabular}}} &
  \multirow{2}{*}{\textbf{\begin{tabular}[c]{@{}c@{}}Sampling\\ Type\end{tabular}}} &
  \multirow{2}{*}{\textbf{\begin{tabular}[c]{@{}c@{}}Sampling\\ Initialization\end{tabular}}} &
  \multicolumn{5}{c}{\textbf{Loss Details}} &
  \multicolumn{3}{c}{\textbf{$4\times$}} &
  \multicolumn{3}{c}{\textbf{$6\times$}} &
  \multicolumn{3}{c}{\textbf{}} &
  \multirow{2}{*}{\textbf{\begin{tabular}[c]{@{}c@{}}Inference\\ Time\end{tabular}}} \\
 &
   &
   &
   &
  \textbf{$\mathcal{L}_{\mathrm{sim}}, \mathcal{L}_{\mathrm{reg}}$} &
  \textbf{$\alpha$} &
  \textbf{$\beta$} &
  \textbf{$\mathcal{L}_{\mathrm{smooth}}$} &
  \textbf{E2E Loss} &
  \textbf{SSIM ($\uparrow$)} &
  \textbf{PSNR ($\uparrow$)} &
  \textbf{NMSE ($\downarrow$)} &
  \textbf{SSIM ($\uparrow$)} &
  \textbf{PSNR ($\uparrow$)} &
  \textbf{NMSE ($\downarrow$)} &
  \textbf{SSIM ($\uparrow$)} &
  \textbf{PSNR ($\uparrow$)} &
  \textbf{NMSE ($\downarrow$)} &
   \\ \hline
Learned &
  vSHARP &
  Adaptive &
  \crossmark &
  $\mathcal{L}_{\mathrm{ssim}} + \mathcal{L}_{1}$ &
  1 &
  1 &
  $\checkmark$ &
  $\checkmark$ &
  $0.927 \pm 0.035$ &
  $36.96 \pm 3.41$ &
  $0.029 \pm 0.015$ &
  $0.897 \pm 0.048$ &
  $35.00 \pm 3.37$ &
  $0.044 \pm 0.021$ &
  $0.868 \pm 0.058$ &
  $33.52 \pm 3.30$ &
  $0.061 \pm 0.027$ &
  12.25 \\
Learned &
  vSHARP &
  Adaptive &
  $\checkmark$ &
  $\mathcal{L}_{\mathrm{ssim}} + \mathcal{L}_{1}$ &
  1 &
  1 &
  $\checkmark$ &
  $\checkmark$ &
  $0.936 \pm 0.031$ &
  $37.90 \pm 3.52$ &
  $0.023 \pm 0.011$ &
  $0.910 \pm 0.042$ &
  $35.99 \pm 3.45$ &
  $0.035 \pm 0.018$ &
  $0.882 \pm 0.051$ &
  $34.40 \pm 3.32$ &
  $0.050 \pm 0.023$ &
  12.18 \\
Learned &
  vSHARP &
  Equispaced &
  NA &
  $\mathcal{L}_{\mathrm{ssim}} + \mathcal{L}_{1}$ &
  1 &
  1 &
  $\checkmark$ &
  $\checkmark$ &
  $0.963 \pm 0.019$ &
  $41.11 \pm 3.58$ &
  $0.011 \pm 0.007$ &
  $0.924 \pm 0.038$ &
  $37.27 \pm 3.52$ &
  $0.027 \pm 0.015$ &
  $0.871 \pm 0.056$ &
  $34.03 \pm 3.22$ &
  $0.054 \pm 0.024$ &
  12.10 \\
Learned &
  VarNet &
  Adaptive &
  \crossmark &
  $\mathcal{L}_{\mathrm{ssim}} + \mathcal{L}_{1}$ &
  1 &
  1 &
  $\checkmark$ &
  $\checkmark$ &
  $0.733 \pm 0.086$ &
  $28.76 \pm 2.93$ &
  $0.171 \pm 0.045$ &
  $0.707 \pm 0.091$ &
  $28.16 \pm 2.92$ &
  $0.196 \pm 0.048$ &
  $0.698 \pm 0.094$ &
  $28.06 \pm 2.94$ &
  $0.200 \pm 0.048$ &
  11.74 \\
DEMONS &
  vSHARP &
  Adaptive &
  \crossmark &
  $\mathcal{L}_{\mathrm{ssim}} + \mathcal{L}_{1}$ &
  1 &
  1 &
  $\checkmark$ &
  NA &
  $0.936 \pm 0.028$ &
  $37.63 \pm 3.32$ &
  $0.024 \pm 0.011$ &
  $0.905 \pm 0.041$ &
  $35.24 \pm 3.25$ &
  $0.041 \pm 0.018$ &
  $0.874 \pm 0.052$ &
  $33.58 \pm 3.15$ &
  $0.059 \pm 0.023$ &
  31.46 \\
DEMONS &
  vSHARP &
  Adaptive &
  \crossmark &
  $\mathcal{L}_{\mathrm{ssim}} + \mathcal{L}_{1}$ &
  1 &
  1 &
  $\checkmark$ &
  NA &
  $0.937 \pm 0.028$ &
  $37.68 \pm 3.34$ &
  $0.024 \pm 0.012$ &
  $0.906 \pm 0.041$ &
  $35.27 \pm 3.24$ &
  $0.041 \pm 0.018$ &
  $0.875 \pm 0.051$ &
  $33.59 \pm 3.14$ &
  $0.059 \pm 0.023$ &
  35.27 \\
DEMONS &
  vSHARP &
  Adaptive &
  \crossmark &
  $\mathcal{L}_{\mathrm{ssim}} + \mathcal{L}_{1}$ &
  1 &
  1 &
  $\checkmark$ &
  NA &
  $0.937 \pm 0.029$ &
  $37.66 \pm 3.40$ &
  $0.024 \pm 0.012$ &
  $0.905 \pm 0.041$ &
  $35.18 \pm 3.19$ &
  $0.041 \pm 0.018$ &
  $0.875 \pm 0.051$ &
  $33.59 \pm 3.15$ &
  $0.059 \pm 0.023$ &
  33.34 \\
DEMONS &
  vSHARP &
  Adaptive &
  $\checkmark$ &
  $\mathcal{L}_{\mathrm{ssim}} + \mathcal{L}_{1}$ &
  1 &
  1 &
  $\checkmark$ &
  NA &
  $0.947 \pm 0.025$ &
  $38.94 \pm 3.50$ &
  $0.018 \pm 0.009$ &
  $0.920 \pm 0.036$ &
  $36.56 \pm 3.42$ &
  $0.031 \pm 0.015$ &
  $0.890 \pm 0.047$ &
  $34.73 \pm 3.27$ &
  $0.046 \pm 0.021$ &
  31.41 \\
DEMONS &
  vSHARP &
  Adaptive &
  $\checkmark$ &
  $\mathcal{L}_{\mathrm{ssim}} + \mathcal{L}_{1}$ &
  1 &
  1 &
  $\checkmark$ &
  NA &
  $0.948 \pm 0.024$ &
  $38.97 \pm 3.38$ &
  $0.018 \pm 0.009$ &
  $0.920 \pm 0.037$ &
  $36.54 \pm 3.41$ &
  $0.031 \pm 0.015$ &
  $0.890 \pm 0.047$ &
  $34.75 \pm 3.27$ &
  $0.046 \pm 0.021$ &
  35.10 \\
DEMONS &
  vSHARP &
  Adaptive &
  $\checkmark$ &
  $\mathcal{L}_{\mathrm{ssim}} + \mathcal{L}_{1}$ &
  1 &
  1 &
  $\checkmark$ &
  NA &
  $0.948 \pm 0.024$ &
  $38.99 \pm 3.46$ &
  $0.018 \pm 0.009$ &
  $0.920 \pm 0.037$ &
  $36.56 \pm 3.41$ &
  $0.031 \pm 0.015$ &
  $0.890 \pm 0.047$ &
  $34.75 \pm 3.27$ &
  $0.046 \pm 0.021$ &
  33.46 \\
DEMONS &
  vSHARP &
  Equispaced &
  NA &
  $\mathcal{L}_{\mathrm{ssim}} + \mathcal{L}_{1}$ &
  1 &
  1 &
  $\checkmark$ &
  NA &
  $0.968 \pm 0.015$ &
  $42.33 \pm 3.60$ &
  $0.009 \pm 0.005$ &
  $0.934 \pm 0.031$ &
  $38.32 \pm 3.59$ &
  $0.021 \pm 0.012$ &
  $0.884 \pm 0.050$ &
  $34.86 \pm 3.34$ &
  $0.045 \pm 0.021$ &
  31.04 \\
DEMONS &
  vSHARP &
  Equispaced &
  NA &
  $\mathcal{L}_{\mathrm{ssim}} + \mathcal{L}_{1}$ &
  1 &
  1 &
  $\checkmark$ &
  NA &
  $0.968 \pm 0.015$ &
  $42.33 \pm 3.60$ &
  $0.009 \pm 0.005$ &
  $0.934 \pm 0.031$ &
  $38.32 \pm 3.59$ &
  $0.021 \pm 0.012$ &
  $0.884 \pm 0.050$ &
  $34.86 \pm 3.34$ &
  $0.045 \pm 0.021$ &
  35.10 \\
DEMONS &
  vSHARP &
  Equispaced &
  NA &
  $\mathcal{L}_{\mathrm{ssim}} + \mathcal{L}_{1}$ &
  1 &
  1 &
  $\checkmark$ &
  NA &
  $0.968 \pm 0.015$ &
  $42.33 \pm 3.60$ &
  $0.009 \pm 0.005$ &
  $0.934 \pm 0.031$ &
  $38.32 \pm 3.59$ &
  $0.021 \pm 0.012$ &
  $0.884 \pm 0.050$ &
  $34.86 \pm 3.34$ &
  $0.045 \pm 0.021$ &
  33.09 \\
Learned &
  vSHARP &
  Optimized &
  \crossmark &
  $\mathcal{L}_{\mathrm{ssim}} + \mathcal{L}_{1}$ &
  1 &
  1 &
  $\checkmark$ &
  $\checkmark$ &
  $0.929 \pm 0.034$ &
  $36.98 \pm 3.44$ &
  $0.028 \pm 0.014$ &
  $0.903 \pm 0.044$ &
  $34.97 \pm 3.30$ &
  $0.043 \pm 0.018$ &
  $0.881 \pm 0.051$ &
  $33.85 \pm 3.24$ &
  $0.055 \pm 0.021$ &
  12.23 \\ \hline
\end{tabular}
}
}
\caption{Reconstruction quantitative results for various configurations on the aorta inference set (under unified sampling settings).}
\label{tab:chapter8:appendix:metrics-unified-aorta-recon}
\end{table}

}

\clearpage

\subsection{Additional Figures}
\begin{figure}[!hbt]

    \centering
\begin{subfigure}[b]{0.48\textwidth}
        \centering
        \parbox[b]{0.35\columnwidth}{%
            \centering
            \includegraphics[height=0.075\textheight]{\thischapter/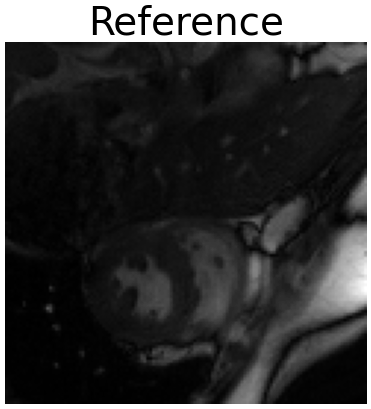}
            \vfill
            \includegraphics[height=0.12\textheight]{\thischapter/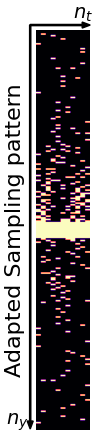}
        }%
        \hfill
        \raisebox{0pt}[0pt][0pt]{%
            \includegraphics[height=0.2\textheight]{\thischapter/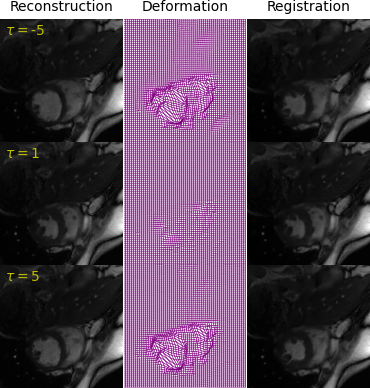}
        }
        \caption{Original setup.}
        \label{fig:chapter8:appendix:example_original_reg}
    \end{subfigure}
    \hfill

    \vspace{10pt} 
\begin{subfigure}[b]{0.48\textwidth}
        \centering
        \parbox[b]{0.35\columnwidth}{%
            \centering
            \includegraphics[height=0.075\textheight]{\thischapter/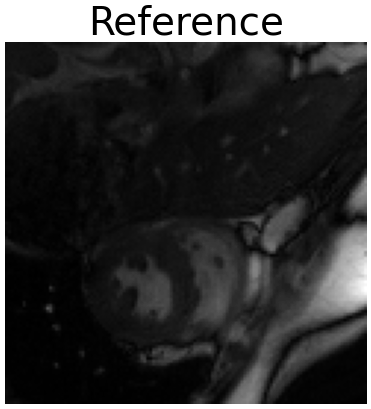}
            \vfill
            \includegraphics[height=0.12\textheight]{\thischapter/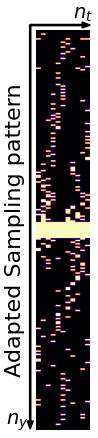}
        }%
        \hfill
        \raisebox{0pt}[0pt][0pt]{%
            \includegraphics[height=0.2\textheight]{\thischapter/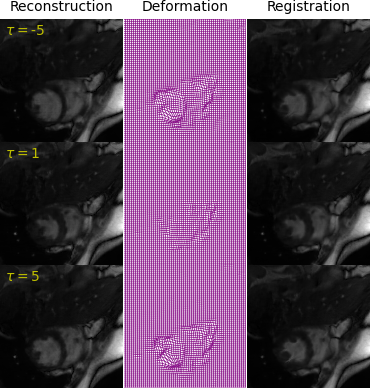}
        }
        \caption{Voxelmorph.}
        \label{fig:chapter8:appendix:example_voxel_reg}
    \end{subfigure}
    \hfill
    \begin{subfigure}[b]{0.48\textwidth}
        \centering
        \parbox[b]{0.35\columnwidth}{%
            \centering
            \includegraphics[height=0.075\textheight]{\thischapter/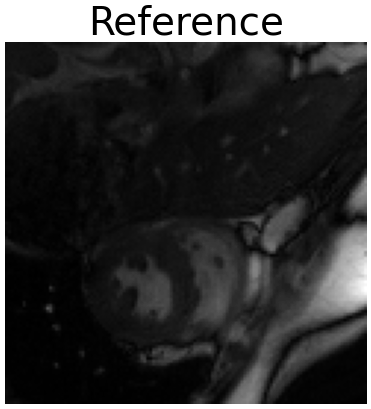}
            \vfill
            \includegraphics[height=0.12\textheight]{\thischapter/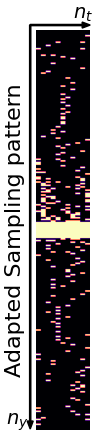}
        }%
        \hfill
        \raisebox{0pt}[0pt][0pt]{%
            \includegraphics[height=0.2\textheight]{\thischapter/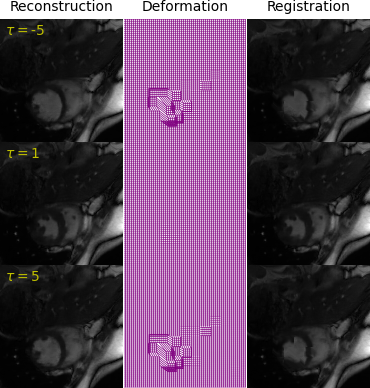}
        }
        \caption{Transmorph.}
        \label{fig:chapter8:appendix:example_vit_reg}
    \end{subfigure}

    \vspace{10pt} 
\begin{subfigure}[b]{0.48\textwidth}
        \centering
        \parbox[b]{0.35\columnwidth}{%
            \centering
            \includegraphics[height=0.075\textheight]{\thischapter/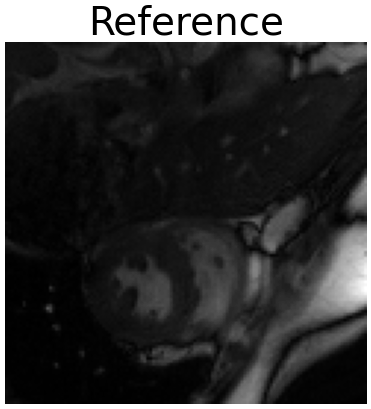}
            \vfill
            \includegraphics[height=0.12\textheight]{\thischapter/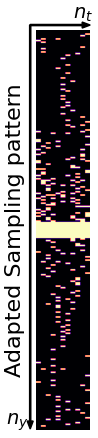}
        }%
        \hfill
        \raisebox{0pt}[0pt][0pt]{%
            \includegraphics[height=0.2\textheight]{\thischapter/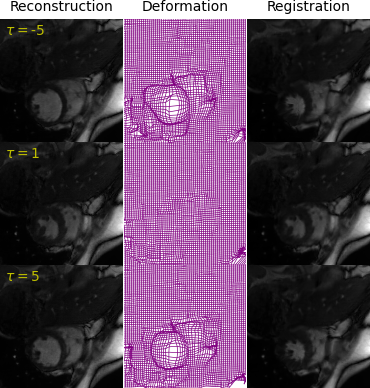}
        }
        \caption{Optical Flow ILK.}
        \label{fig:chapter8:appendix:example_ofilk_reg}
    \end{subfigure}
    \hfill
    \begin{subfigure}[b]{0.48\textwidth}
        \centering
        \parbox[b]{0.35\columnwidth}{%
            \centering
            \includegraphics[height=0.075\textheight]{\thischapter/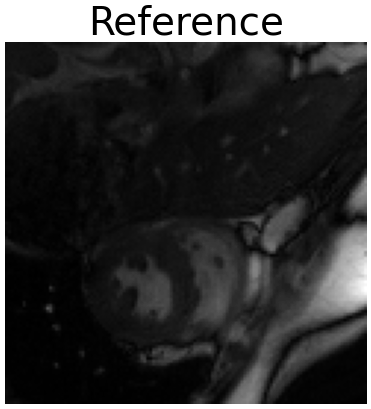}
            \vfill
            \includegraphics[height=0.12\textheight]{\thischapter/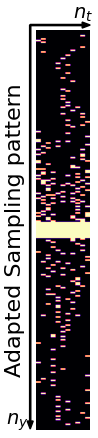}
        }%
        \hfill
        \raisebox{0pt}[0pt][0pt]{%
            \includegraphics[height=0.2\textheight]{\thischapter/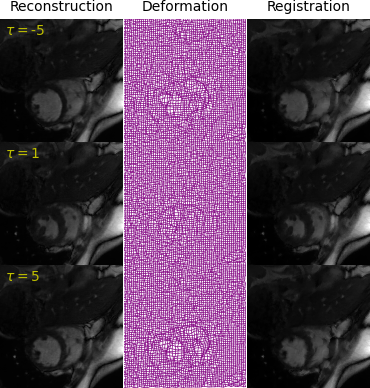}
        }
        \caption{DEMONS.}
        \label{fig:chapter8:appendix:example_demons_reg}
    \end{subfigure}
    
    \caption{Example results I for a cardiac cine case, shown at various temporal frames (\( \tau \)) relative to the reference image, for different choices of registration network at $R = 8$.}
    \label{fig:chapter8:appendix:example_reg1}
\end{figure}

\clearpage
\begin{figure}[!hbt]

    \centering
\begin{subfigure}[b]{0.48\textwidth}
        \centering
        \parbox[b]{0.35\columnwidth}{%
            \centering
            \includegraphics[height=0.075\textheight]{\thischapter/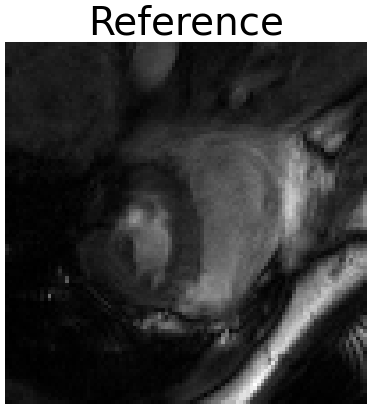}
            \vfill
            \includegraphics[height=0.12\textheight]{\thischapter/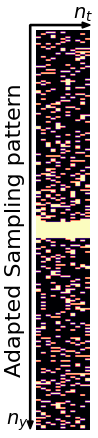}
        }%
        \hfill
        \raisebox{0pt}[0pt][0pt]{%
            \includegraphics[height=0.2\textheight]{\thischapter/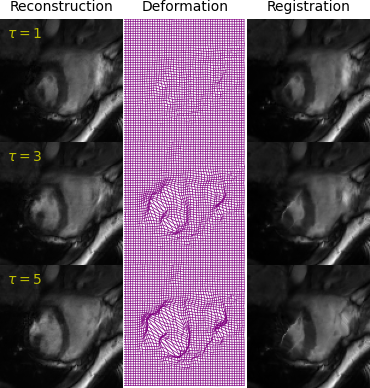}
        }
        \caption{Original setup.}
        \label{fig:chapter8:appendix:example_original_reg2}
    \end{subfigure}
    \hfill

    \vspace{10pt} 
\begin{subfigure}[b]{0.48\textwidth}
        \centering
        \parbox[b]{0.35\columnwidth}{%
            \centering
            \includegraphics[height=0.075\textheight]{\thischapter/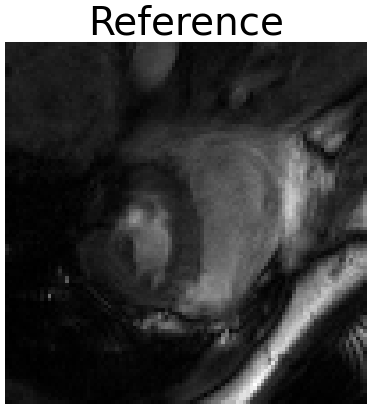}
            \vfill
            \includegraphics[height=0.12\textheight]{\thischapter/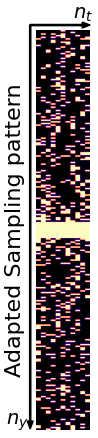}
        }%
        \hfill
        \raisebox{0pt}[0pt][0pt]{%
            \includegraphics[height=0.2\textheight]{\thischapter/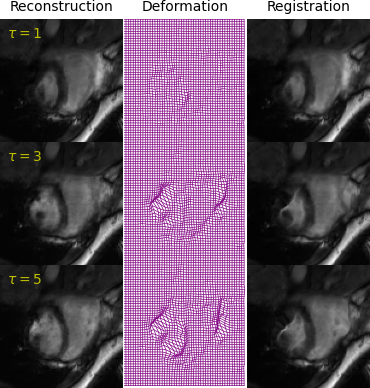}
        }
        \caption{Voxelmorph.}
        \label{fig:chapter8:appendix:example_voxel_reg2}
    \end{subfigure}
    \hfill
    \begin{subfigure}[b]{0.48\textwidth}
        \centering
        \parbox[b]{0.35\columnwidth}{%
            \centering
            \includegraphics[height=0.075\textheight]{\thischapter/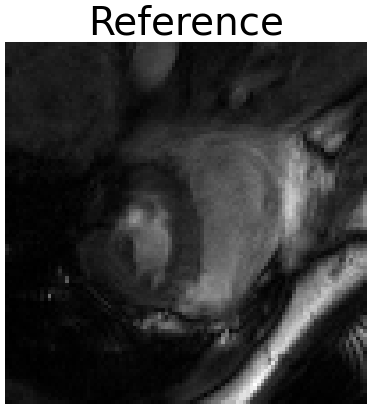}
            \vfill
            \includegraphics[height=0.12\textheight]{\thischapter/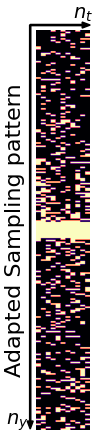}
        }%
        \hfill
        \raisebox{0pt}[0pt][0pt]{%
            \includegraphics[height=0.2\textheight]{\thischapter/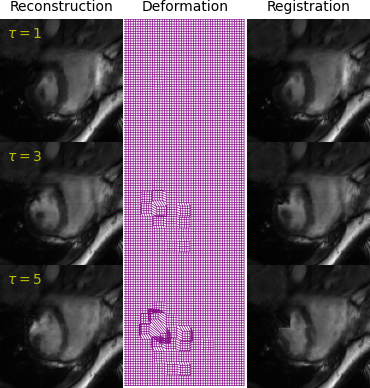}
        }
        \caption{Transmorph.}
        \label{fig:chapter8:appendix:example_vit_reg2}
    \end{subfigure}

    \vspace{10pt} 
\begin{subfigure}[b]{0.48\textwidth}
        \centering
        \parbox[b]{0.35\columnwidth}{%
            \centering
            \includegraphics[height=0.075\textheight]{\thischapter/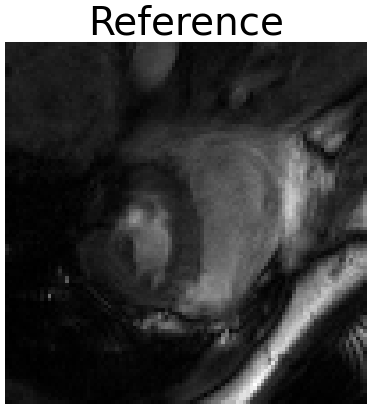}
            \vfill
            \includegraphics[height=0.12\textheight]{\thischapter/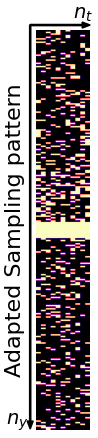}
        }%
        \hfill
        \raisebox{0pt}[0pt][0pt]{%
            \includegraphics[height=0.2\textheight]{\thischapter/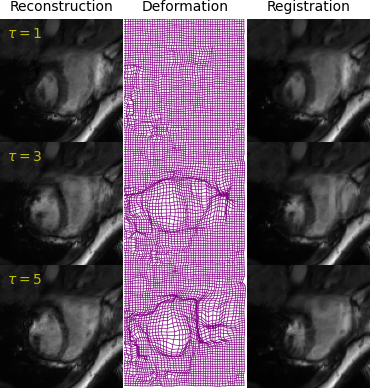}
        }
        \caption{Oprical Flow ILK.}
        \label{fig:chapter8:appendix:example_ofilk_reg2}
    \end{subfigure}
    \hfill
    \begin{subfigure}[b]{0.48\textwidth}
        \centering
        \parbox[b]{0.35\columnwidth}{%
            \centering
            \includegraphics[height=0.075\textheight]{\thischapter/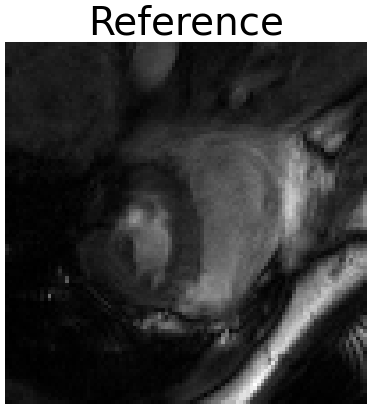}
            \vfill
            \includegraphics[height=0.12\textheight]{\thischapter/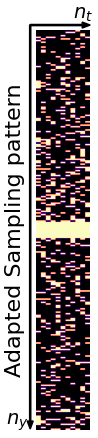}
        }%
        \hfill
        \raisebox{0pt}[0pt][0pt]{%
            \includegraphics[height=0.2\textheight]{\thischapter/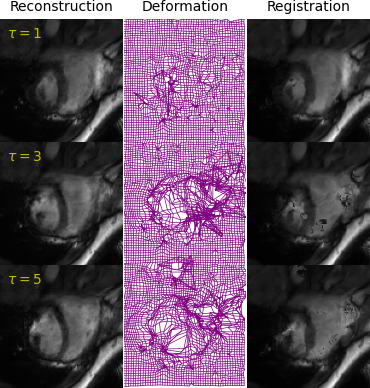}
        }
        \caption{Optical Flow TVL$_1$.}
        \label{fig:chapter8:appendix:example_oftvl1_reg2}
    \end{subfigure}
    
    \caption{Example results II for a cardiac cine case, shown at various temporal frames (\( \tau \)) relative to the reference image, for different choices of registration network at $R = 4$.}
    \label{fig:chapter8:appendix:example_reg2}
\end{figure}

\clearpage
\begin{figure}[!bht]
    \centering
        \vspace{50pt}
        \begin{subfigure}[b]{0.48\textwidth}
        \centering
        \parbox[b]{0.35\columnwidth}{%
            \centering
            \includegraphics[height=0.12\textheight]{\thischapter/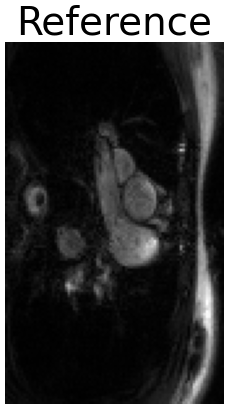}
            \vfill
            \includegraphics[height=0.12\textheight]{\thischapter/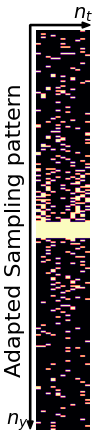}
        }%
        \hfill
        \raisebox{0pt}[0pt][0pt]{%
            \includegraphics[height=0.3\textheight]{\thischapter/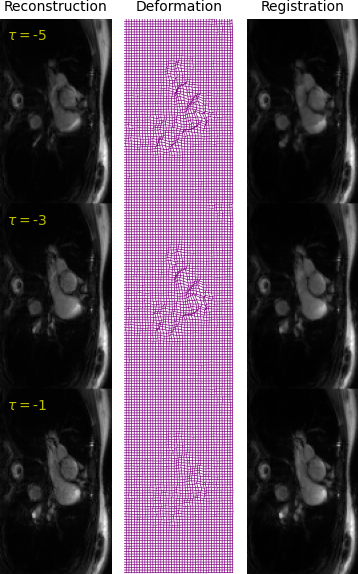}
        }
        \caption{Original setup.}
        \label{fig:chapter8:appendix:example_original_reg3}
    \end{subfigure}
    \hfill
    \vspace{40pt} 
\begin{subfigure}[b]{0.48\textwidth}
        \centering
        \parbox[b]{0.35\columnwidth}{%
            \centering
            \includegraphics[height=0.12\textheight]{\thischapter/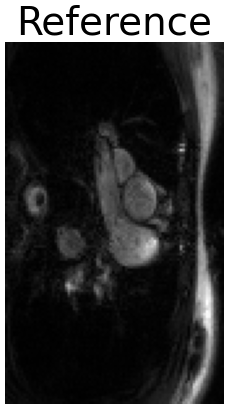}
            \vfill
            \includegraphics[height=0.12\textheight]{\thischapter/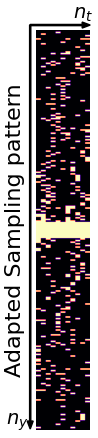}
        }%
        \hfill
        \raisebox{0pt}[0pt][0pt]{%
            \includegraphics[height=0.3\textheight]{\thischapter/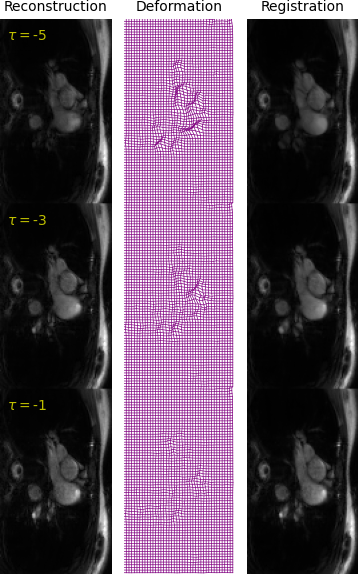}
        }
        \caption{Voxelmorph.}
        \label{fig:chapter8:appendix:example_voxel_reg3}
    \end{subfigure}
    \hfill
    \begin{subfigure}[b]{0.48\textwidth}
        \centering
        \parbox[b]{0.35\columnwidth}{%
            \centering
            \includegraphics[height=0.12\textheight]{\thischapter/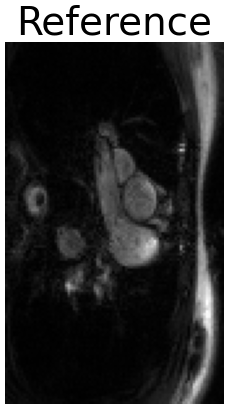}
            \vfill
            \includegraphics[height=0.12\textheight]{\thischapter/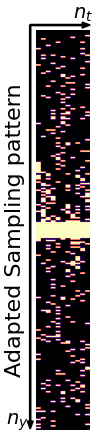}
        }%
        \hfill
        \raisebox{0pt}[0pt][0pt]{%
            \includegraphics[height=0.3\textheight]{\thischapter/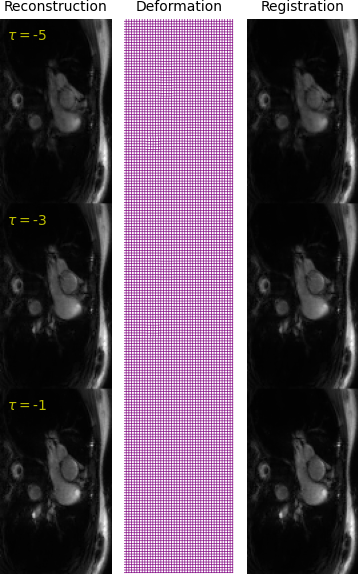}
        }
        \caption{Transmorph.}
        \label{fig:chapter8:appendix:example_vit_reg3}
    \end{subfigure}
    
    \caption{Example results III for an aorta case, shown at various temporal frames (\( \tau \)) relative to the reference image, for different choices of registration network at $R = 6$.}
    \label{fig:chapter8:appendix:example_reg3}
\end{figure}

\clearpage
\begin{figure}[ht]

    \centering
\begin{subfigure}[b]{0.48\textwidth}
        \centering
        \parbox[b]{0.35\columnwidth}{%
            \centering
            \includegraphics[height=0.075\textheight]{\thischapter/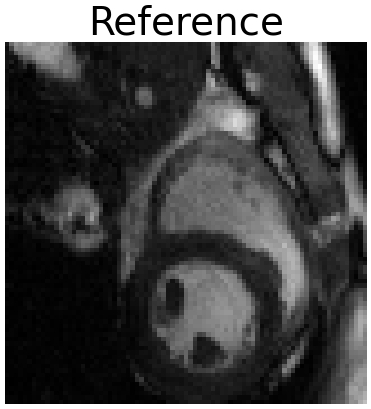}
            \vfill
            \includegraphics[height=0.12\textheight]{\thischapter/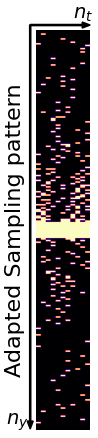}
        }%
        \hfill
        \raisebox{0pt}[0pt][0pt]{%
            \includegraphics[height=0.2\textheight]{\thischapter/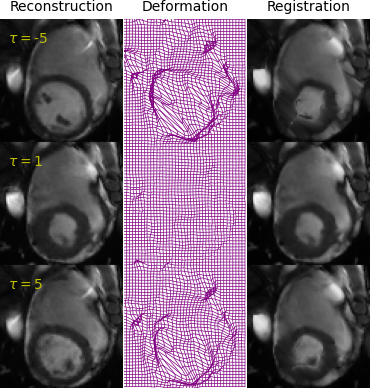}
        }
        \caption{Case I (cine): vSHARP $R = 8$.}
        \label{fig:chapter8:appendix:example_vsharp_2}
    \end{subfigure}
    \hfill
    \begin{subfigure}[b]{0.48\textwidth}
        \centering
        \parbox[b]{0.35\columnwidth}{%
            \centering
            \includegraphics[height=0.075\textheight]{\thischapter/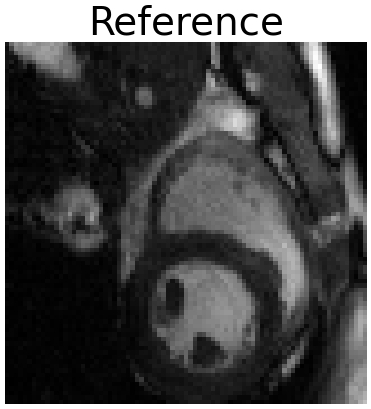}
            \vfill
            \includegraphics[height=0.12\textheight]{\thischapter/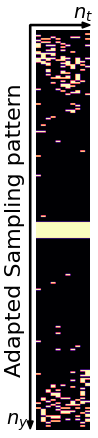}
        }%
        \hfill
        \raisebox{0pt}[0pt][0pt]{%
            \includegraphics[height=0.2\textheight]{\thischapter/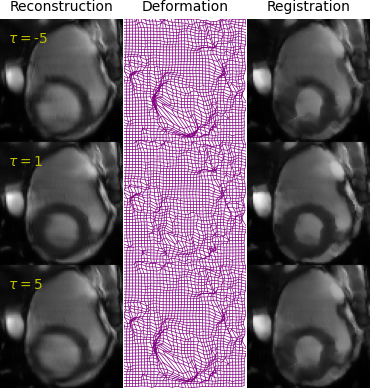}
        }
        \caption{Case I (cine): VarNet, $R = 8$.}
        \label{fig:chapter8:appendix:example_varnet_2}
    \end{subfigure}

    \vspace{10pt} 
    \begin{subfigure}[b]{0.48\textwidth}
        \centering
        \parbox[b]{0.35\columnwidth}{%
            \centering
            \includegraphics[height=0.075\textheight]{\thischapter/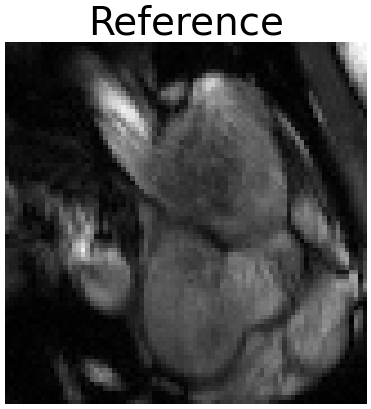}
            \vfill
            \includegraphics[height=0.12\textheight]{\thischapter/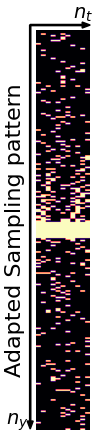}
        }%
        \hfill
        \raisebox{0pt}[0pt][0pt]{%
            \includegraphics[height=0.2\textheight]{\thischapter/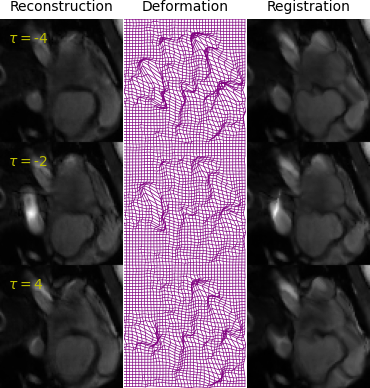}
        }
        \caption{Case II (cine): vSHARP, $R = 6$.}
        \label{fig:chapter8:appendix:example_vsharp}
    \end{subfigure}
    \hfill
    \begin{subfigure}[b]{0.48\textwidth}
        \centering
        \parbox[b]{0.35\columnwidth}{%
            \centering
            \includegraphics[height=0.075\textheight]{\thischapter/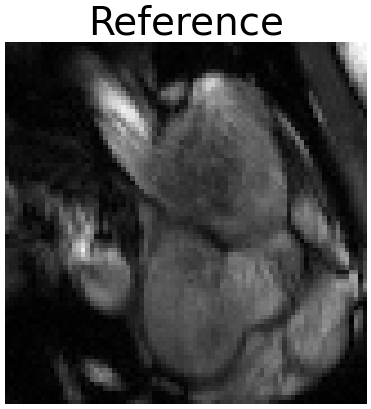}
            \vfill
            \includegraphics[height=0.12\textheight]{\thischapter/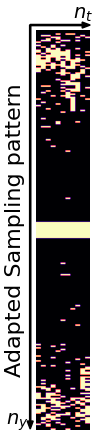}
        }%
        \hfill
        \raisebox{0pt}[0pt][0pt]{%
            \includegraphics[height=0.2\textheight]{\thischapter/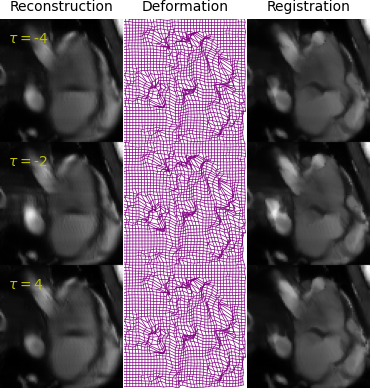}
        }
        \caption{Case II (cine): VarNet, $R = 6$.}
        \label{fig:chapter8:appendix:example_varnet}
    \end{subfigure}
    
    \vspace{10pt} 

    \begin{subfigure}[b]{0.48\textwidth}
        \centering
        \parbox[b]{0.35\columnwidth}{%
            \centering
            \includegraphics[height=0.075\textheight]{\thischapter/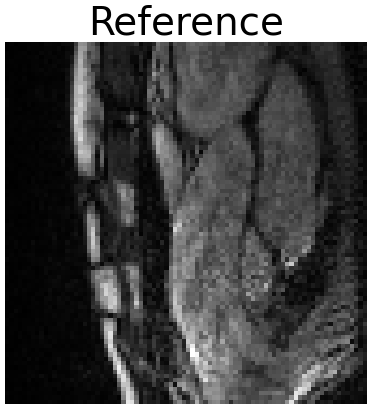}
            \vfill
            \includegraphics[height=0.12\textheight]{\thischapter/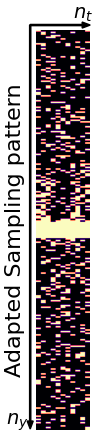}
        }%
        \hfill
        \raisebox{0pt}[0pt][0pt]{%
            \includegraphics[height=0.2\textheight]{\thischapter/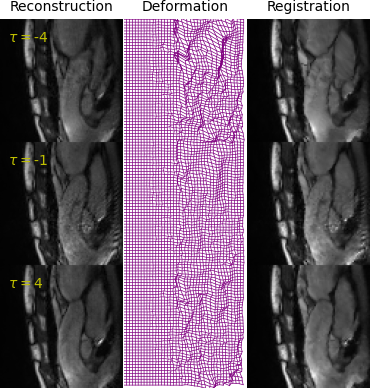}
        }
        \caption{Case III (aorta): vSHARP, $R = 4$.}
        \label{fig:chapter8:appendix:example_vsharp_aorta}
    \end{subfigure}
    \hfill
    \begin{subfigure}[b]{0.48\textwidth}
        \centering
        \parbox[b]{0.35\columnwidth}{%
            \centering
            \includegraphics[height=0.075\textheight]{\thischapter/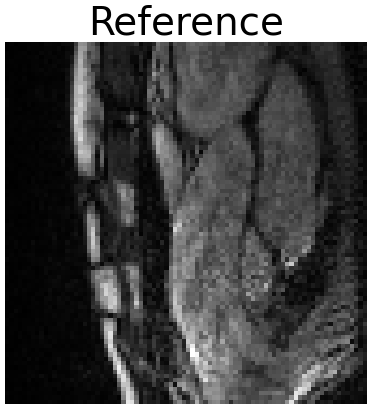}
            \vfill
            \includegraphics[height=0.12\textheight]{\thischapter/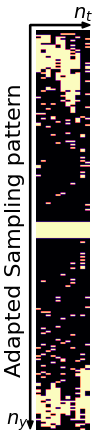}
        }%
        \hfill
        \raisebox{0pt}[0pt][0pt]{%
            \includegraphics[height=0.2\textheight]{\thischapter/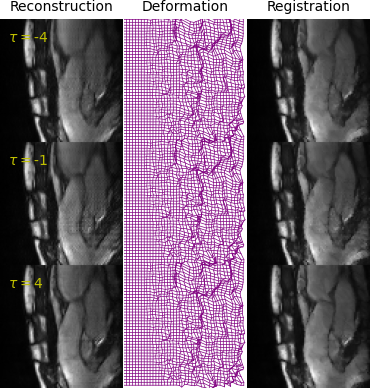}
        }
        \caption{Case III (aorta): VarNet, $R = 4$.}
        \label{fig:chapter8:appendix:example_varnet_aorta}
    \end{subfigure}
    
    \caption{Example results for three cases, for shown at various temporal frames (\( \tau \)) relative to the reference image, for different choices of reconstruction network at different accelerations.}
    \label{fig:chapter8:appendix:example_recon}
\end{figure}

\clearpage

\begin{figure}[ht]

    \centering
\begin{subfigure}[b]{0.48\textwidth}
        \centering
        \parbox[b]{0.35\columnwidth}{%
            \centering
            \includegraphics[height=0.075\textheight]{\thischapter/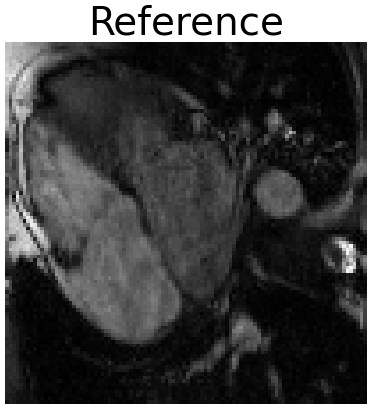}
            \vfill
            \includegraphics[height=0.12\textheight]{\thischapter/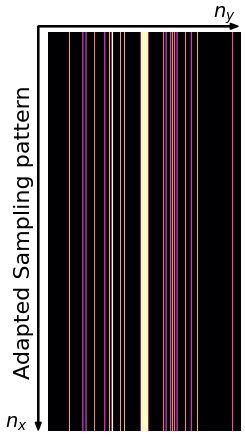}
        }%
        \hfill
        \raisebox{0pt}[0pt][0pt]{%
            \includegraphics[height=0.2\textheight]{\thischapter/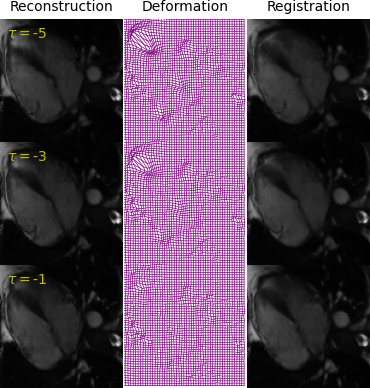}
        }
        \caption{Unified sampling.}
        \label{fig:chapter8:appendix:example_original_samp1}
    \end{subfigure}
    \hfill
    \begin{subfigure}[b]{0.48\textwidth}
        \centering
        \parbox[b]{0.35\columnwidth}{%
            \centering
            \includegraphics[height=0.075\textheight]{\thischapter/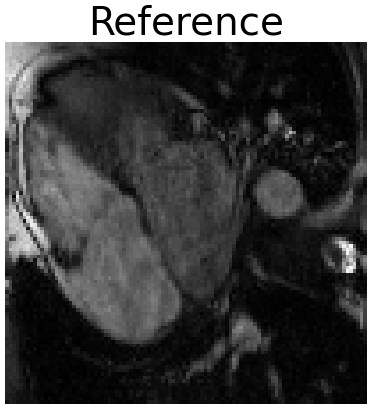}
            \vfill
            \includegraphics[height=0.12\textheight]{\thischapter/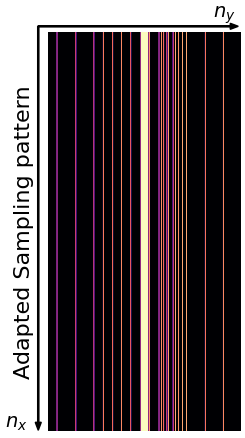}
        }%
        \hfill
        \raisebox{0pt}[0pt][0pt]{%
            \includegraphics[height=0.2\textheight]{\thischapter/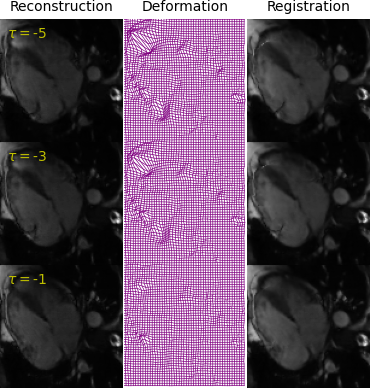}
        }
        \caption{Unified sampling with initialization.}
        \label{fig:chapter8:appendix:example_original_samp_init1}
    \end{subfigure}

    \vspace{10pt} 
\begin{subfigure}[b]{0.48\textwidth}
        \centering
        \parbox[b]{0.35\columnwidth}{%
            \centering
            \includegraphics[height=0.075\textheight]{\thischapter/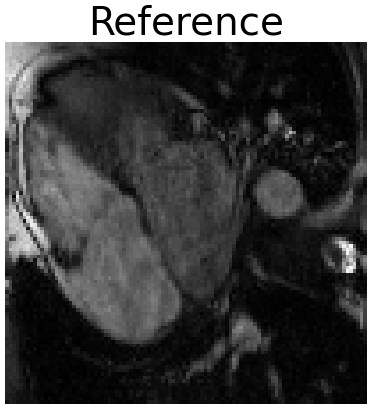}
            \vfill
            \includegraphics[height=0.12\textheight]{\thischapter/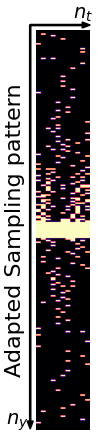}
        }%
        \hfill
        \raisebox{0pt}[0pt][0pt]{%
            \includegraphics[height=0.2\textheight]{\thischapter/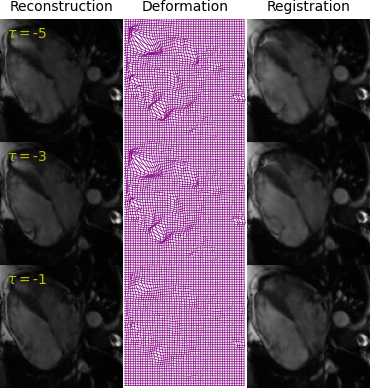}
        }
        \caption{Phase-specific sampling.}
        \label{fig:chapter8:appendix:example_original_dyn_samp1}
    \end{subfigure}
    \hfill
    \begin{subfigure}[b]{0.48\textwidth}
        \centering
        \parbox[b]{0.35\columnwidth}{%
            \centering
            \includegraphics[height=0.075\textheight]{\thischapter/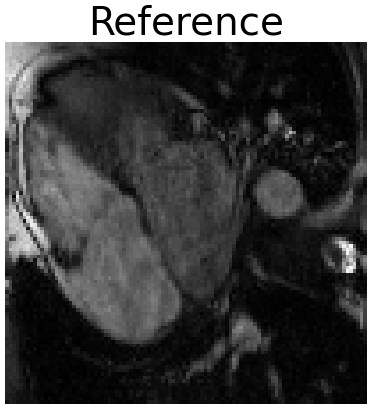}
            \vfill
            \includegraphics[height=0.12\textheight]{\thischapter/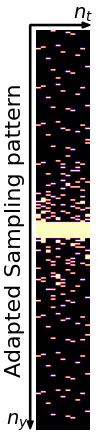}
        }%
        \hfill
        \raisebox{0pt}[0pt][0pt]{%
            \includegraphics[height=0.2\textheight]{\thischapter/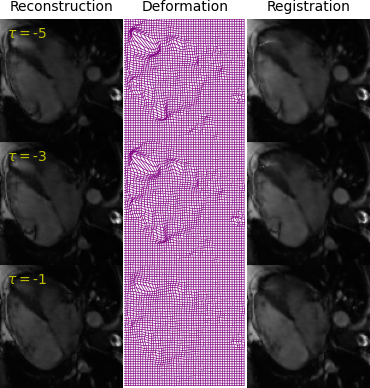}
        }
        \caption{Phase-specific sampling with initialization.}
        \label{fig:chapter8:appendix:example_original_samp_dyn_init1}
    \end{subfigure}

    \caption{Example results I for a cardiac cine case, shown at various temporal frames (\( \tau \)) relative to the reference image, for different choices of adaptive sampling at $R = 8$.}
    \label{fig:chapter8:appendix:example_samp1}
\end{figure}

\clearpage
\begin{figure}[ht]

    \centering
\begin{subfigure}[b]{0.48\textwidth}
        \centering
        \parbox[b]{0.35\columnwidth}{%
            \centering
            \includegraphics[height=0.075\textheight]{\thischapter/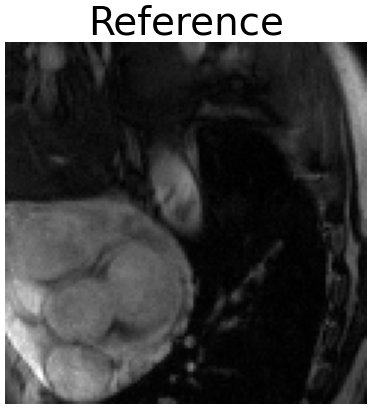}
            \vfill
            \includegraphics[height=0.12\textheight]{\thischapter/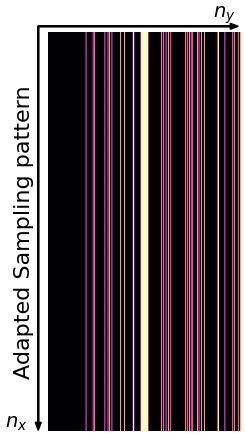}
        }%
        \hfill
        \raisebox{0pt}[0pt][0pt]{%
            \includegraphics[height=0.2\textheight]{\thischapter/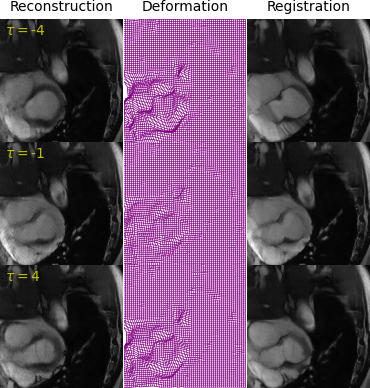}
        }
        \caption{Unified sampling.}
        \label{fig:chapter8:appendix:example_original_samp2}
    \end{subfigure}
    \hfill
    \begin{subfigure}[b]{0.48\textwidth}
        \centering
        \parbox[b]{0.35\columnwidth}{%
            \centering
            \includegraphics[height=0.075\textheight]{\thischapter/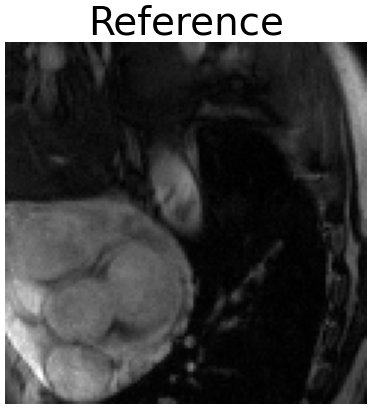}
            \vfill
            \includegraphics[height=0.12\textheight]{\thischapter/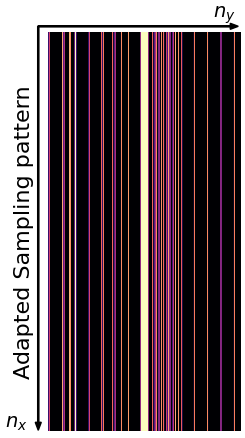}
        }%
        \hfill
        \raisebox{0pt}[0pt][0pt]{%
            \includegraphics[height=0.2\textheight]{\thischapter/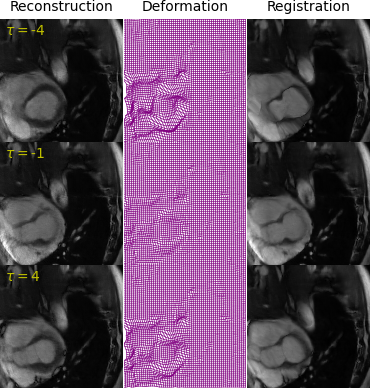}
        }
        \caption{Unified sampling with initialization.}
        \label{fig:chapter8:appendix:example_original_samp_init2}
    \end{subfigure}

    \vspace{10pt} 
\begin{subfigure}[b]{0.48\textwidth}
        \centering
        \parbox[b]{0.35\columnwidth}{%
            \centering
            \includegraphics[height=0.075\textheight]{\thischapter/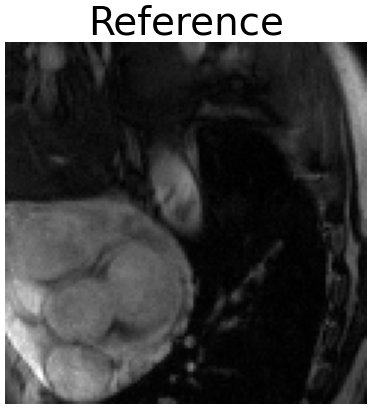}
            \vfill
            \includegraphics[height=0.12\textheight]{\thischapter/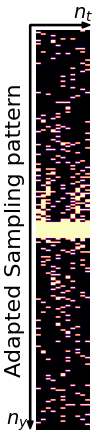}
        }%
        \hfill
        \raisebox{0pt}[0pt][0pt]{%
            \includegraphics[height=0.2\textheight]{\thischapter/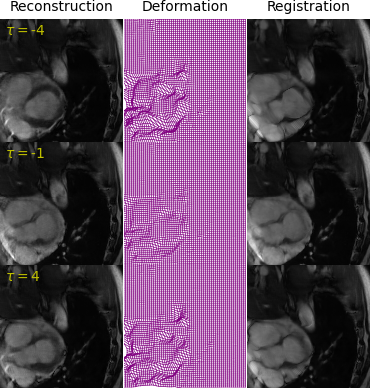}
        }
        \caption{Phase-specific sampling.}
        \label{fig:chapter8:appendix:example_original_dyn_samp2}
    \end{subfigure}
    \hfill
    \begin{subfigure}[b]{0.48\textwidth}
        \centering
        \parbox[b]{0.35\columnwidth}{%
            \centering
            \includegraphics[height=0.075\textheight]{\thischapter/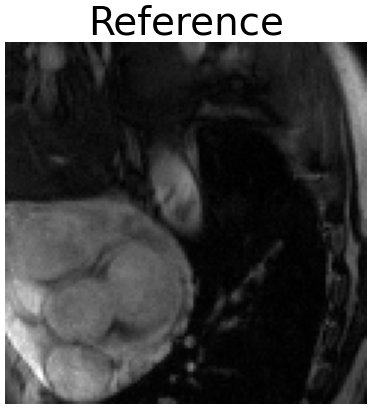}
            \vfill
            \includegraphics[height=0.12\textheight]{\thischapter/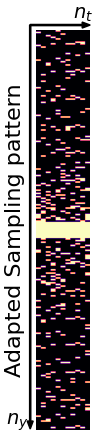}
        }%
        \hfill
        \raisebox{0pt}[0pt][0pt]{%
            \includegraphics[height=0.2\textheight]{\thischapter/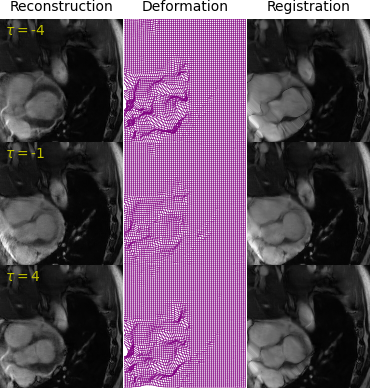}
        }
        \caption{Phase-specific sampling with initialization.}
        \label{fig:chapter8:appendix:example_original_samp_dyn_init2}
    \end{subfigure}

    \caption{Example results II for a cardiac cine case, shown at various temporal frames (\( \tau \)) relative to the reference image, for different choices of adaptive sampling at $R = 6$.}
    \label{fig:chapter8:appendix:example_samp2}
\end{figure}

\clearpage

\clearpage

\clearpage
\begin{figure}[ht]

    \centering

    \begin{subfigure}[b]{0.48\textwidth}
        \centering
        \parbox[b]{0.35\columnwidth}{%
            \centering
            \includegraphics[height=0.075\textheight]{\thischapter/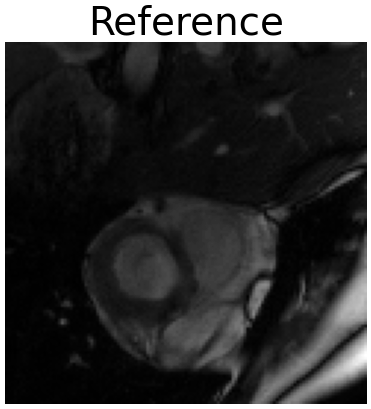}
            \vfill
            \includegraphics[height=0.12\textheight]{\thischapter/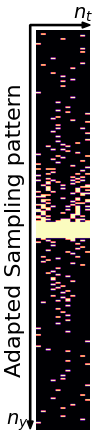}
        }%
        \hfill
        \raisebox{0pt}[0pt][0pt]{%
            \includegraphics[height=0.2\textheight]{\thischapter/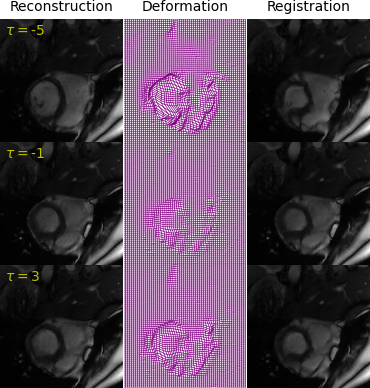}
        }
        \caption{Joint training.}
        \label{fig:chapter8:appendix:example_joint}
    \end{subfigure}
    \hfill
    \begin{subfigure}[b]{0.48\textwidth}
        \centering
        \parbox[b]{0.35\columnwidth}{%
            \centering
            \includegraphics[height=0.075\textheight]{\thischapter/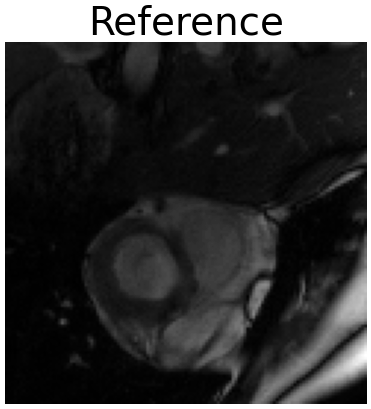}
            \vfill
            \includegraphics[height=0.12\textheight]{\thischapter/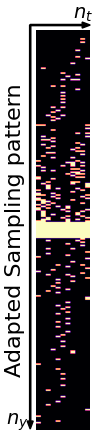}
        }%
        \hfill
        \raisebox{0pt}[0pt][0pt]{%
            \includegraphics[height=0.2\textheight]{\thischapter/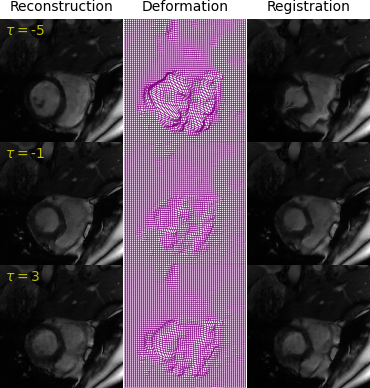}
        }
        \caption{Decoupled training}
        \label{fig:chapter8:appendix:example_dec}
    \end{subfigure}
    \vspace{10pt} 

    \begin{subfigure}[b]{0.48\textwidth}
        \centering
        \parbox[b]{0.35\columnwidth}{%
            \centering
            \includegraphics[height=0.075\textheight]{\thischapter/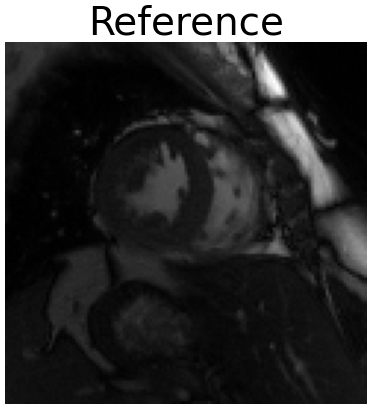}
            \vfill
            \includegraphics[height=0.12\textheight]{\thischapter/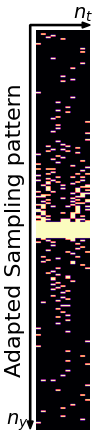}
        }%
        \hfill
        \raisebox{0pt}[0pt][0pt]{%
            \includegraphics[height=0.2\textheight]{\thischapter/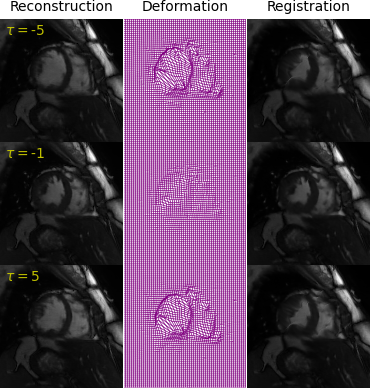}
        }
        \caption{Joint training.}
        \label{fig:chapter8:appendix:example_joint2}
    \end{subfigure}
    \hfill
    \begin{subfigure}[b]{0.48\textwidth}
        \centering
        \parbox[b]{0.35\columnwidth}{%
            \centering
            \includegraphics[height=0.075\textheight]{\thischapter/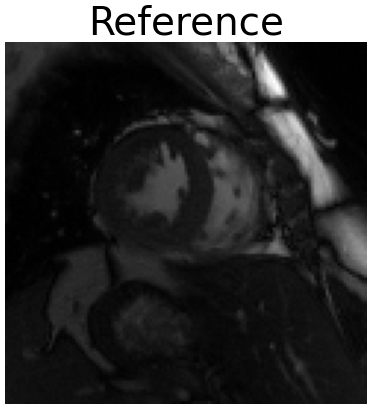}
            \vfill
            \includegraphics[height=0.12\textheight]{\thischapter/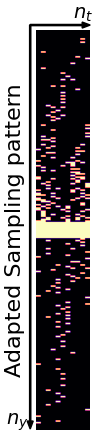}
        }%
        \hfill
        \raisebox{0pt}[0pt][0pt]{%
            \includegraphics[height=0.2\textheight]{\thischapter/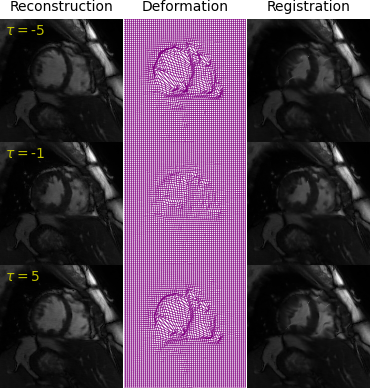}
        }
        \caption{Decoupled training}
        \label{fig:chapter8:appendix:example_dec2}
    \end{subfigure}

    \caption{Example results for two cases, shown at various temporal frames (\( \tau \)) relative to the reference image, comparing joint vs decoupled training setups at $R = 8$.}
    \label{fig:chapter8:appendix:example_jointdecoupled}
\end{figure}

\clearpage
\begin{figure}[ht]

    \centering

    \begin{subfigure}[b]{0.48\textwidth}
        \centering
        \parbox[b]{0.35\columnwidth}{%
            \centering
            \includegraphics[height=0.075\textheight]{\thischapter/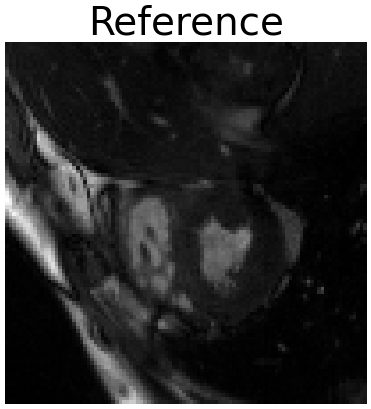}
            \vfill
            \includegraphics[height=0.12\textheight]{\thischapter/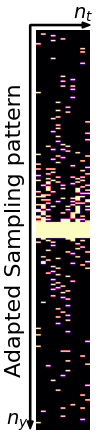}
        }%
        \hfill
        \raisebox{0pt}[0pt][0pt]{%
            \includegraphics[height=0.2\textheight]{\thischapter/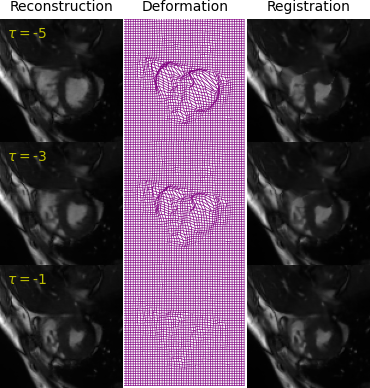}
        }
        \caption{All loses.}
        \label{fig:chapter8:appendix:example_all_losses}
    \end{subfigure}
    \hfill
    \begin{subfigure}[b]{0.48\textwidth}
        \centering
        \parbox[b]{0.35\columnwidth}{%
            \centering
            \includegraphics[height=0.075\textheight]{\thischapter/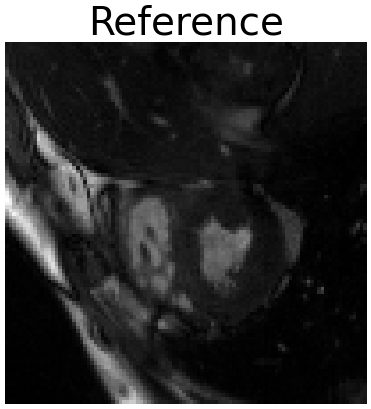}
            \vfill
            \includegraphics[height=0.12\textheight]{\thischapter/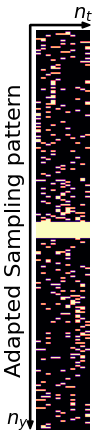}
        }%
        \hfill
        \raisebox{0pt}[0pt][0pt]{%
            \includegraphics[height=0.2\textheight]{\thischapter/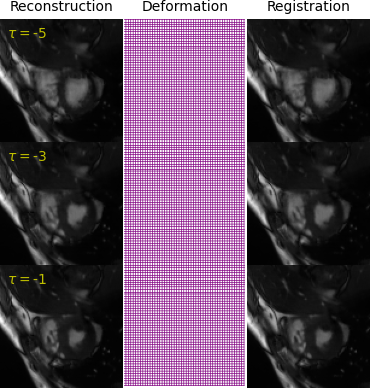}
        }
        \caption{\(\mathcal{L}_{\text{reg}} = \mathcal{L}_{\text{sim}} = \mathcal{L}_{\text{ssim2D}}\)}
        \label{fig:chapter8:appendix:example_ssim_only}
    \end{subfigure}
    \vspace{10pt} 

    \begin{subfigure}[b]{0.48\textwidth}
        \centering
        \parbox[b]{0.35\columnwidth}{%
            \centering
            \includegraphics[height=0.075\textheight]{\thischapter/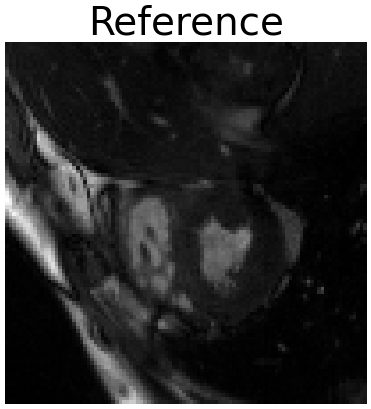}
            \vfill
            \includegraphics[height=0.12\textheight]{\thischapter/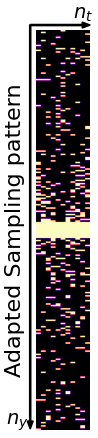}
        }%
        \hfill
        \raisebox{0pt}[0pt][0pt]{%
            \includegraphics[height=0.2\textheight]{\thischapter/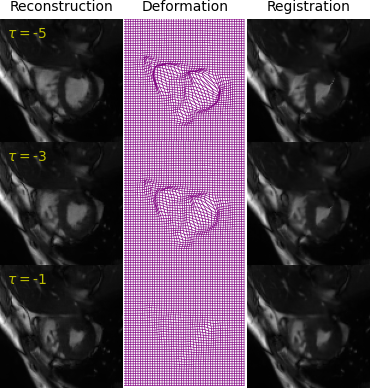}
        }
        \caption{\(\mathcal{L}_{\text{reg}} = \mathcal{L}_{\text{sim}} = \mathcal{L}_{1}\)}
        \label{fig:chapter8:appendix:example_l1}
    \end{subfigure}
    \hfill
    \begin{subfigure}[b]{0.48\textwidth}
        \centering
        \parbox[b]{0.35\columnwidth}{%
            \centering
            \includegraphics[height=0.075\textheight]{\thischapter/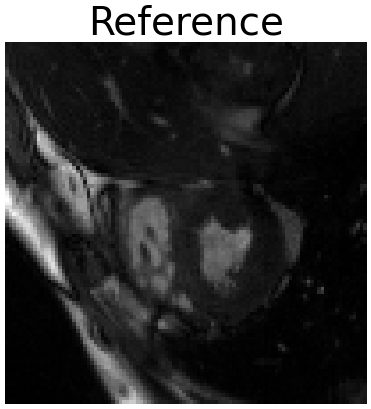}
            \vfill
            \includegraphics[height=0.12\textheight]{\thischapter/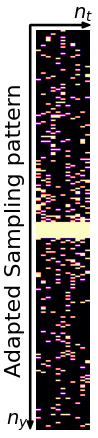}
        }%
        \hfill
        \raisebox{0pt}[0pt][0pt]{%
            \includegraphics[height=0.2\textheight]{\thischapter/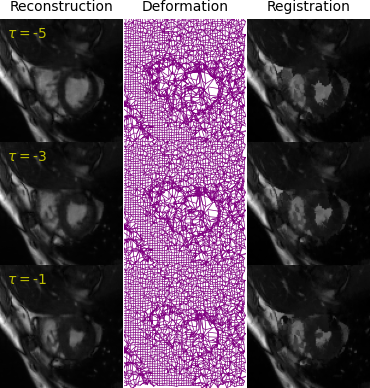}
        }
        \caption{No smooth loss.}
        \label{fig:chapter8:appendix:example_no_smooth}
    \end{subfigure}

    \caption{Example results for a case, shown at various temporal frames (\( \tau \)) relative to the reference image, comparing different loss choice setups at $R = 8$.}
    \label{fig:chapter8:appendix:example_loss_fun}
\end{figure}

\end{document}